\documentclass[a4paper,11pt]{article}

\usepackage[utf8]{inputenc}
\usepackage{jheppub}
\usepackage{amssymb, latexsym, amsmath, amsthm}
\usepackage{mathtools}
\usepackage{gensymb}
\usepackage{graphicx}
\usepackage{array}
\usepackage{rotating,float,caption}
\usepackage{multirow}
\usepackage{braket}
\usepackage{bm}

\usepackage{color}

\definecolor{mygreen}{rgb}{0.0,0.5,0.0}

\DeclareMathAlphabet\mathbfcal{OMS}{cmsy}{b}{n}

\newcommand{\ampnum}[2]{\mathcal{A}_{#1}^{#2}}
\newcommand{\ampnumstraight}[2]{A_{#1}^{#2}}
\newcommand{\anum}[2]{a_{#1}^{#2}}
\newcommand{\me}[2]{\mathcal{M}_{#1}^{#2}}
\newcommand{\ketampdep}[3]{\ket{\ampnum{#1}{#2}(#3)}}
\newcommand{\braampdep}[3]{\bra{\ampnum{#1}{#2}(#3)}}
\newcommand{\ketamp}[2]{\ket{\ampnum{#1}{#2}}}
\newcommand{\braamp}[2]{\bra{\ampnum{#1}{#2}}}
\newcommand{\Iop}[2]{\bm{I}^{(#1)}\left(#2\right)}
\newcommand{\Iopd}[2]{\bm{I}^{(#1),\dagger}\left(#2\right)}
\newcommand{\Hop}[2]{\bm{H}^{(#1)}\left(#2\right)}
\newcommand{\Hopd}[2]{\bm{H}^{(#1),\dagger}\left(#2\right)}

\newcommand{\ourIop}[3]{\mathcal{I}^{(#1)}_{#2}\left(#3\right)}

\newcommand{\ourHopnodep}[2]{\mathcal{H}^{(#1)}_{#2}}
\newcommand{\ourHop}[3]{\mathcal{H}^{(#1)}_{#2}\left(#3\right)}

\newcommand{\Itwoepsiloncoeff}{\text{e}^{-\epsilon\gamma_E}\dfrac{\Gamma(1-2\epsilon)}{\Gamma(1-\epsilon)}\left(\betaoe+K\right)}

\newcommand{\Jcol}[1]{\mathbfcal{J}^{(#1)}}

\newcommand{\J}[1]{\mathcal{J}_{2}^{(#1)}}

\newcommand{\XFFint}[3]{\mathcal{#1}_{#2}^{#3}}
\newcommand{\XIFint}[4]{\mathcal{#1}_{#2,#4}^{#3}}
\newcommand{\XIIint}[4]{\mathcal{#1}_{#2,#4}^{#3}}

\newcommand{\Gammaone}[2]{\Gamma^{(1)}_{#1}\left(#2\right)}
\newcommand{\BGammaone}[2]{\mathbf{\Gamma}^{(1)}_{#1}\left(#2\right)}
\newcommand{\BGammaonenodep}[1]{\mathbf{\Gamma}^{(1)}_{#1}}

\newcommand{\Gammatwo}[2]{\overline{\Gamma}^{(2)}_{#1}\left(#2\right)}
\newcommand{\BGammatwo}[2]{\overline{\mathbf{\Gamma}}^{(2)}_{#1}\left(#2\right)}

\newcommand{\deltaone}{\delta_1}
\newcommand{\deltatwo}{\delta_2}

\newcommand{\betaoe}{\dfrac{\beta_0}{\epsilon}}
\newcommand{\boe}{\dfrac{b_0}{\epsilon}}

\newcommand{\QQs}[1]{\left(\dfrac{|s_{#1}|}{\mu_r^2}\right)^{-\epsilon}}

\newcommand{\as}{\alpha_{s}}
\newcommand{\e}{\epsilon}
\newcommand{\lb}{\left\lbrace}
\newcommand{\rb}{\right\rbrace}
\newcommand{\pset}{\lb p \rb_n}

\newcommand{\dd}{\text{d}}
\newcommand{\poles}{\mathcal{P}oles}
\newcommand{\T}{\bm{T}}
\newcommand{\dr}[1]{{\dfrac{\dd #1}{#1}}}
\newcommand{\sigpart}[2]{{\hat{\sigma}^{#1}_{#2}}}
\newcommand{\ceps}{\overline{C}(\epsilon)}
\newcommand{\coeff}{\left(\dfrac{\as\ceps}{2\pi}\right)}
\newcommand{\coeffLO}{\mathcal{N}_{LO}}
\newcommand{\coeffVNLO}{\mathcal{N}^{V}_{NLO}}
\newcommand{\coeffRNLO}{\mathcal{N}^{R}_{NLO}}
\newcommand{\coeffVVNNLO}{\mathcal{N}^{VV}_{NNLO}}
\newcommand{\coeffRVNNLO}{\mathcal{N}^{RV}_{NNLO}}
\newcommand{\coeffRRNNLO}{\mathcal{N}^{RR}_{NNLO}}
\newcommand{\jet}[3]{J_{#1}^{(#2)}(#3)}
\newcommand{\dphi}[1]{\dd\Phi_{#1}}
\newcommand{\wt}[1]{\widetilde{#1}}

\newcommand{\dsigNNLO}[1]{\dd\sigpart{}{#1,NNLO}}

\newcommand{\dsigV}[1]{\dd\sigpart{V}{#1,NLO}}
\newcommand{\dsigR}[1]{\dd\sigpart{R}{#1,NLO}}
\newcommand{\dsigTNLO}[1]{\dd\sigpart{T}{#1,NLO}}
\newcommand{\dsigSNLO}[1]{\dd\sigpart{S}{#1,NLO}}

\newcommand{\dsigVV}[1]{\dd\sigpart{VV}{#1,NNLO}}
\newcommand{\dsigRV}[1]{\dd\sigpart{RV}{#1,NNLO}}
\newcommand{\dsigRR}[1]{\dd\sigpart{RR}{#1,NNLO}}
\newcommand{\dsigUNNLO}[1]{\dd\sigpart{U}{#1,NNLO}}
\newcommand{\dsigTNNLO}[1]{\dd\sigpart{T}{#1,NNLO}}
\newcommand{\dsigSNNLO}[1]{\dd\sigpart{S}{#1,NNLO}}
\newcommand{\dsigSNNLOspe}[2]{\dd\sigpart{S,#2}{#1,NNLO}}
\newcommand{\dsigVSNNLO}[1]{\dd\sigpart{VS}{#1,NNLO}}
\newcommand{\dsigUNNLOshort}[1]{\dd\sigpart{U,#1}{}}
\newcommand{\dsigTNNLOshort}[1]{\dd\sigpart{T,#1}{}}
\newcommand{\dsigSNNLOshort}[1]{\dd\sigpart{S,#1}{}}

\newcommand{\dsigMFNLO}[1]{\dd\sigpart{MF}{#1,NLO}}
\newcommand{\dsigMFNNLO}[2]{\dd\sigpart{MF,#2}{#1,NNLO}}

\newcommand{\ins}[1]{\mathcal{I}ns\left[#1\right]}
\newcommand{\insdouble}[1]{\mathcal{I}ns_{2}\left[#1\right]}
\newcommand{\LC}{\mathcal{LC}}

\DeclareMathOperator{\tr}{Tr}

\preprint{{\raggedleft%
		CERN-TH-2022-047\\
		IPPP/22/15\\
		KA-TP-05-2022\\ 
		P3H-22-030\\
		ZU-TH 10/22\\
}}

\title{Automation of antenna subtraction in colour space: gluonic processes}

\author{X.\ Chen$^{a,b}$, T.\ Gehrmann$^{c}$, E.W.N.\ Glover$^d$, A.\ Huss$^e$, M.\ Marcoli$^c$}

\affiliation{
	$^a$Institute for Theoretical Physics, Karlsruhe Institute of Technology, 76131 Karlsruhe, Germany\\
	$^b$Institute for Astroparticle Physics, Karlsruhe Institute of Technology,\\ 76344 Eggenstein-Leopoldshafen, Germany\\
	$^c$Physik-Institut, Universit\"at Z\"urich, Winterthurerstrasse 190, 8057 Z\"urich, Switzerland\\
	$^d$Institute for Particle Physics Phenomenology, Department of Physics, University of Durham, Durham, DH1 3LE, UK\\
	$^e$Theoretical Physics Department, CERN, 1211 Geneva 23, Switzerland}

\emailAdd{xuan.chen@kit.edu}
\emailAdd{thomas.gehrmann@uzh.ch}
\emailAdd{e.w.n.glover@durham.ac.uk}
\emailAdd{alexander.huss@cern.ch}
\emailAdd{mmarcoli@physik.uzh.ch}

\abstract{
	We present the colourful antenna subtraction method, a reformulation of the antenna subtraction scheme for next-to-next-to-leading order (NNLO) calculations in QCD. The aim of this new approach is to achieve a general and process-independent construction of the subtraction infrastructure at NNLO. We rely on the predictability of the infrared singularity structure of one- and two-loop amplitudes in colour space to generate virtual subtraction terms and, subsequently, we define an automatable procedure to systematically infer the expression of the real subtraction terms, guided by the correspondence between unintegrated and integrated antenna functions. To demonstrate the applicability of the described approach, we compute the full colour NNLO correction to gluonic three-jet production $pp(gg)\to ggg$, in the gluons-only assumption.
}

\begin{document} 
	
	\maketitle
	
	\section{Introduction}
		
Precision measurements of benchmark cross sections are an important pillar of the LHC physics program. In combining these measurements 
with equally precise theory predictions, fundamental Standard Model parameters are measured to high accuracy, and indirect constraints on 
physics beyond the Standard Model are obtained. The success of the LHC precision physics program relies crucially on 
a close interplay between theory and experiment, which calls in particular for highly accurate theoretical predictions. 
These predictions are obtained through a perturbation theory expansion to sufficiently high order. 
The workflow for the computation of next-to-leading order (NLO) perturbative corrections in QCD and in the electroweak theory combines 
the automated generation of one-loop virtual corrections~\cite{Ellis:2011cr} with multi-purpose event generator 
programs~\cite{Alioli:2010xd,Alwall:2014hca,Kallweit:2015dum,Bellm:2015jjp}, enabling NLO-accurate predictions for 
any collider process.  

Calculations at next-to-next-to-leading order (NNLO) and beyond are performed on a case-by-case basis~\cite{Heinrich:2020ybq} and are mostly 
limited to final-states corresponding to an underlying two-to-two scattering kinematics. This limitation to low-multiplicity processes arises 
from two causes: missing two-loop virtual corrections to higher-multiplicity scattering amplitudes and computational complexity of the real 
radiation corrections. The real radiation contributions develop infrared singularities related to soft and collinear particle emissions, which 
become explicit only after phase space integration, and which require an infrared subtraction method for the extraction of
the singular contributions, thereby enabling the numerical implementation of the finite remainders.  
Several subtraction methods have been developed for NNLO 
calculations~\cite{Catani:2007vq,Gehrmann-DeRidder:2005btv,Currie:2013vh,Czakon:2010td,Boughezal:2011jf,Gaunt:2015pea,DelDuca:2016ily,Caola:2017dug}. Implementations using these methods are largely made on a process-by-process basis, and most methods 
scale either poorly or not at all to higher multiplicities.  

Important progress has been made most recently on the derivation of two-loop $2\to 3$ scattering 
amplitudes~\cite{Abreu:2019odu,Abreu:2021oya,Abreu:2021asb,Chawdhry:2020for,Chawdhry:2021mkw,Badger:2021nhg,Badger:2021imn,Badger:2021ega,Badger:2022ncb}, which already entered the calculations for 
three-photon production~\cite{Chawdhry:2019bji,Kallweit:2020gcp}, 
diphoton-plus-jet production~\cite{Chawdhry:2021hkp,Badger:2021ohm} 
and three-jet production~\cite{Czakon:2021mjy}. Among these, only three-jet production contains three final-state QCD objects at Born level and 
displays the full complexity of real radiation corrections to $2\to 3$ processes. Its NNLO 
infrared subtraction has been performed with a sector-improved 
residue subtraction approach~\cite{Czakon:2010td,Czakon:2014oma}. The large variety of phenomenologically 
relevant high-multiplicity processes for with NNLO predictions will be in future demand highlights the importance to automate the 
workflow of infrared subtraction leading to an algorithmic construction of the subtraction terms in a form directly suitable for numerical implementation.

It is the objective of this paper to enable the automation of
 the NNLO antenna subtraction method~\cite{Gehrmann-DeRidder:2005btv,Currie:2013vh}, which is based up to now on the identification of 
 single and double real radiation patterns in colour-ordered subprocess contributions. This method has been applied successfully in 
 computing NNLO corrections to a variety of hadron-collider processes~\cite{Gehrmann-DeRidder:2015wbt,Chen:2016zka,Currie:2017eqf,Gehrmann-DeRidder:2017mvr,Cruz-Martinez:2018rod,Chen:2019zmr,Gauld:2019yng,Gehrmann:2020oec,Gauld:2021ule}. 
 Working towards a fully automated workflow, we 
 propose a complete reformulation, named colourful antenna subtraction, based on a colour-space representation of parton-level
 subprocesses and subtraction terms. We fully formulate the new method for gluonic processes, and demonstrate its implementation 
 by generating and computing the NNLO gluons-only contribution to three-jet production at hadron colliders.

	The paper is structured as follows. In section~\ref{sec:col_space} we introduce our notation and summarize the colour space techniques. In sections~\ref{sec:subNLO} and~\ref{sec:subNNLO} we delineate the application of the colourful antenna subtraction method for the construction of the subtraction terms at NLO and NNLO. The proof-of-principle application to gluonic three-jet production at NNLO is presented in section~\ref{sec:ggggg}, where we test the behaviour of the generated subtraction terms in single and double unresolved limits and we report the computation of the NNLO correction for a selection of observables.  
	We conclude in section~\ref{sec:conc} with an overview of the remaining steps required to include quarks in this new approach.
	
	\section{Colour space}\label{sec:col_space}
	
	\subsection{Gluon amplitudes in colour space}\label{sec:gluonamp}
	
	We summarize here the colour space formalism (also called colour dipole formalism) and the associated notation. We choose to treat the QCD amplitudes as objects in colour space since the singularity structure of QCD loop amplitudes is best described working in this framework, as discussed in~\cite{Catani:1996jh,Catani:1998bh,Becher:2009qa,DelDuca:2016ily}. In colour space, a $\ell$-loop amplitude with $n$ external partons is represented by an abstract vector $\ket{\ampnum{n}{\ell}(\lb p\rb_n)}$. If a set of generating vectors $\lb \mathbfcal{C}_{n,i}^{\ell} \rb$ is defined, which span the $n$-parton colour space, any amplitude can be decomposed as:
	\begin{equation}\label{coldec}
		\ket{\ampnum{n}{\ell}(\lb p\rb_n)}=\sum_{i\in I^{\ell}} \mathbfcal{C}_{n,i}^{\ell} \, A^{\ell}_{n,i}(\lb p\rb_n),
	\end{equation}
	where $I^{\ell}$ indicates a suitable subset of generating vectors. The scalar quantities $A^{\ell}_{n,i}(\lb p\rb_n)$ are colour-ordered partial amplitudes. In equation~\eqref{coldec}, the dependence on the helicities of the external partons is implicit and in the following a sum over helicity configurations is always assumed when squared quantities are considered. For $\ell\geq1$, the dependence on the renormalization scale $\mu_r$ is understood. Our convention is to strip the partial amplitudes of overall coefficients such as couplings and incoming particles average factors, which are inserted later at the cross section level. In particular, we strip an $\ell$-loop amplitude of an overall factor $\left(\tfrac{\alpha_s \bar{C}(\e)}{2\pi}\right)^\ell$ with respect to the corresponding tree-level amplitude, where $\overline{C}(\e)=(4\pi)^\e e^{-\gamma_E\e}$.
	
	We focus on amplitudes involving only gluons, for which a convenient choice of generating vectors is given by the so-called trace basis or Chan-Paton basis, which consists of traces of $SU(N_c)$ generators in the fundamental representation~\cite{Paton:1969je,Mangano:1987xk}. For tree-level amplitudes, the basis corresponds to:
	\begin{equation}
		\mathbfcal{C}_{n,\sigma}^{0}=\tr(\T^{a_{\sigma(1)}}\dots \T^{a_{\sigma(n)}}),\quad\text{with}\quad\sigma\in S_n/Z_n
	\end{equation}
	where $S_n/Z_n$ represents the group of non-cyclic permutations of $n$ objects. The $SU(N_c)$ generators are normalized according to:
	\begin{equation}
		\tr(\T^a\T^b)=\dfrac{1}{2}\delta^{ab}.
	\end{equation}
	With this choice, a tree-level amplitude with $n$ external gluons is given by~\cite{Berends:1987cv}:
	\begin{equation}
		\ket{\ampnum{n}{0}(\lb p\rb_n)}=\sum_{\sigma\in S_n/Z_n}\tr(\T^{a_{\sigma(1)}}\dots \T^{a_{\sigma(n)}})A_n^{0}(\sigma(p_1),\dots,\sigma(p_n)),
	\end{equation}
	and the corresponding squared matrix element $\me{n}{0}$ can be computed as its squared norm in colour space:
	\begin{eqnarray}\label{treeME}
		\me{n}{0}(\lb p \rb_n)&=&\braket{\ampnum{n}{0}(\lb p \rb_n)|\ampnum{n}{0}(\lb p\rb_n)}\nonumber\\ 
		&=&\Bigg.\sum_{\sigma,\sigma'\in S_n/Z_n}\left(\mathbfcal{C}^{0}_{n,\sigma}\right)^{\dagger}\mathbfcal{C}^{0}_{n,\sigma'} A_n^{0}(\sigma(\lb p\rb_n))^{\dagger}A_n^{0}(\sigma'(\lb p\rb_n)),
	\end{eqnarray}
	where a sum over colour indices is assumed.
	
	For one-loop amplitudes with gluons only, both as external and internal particles, we have an analogous colour decomposition~\cite{Bern:1990ux}:
	\begin{equation}
		\ket{\ampnum{n}{1}(\lb p\rb_n)}=\sum_{c=1}^{\lfloor n/2\rfloor+1}\sum_{\sigma\in S_n/S_{n,c}}\mathbfcal{C}^{1}_{n,c,\sigma}A^{1}_{n,c}(\sigma(p_1),\dots,\sigma(p_n)),
	\end{equation}
	where the generating one-loop vectors in colour space are given by:
	\begin{eqnarray}
		\mathbfcal{C}^{1}_{n,1,\sigma}&=&N_c\,\mathbfcal{C}^{0}_{n,\sigma}=N_c\tr(\T^{a_{\sigma(1)}}\dots \T^{a_{\sigma(n)}}),\nonumber\\
		\mathbfcal{C}^{1}_{n,c,\sigma}&=&\tr(\T^{a_{\sigma(1)}}\dots \T^{a_{\sigma(c-1)}})\tr(\T^{a_{\sigma(c)}}\dots \T^{a_{\sigma(n)}}),\quad\text{for }c>1,
	\end{eqnarray}
	and $S_{n,c}$ represents the subgroup of $S_n$ which leaves the trace structure of $\mathbfcal{C}^{1}_{n,c}$ unaffected. We notice that $\mathbfcal{C}^{1}_{n,2,\sigma}=0$, since the $SU(N_c)$ generators are traceless. The squared one-loop matrix element $\me{n}{1}$ is given by: 
	\begin{eqnarray}\label{loop1l}
		\me{n}{1}(\lb p \rb_n)&=&\braket{\ampnum{n}{0}(\lb p \rb_n)|\ampnum{n}{1}(\lb p\rb_n)}+\braket{\ampnum{n}{1}(\lb p \rb_n)|\ampnum{n}{0}(\lb p\rb_n)}\nonumber\\
		&=&2\text{Re}\Bigg\lbrace\Bigg.\sum_{c=1}^{\lfloor n/2\rfloor+1}\sum_{\substack{\sigma\in S_n/Z_n \\ \sigma'\in S_n/S_{n,c}}}\left(\mathbfcal{C}^{0}_{n,\sigma}\right)^{\dagger}\mathbfcal{C}^{1}_{n,c,\sigma}A_n^{0}(\sigma(\lb p\rb_n))^{\dagger}A_{n,c}^{1}(\sigma'(\lb p\rb_n))\Bigg.\Bigg\rbrace.,
	\end{eqnarray}
	We note that, both for $\ell=0$ and $\ell=1$, the colour-ordered partial amplitudes are not all independent quantities. Several relations can be defined among them such as reflection identities or decoupling equations~\cite{Berends:1987me,Mangano:1987xk,Kleiss:1988ne,Bern:1990ux}.

	\subsection{Gluon exchange in colour space}
	
	The coherent emission of a gluon between a dipole formed by parton $i$ and parton $j$ in a tree-level amplitude is described in colour space by colour correlators as
	\begin{equation}\label{cc}
		\braampdep{n}{0}{\lb p \rb_n}\T_i\cdot\T_j\ketampdep{n}{0}{\lb p \rb_n},
	\end{equation}
	where $\T_i\cdot\T_j=T^a_i T^a_j$. These properties hold:
	\begin{eqnarray}
		\label{id_dip_1}\T_i\cdot\T_j&=&\T_j\cdot\T_i,\\
		\label{id_dip_2}\T_i\cdot\T_i&=&\T_i^2=C_i\,\mathbf{Id},
	\end{eqnarray}
	where $C_i$ is the Casimir coefficient for the $SU(N_c)$ representation associated to parton $i$ and $\mathbf{Id}$ represents the identity operator in colour space. For gluons we have $C_g=C_A$.
	
	For tree amplitudes with only gluons in the trace basis representation, it is possible to compute colour correlators in~\eqref{cc} using:
	\begin{eqnarray}\label{cdip}
		\braket{\mathbfcal{C}_{n,\sigma}^{0}|\T_i\cdot\T_j|\mathbfcal{C}_{n,\sigma'}^{0}}&=&\left(\tr(\T^{a_{\sigma(1)}}\dots \T^{a_{\sigma(n)}})\right)^{\dagger}\T_i\cdot\T_j\tr(\T^{a_{\sigma'(1)}}\dots \T^{a_{\sigma'(n)}})\nonumber\\
		&=&\left(\tr(\T^{a_{\sigma(1)}}\dots \T^{a_i}\dots \T^{a_j}\dots \T^{a_{\sigma(n)}})\right)^{\dagger}\,\left(T_i\right)^a_{a_i b_i}\left(T_j\right)^a_{a_j b_j}\nonumber\\
		&&\hspace{3.cm}\times\tr(\T^{a_{\sigma'(1)}}\dots \T^{b_i}\dots \T^{b_j}\dots \T^{a_{\sigma'(n)}}),
	\end{eqnarray}
	where, for either incoming or outgoing gluons, $\T_i$ is in the adjoint representation:
	\begin{equation}
		\left(T_i\right)^a_{b c} = i f_{bac}.
	\end{equation}
	
	In fact, for the purpose of an NNLO calculation, the single colour dipole insertion at tree-level in~\eqref{cc} is not sufficient. The computation of colour correlators representing the exchange of a gluon at one-loop level
	\begin{equation}\label{cc1l}
		\braampdep{n}{1}{\lb p \rb_n}(\T_i\cdot\T_j)\ketampdep{n}{0}{\lb p \rb_n}+\braampdep{n}{0}{\lb p \rb_n}(\T_i\cdot\T_j)\ketampdep{n}{1}{\lb p \rb_n}
	\end{equation}
	and the exchange of two gluons at tree-level
	\begin{equation}\label{cc2}
		\braampdep{n}{0}{\lb p \rb_n}(\T_i\cdot\T_j)(\T_k\cdot\T_l)\ketampdep{n}{0}{\lb p \rb_n}.
	\end{equation}
	is necessary. It is straightforward~\cite{DelDuca:2016ily} to extend~\eqref{cdip} to compute~\eqref{cc1l} and~\eqref{cc2} when both $k$ and $l$ are different from $i$ or $j$, namely the gluons are exchanged between two separated dipoles. When a radiator in the second dipole coincides with one in the first dipole, the two gluons are emitted from the same leg, and the colour algebra must be computed accordingly. The same applies if both partons in the second dipole coincide with the ones in the first. For example, if $k=i$ and $l\ne j$ one has:
	\begin{eqnarray}\label{cdipdip}
\lefteqn{\left(\tr(\T^{a_{\sigma(1)}}\dots \T^{a_{\sigma(n)}})\right)^{\dagger}(\T_i\cdot\T_j)(\T_i\cdot\T_l)\tr(\T^{a_{\sigma'(1)}}\dots \T^{a_{\sigma'(n)}})}\nonumber\\
		&=&\left(\tr(\T^{a_{\sigma(1)}}\dots \T^{a_i}\dots \T^{a_j}\dots \T^{a_l}\dots \T^{a_{\sigma(n)}})\right)^{\dagger}
		\left[\left(T_i\right)^a_{a_i c}\left(T_j\right)^a_{a_j b_j}\right]\left[\left(T_i\right)^b_{c b_i}\left(T_l\right)^b_{a_l b_l}\right]\nonumber\\
		&&\quad\times\tr(\T^{a_{\sigma'(1)}}\dots \T^{b_i}\dots \T^{b_l}\dots \T^{b_j}\dots \T^{a_{\sigma'(n)}}).
	\end{eqnarray}
	
	Finally, we point out that each state $\ketampdep{n}{\ell}{\lb p \rb_n}$ is a colour singlet and so by colour conservation:
	\begin{equation}\label{colour_conservation}
		\sum_{i=1}^{n}\T_i\ketampdep{n}{\ell}{\lb p \rb_n}=0.
	\end{equation}
	Since in what follows we always consider colour singlet states, we can employ the previous identity as $\sum_{j\ne i}\T_j=-\T_i$.
	
	\subsection{Leading and subleading colour}\label{sec:LCandSLC}
	
	When squared matrix elements like~\eqref{treeME} and~\eqref{loop1l}, or colour operator insertions such as~\eqref{cc},~\eqref{cc1l} and~\eqref{cc2} are computed, the result is a real function of the external momenta and possibly the renormalization scale. At tree-level, the general form of such a function is given by:
	\begin{equation}\label{func0}
		f_0\left(\set{p}_n\right)=\sum_{\sigma,\sigma'\in S_n/Z_n} c^{0}_{n}(\sigma,\sigma')\,\anum{n}{0}(\sigma,\sigma';\pset), 
	\end{equation}
	where the coefficients $c^{0}_{n}(\sigma,\sigma')$ are colour factors which depend on $N_c$ and the number of gluons $n$ and 
	\begin{equation}
		\anum{n}{0}(\sigma,\sigma';\pset)=
		\begin{cases}
			\left|\ampnumstraight{n}{0}(\sigma(\pset))\right|^2\quad&\text{if }\sigma=\sigma',\\
			\ampnumstraight{n}{0}(\sigma(\pset))^{\dagger}\ampnumstraight{n}{0}(\sigma'(\pset))\quad&\text{if }\sigma\neq\sigma'.
		\end{cases}
	\end{equation}
	This quantity represents the squared interference of two colour-ordered partial amplitudes, with generic colour orderings dictated by $\sigma$ and $\sigma'$. When $\sigma=\sigma'$, we have squared coherent partial amplitudes, while we refer to the case $\sigma\neq\sigma'$ as incoherent interference. We notice that even in the case $\sigma\neq\sigma'$, $\anum{n}{0}(\sigma,\sigma';\pset)$ is a real quantity, since a sum over helicities is assumed. Indeed, if we consider a helicity configuration for the external particles $\{h\}_n$, the opposite configuration where the helicities are all swapped, denoted by $\{-h\}_n$ also appears in the sum. Restoring the helicity dependence in partial amplitudes, by charge conjugation at tree level we have:
	\begin{equation}
		\ampnumstraight{n}{0}(\sigma(\pset,\{-h\}_n))=\ampnumstraight{n}{0}(\sigma(\pset,\{h\}_n))^{\dagger}, 
	\end{equation}
	and so
	\begin{eqnarray}
	\lefteqn{\ampnumstraight{n}{0}(\sigma(\pset,\{h\}_n))^{\dagger}\ampnumstraight{n}{0}(\sigma'(\pset,\{h\}_n))+\ampnumstraight{n}{0}(\sigma(\pset,\{-h\}_n))^{\dagger}\ampnumstraight{n}{0}(\sigma'(\pset,\{-h\}_n))}\nonumber\\
		&=&\ampnumstraight{n}{0}(\sigma(\pset,\{h\}_n))^{\dagger}\ampnumstraight{n}{0}(\sigma'(\pset,\{h\}_n))+\ampnumstraight{n}{0}(\sigma(\pset,\{h\}_n))\ampnumstraight{n}{0}(\sigma'(\pset,\{h\}_n))^{\dagger}\nonumber\\
		&=&2\text{Re}\left[\ampnumstraight{n}{0}(\sigma(\pset,\{h\}_n))^{\dagger}\ampnumstraight{n}{0}(\sigma'(\pset,\{h\}_n))\right],
	\end{eqnarray}
	which implies that the sum over helicities is indeed real. Analogously, at one loop we have:
	\begin{equation}\label{func1}
		f_1\left(\set{p}_n\right)=\sum_{c=1}^{\lfloor n/2\rfloor+1}\sum_{\substack{\sigma\in S_n/S_{n,c} \\ \sigma' \in S_n/Z_n}} c^{1}_{n,c}(\sigma,\sigma')\,\anum{n,c}{1}(\sigma,\sigma';\pset), 
	\end{equation}
	where
	\begin{equation}
		\anum{n,c}{1}(\sigma,\sigma';\pset)=2\text{Re}\left[\ampnumstraight{n}{0}(\sigma(\pset))^{\dagger}\ampnumstraight{n,c}{1}(\sigma'(\pset))\right].
	\end{equation}

	The $N_c$ dependence of~\eqref{func0} ad~\eqref{func1} is in general given by:
	\begin{equation}
		f_{\ell}=\dfrac{P(N_c)}{N_c^m},
	\end{equation}
	where $P(N_c)$ is a polynomial in $N_c$. The degree of $P(N_c)$ and the integer $m$ depend on $\ell$ and the number of gluons $n$. It is indeed possible to predict which is the highest power of $N_c$ appearing in a large $N_c$ expansion of the squared quantities, as shown in Table~\ref{tab:Nc_pow}.
	
	\begin{table}[h]
		\centering
		\begin{tabular}{c|c}
			\hline
			$f$ & highest $N_c$ power \\
			\hline
			$\me{n}{0}$ &  $N_c^{n}$\\
			$\me{n}{1}$ &  $N_c^{n+1}$\\
			$\braket{\ampnum{n}{0}|(\T_i\cdot \T_j)|\ampnum{n}{0}}$ & $N_c^{n+1}$\\
			$\braket{\ampnum{n}{1}|(\T_i\cdot \T_j)|\ampnum{n}{0}} + \braket{\ampnum{n}{0}|(\T_i\cdot \T_j)|\ampnum{n}{1}}$ & $N_c^{n+2}$\\
			$\braket{\ampnum{n}{0}|(\T_i\cdot \T_j)(\T_k\cdot \T_l)|\ampnum{n}{0}}$ &  $N_c^{n+2}$\\
		\end{tabular}
		\caption{Highest power of $N_c$ appearing in the squared quantities.}\label{tab:Nc_pow}
	\end{table}

	The \textit{leading colour} approximation is defined retaining only the terms coming with the power of $N_c$ indicated in Table~\ref{tab:Nc_pow}. Our convention is actually to incorporate an overall factor $\frac{N_c^2-1}{Nc^2}$ in the definition of the leading colour contribution since this factor is ubiquitous in our calculations and partially retains subleading terms without affecting the complexity. We define the operator $\mathcal{LC}(\cdot)$ which extracts the leading colour part of a given quantity. At leading colour,~\eqref{func0} and~\eqref{func1} have a particularly simple form, since only the coherent contributions with $\sigma=\sigma'$ survive. This convenient behaviour follows from the fact that the trace basis behaves as an actual orthogonal basis at leading colour:
	\begin{equation}\label{LC}
		\LC\left[\left(\tr(\T^{a_{\sigma(1)}}\dots \T^{a_{\sigma(n)}})\right)^{\dagger}\tr(\T^{a_{\sigma'(1)}}\dots \T^{a_{\sigma'(n)}})\right]=\dfrac{N_c^{n-2}(N_c^2-1)}{2^n}\delta_{\sigma\sigma'}.
	\end{equation}
	which is a direct consequence of Fierz identity for $SU(N_c)$ generators in the fundamental representation:
	\begin{equation}\label{Fierz}
		T^{a}_{ij}T^{a}_{kl}=\dfrac{1}{2}\left(\delta_{il}\delta_{jk}-\dfrac{1}{N_c}\delta_{ij}\delta_{kl}\right).
	\end{equation}
	Using~\eqref{LC}, it is straightforward to obtain the expression for squared matrix elements at leading colour:
	\begin{equation}\label{treeMELC}
		\LC\left[\me{n}{0}(\lb p \rb_n)\right]=(N_c)^{n-2}\dfrac{(N_c^2-1)}{2^n}\sum_{\sigma\in S_n/Z_n}\left|A_n^{0}( \sigma(\lb p\rb_n))\right|^2
	\end{equation}
	and
	\begin{equation}\label{loop1lLC}
		\LC\left[\me{n}{1}(\lb p \rb_n)\right]=(N_c)^{n-1}\dfrac{(N_c^2-1)}{2^n}\sum_{\sigma\in S_n/Z_n}2\text{Re}\left[A_n^{0}( \sigma(\lb p\rb_n))^{\dagger}A_{n,1}^{1}( \sigma(\lb p\rb_n))\right].
	\end{equation}
	This simplicity is the foundation of the efficient application of the antenna subtraction in the leading colour approximation. The antenna functions are directly derived from squared matrix elements~\cite{Gehrmann-DeRidder:2005svg,Gehrmann-DeRidder:2005alt,Gehrmann-DeRidder:2005btv} and are therefore well suited to describe the infrared behaviour of matrix elements when these are expressed as a sum of coherent squared partial amplitudes, as it happens in the leading colour approximation.
	The same simple structure arises in colour correlators too:
	\begin{eqnarray}
		\LC\left[\braket{\mathbfcal{C}_{n,\sigma}^{0}|\T_i\cdot\T_j|\mathbfcal{C}_{n,\sigma'}^{0}}\right]&=&-\delta_{\sigma\sigma'}\chi_{\sigma}(i,j)\left(\dfrac{N_c}{2}\right)\dfrac{N_c^{n-2}(N_c^2-1)}{2^n},\\
		\LC\left[\braket{\mathbfcal{C}_{n,1,\sigma}^{1}|\T_i\cdot\T_j|\mathbfcal{C}_{n,\sigma'}^{0}}\right]&=&N_c\,\LC\left[\braket{\mathbfcal{C}_{n,\sigma}^{0}|\T_i\cdot\T_j|\mathbfcal{C}_{n,\sigma'}^{0}}\right]
	\end{eqnarray}
	and
	\begin{equation}\label{cdipdip2}
		\LC\left[\braket{\mathbfcal{C}_{n,\sigma}^{0}|(\T_i\cdot\T_j)(\T_k\cdot\T_l)|\mathbfcal{C}_{n,\sigma'}^{0}}\right]=\delta_{\sigma\sigma'}\chi_{\sigma}(i,j)\chi_{\sigma}(k,l)\left(\dfrac{N_c}{2}\right)^{2}\dfrac{N_c^{n-2}(N_c^2-1)}{2^n},
	\end{equation}
	where $\chi_{\sigma}(i,j)$ is $1$ if partons $i$ and $j$ appear adjacent to each other in the colour ordering represented by $\sigma$ and $0$ otherwise. 
	
	At the subleading colour level, the emergence of incoherent interferences between different colour orderings spoils the pattern observed at leading colour. As pointed out in~\cite{Currie:2013vh}, this represented an obstacle to the application of the antenna subtraction method beyond leading colour. When the number of considered partons at the Born level $n_p$ is small, in particular for $n_p<4$, relations among partial amplitudes can be conveniently exploited to convert incoherent interferences into combinations of squared coherent partial amplitudes and proceed as in the leading colour case. However, for $n_p\geq4$ this is not possible any more and the treatment of the subleading colour part requires a significant effort. As we show in the rest of this paper, working in colour space allows us to retain the full $N_c$ dependence in a straightforward way and to consistently approach leading and subleading colour contributions with the same techniques.
	
	In the following sections, we drop the explicit dependence of the amplitudes on the momenta $\lb p \rb_n$ to ease the notation. Moreover, in general we consider Born-level \mbox{$(n+2)$-parton} amplitudes, to keep $n$ as the number of final-state partons. Indices $1$ and $2$ correspond to initial-state partons, while $i\geq3$ indicates a final-state parton.

	\section{Colourful antenna subtraction at NLO}\label{sec:subNLO}
	
	The NLO QCD correction to an $n$-jet partonic cross section with parton species $a$ and $b$ in the initial state is given by:
	\begin{equation}\label{NLOcs}
		\dd\sigpart{}{ab,NLO}=\int_{n}\left(\dd\sigpart{V}{ab,NLO}+\dd\sigpart{MF}{ab,NLO}\right)+\int_{n+1}\dd\sigpart{R}{ab,NLO},
	\end{equation}
	where the symbol $\int_n$ indicates an integration over the $n$ final state particles. $\dsigV{ab}$ and $\dsigR{ab}$ respectively represent the virtual and real corrections, while $\dsigMFNLO{ab}$ is the NLO mass factorization counterterm. The NLO cross section in~\eqref{NLOcs}, despite being well defined and finite, is not suitable for numerical integration in this form. The virtual correction and the mass factorization contribution contain explicit $\e$-poles and the real correction diverges in soft and collinear infrared (IR) limits. These singularities cancel in the final result, but a proper subtraction procedure is needed to separately remove the singularities in the real and virtual corrections and make both integrals in~\eqref{NLOcs} computable with numerical methods. 
	
	In the context of antenna subtraction, this is achieved constructing a real subtraction term $\dsigSNLO{ab}$~\cite{Currie:2013vh}, which locally removes the singular behaviour of $\dsigR{ab}$ in the IR limits and can be analytically integrated over the phase space of the unresolved radiation. This latter feature is required to obtain from $\dsigSNLO{ab}$ the virtual subtraction term $\dsigTNLO{ab}$, which cancels the explicit poles of the virtual correction and contains the mass factorization contribution. The NLO cross section can then be reformulated as:
	\begin{equation}\label{NLOcssub}
		\dd\sigpart{}{ab,NLO}=\int_{n}\left[\dd\sigpart{V}{ab,NLO}-\dsigTNLO{ab}\right]+\int_{n+1}\left[\dd\sigpart{R}{ab,NLO}-\dsigSNLO{ab}\right],
	\end{equation}
	with 
	\begin{equation}\label{sigTNLO}
		\dsigTNLO{ab}=-\int_1\dsigSNLO{ab}-\dsigMFNLO{ab}.
	\end{equation}
	Both contributions in~\eqref{NLOcssub} are now free of IR singularities and suitable for a numerical integration through Monte Carlo methods. The procedure we have just depicted represents the traditional antenna subtraction approach: the real subtraction term is constructed first, studying the behaviour of the real radiation matrix elements in soft and collinear limits, and then it is integrated and combined with the mass factorization contribution to obtain the virtual subtraction term. 
	
	In the following we give an overview of the colourful antenna subtraction approach at NLO. General subtraction schemes for automated NLO calculations have long been available \cite{Frixione:1995ms,Catani:1996jh}, so the purpose of this section is mainly to introduce important concepts concepts behind the new approach, which will be crucial for its application at NNLO. Indeed, despite posing a significantly simpler task with respect to NNLO, some key observations can already be made at the NLO level. The main idea behind the colourful antenna approach is to exploit the predictability of the singularity structure of virtual amplitudes in colour space to directly construct the virtual subtraction term in a general way and subsequently derive the real subtraction term with a systematic procedure.  
	
	\subsection{IR singularity structure at one loop}
	
	The singularity structure of renormalized $(n+2)$-parton one-loop amplitudes in QCD can be described in colour space with~\cite{Catani:1998bh}:
	\begin{equation}\label{1l-sing}
		\ketamp{n+2}{1}=\Iop{1}{\epsilon,\mu_r^2}\ketamp{n+2}{0}+\ketampdep{n+2}{1,\text{fin}}{\mu_r^2},
	\end{equation}
	where $\mu_r$ is the renormalization scale, $\ketampdep{n+2}{1,\text{fin}}{\mu_r^2}$ is a finite remainder and $\Iop{1}{\e,\mu_r^2}$ is the Catani's IR insertion operator given by~\cite{Catani:1998bh}:
	\begin{equation}\label{I1}
		\Iop{1}{\epsilon,\mu_r^2}=\dfrac{1}{2}\dfrac{e^{\e\gamma_E}}{\Gamma(1-\e)}\sum_{i=1}^{n+2}\dfrac{1}{\T_i^2}\mathcal{V}_{i}(\e)\sum_{j\ne i}(\T_i\cdot \T_j)\left(\dfrac{-s_{ij}}{\mu_r^2}\right)^{-\e}\,.
	\end{equation}
	The singular functions $\mathcal{V}_{i}(\e)$ contain double and single $\e$-poles. The previous expression can be rewritten as
	\begin{eqnarray}\label{I1v2}
		\Iop{1}{\epsilon,\mu_r^2}&=&\dfrac{1}{2}\sum_{i=1}^{n+2}\sum_{j\ne i}\left(\T_i\cdot\T_j\right)\ourIop{1}{ij}{\e,\mu_r^2}\nonumber\\
		&=&\sum_{(i,j)}\left(\T_i\cdot\T_j\right)\ourIop{1}{ij}{\e,\mu_r^2},
	\end{eqnarray}
	where in the last line the sum runs over pairs of partons. For the gluons-only case that is considered here, 
	we only need the expression of $\ourIop{1}{i_gj_g}{\e,\mu_r^2}$ at $N_f=0$:
	\begin{equation}\label{Igg}
		\ourIop{1}{i_gj_g}{\e,\mu_r^2}=\dfrac{e^{\e\gamma_E}}{\Gamma(1-\e)}\left[\dfrac{1}{\e^2}+\dfrac{b_0}{\e}\right]\left(\dfrac{-s_{ij}}{\mu_r^2}\right)^{-\e},
	\end{equation}
	where $b_0$ is the gluon component of the one-loop QCD $\beta$-function:
	\begin{equation}
		b_0=\dfrac{11}{6}\,.
	\end{equation}
	
	Using~\eqref{1l-sing} it is possible to extract the poles of one-loop matrix elements in the following way:
	\begin{eqnarray}
		\poles\left(\me{n+2}{1}\right)&=&\poles\left(\braket{\ampnum{n+2}{0}|\ampnum{n+2}{1}}+\braket{\ampnum{n+2}{1}|\ampnum{n+2}{0}}\right)\nonumber\\
		&=&\poles(\braket{\ampnum{n+2}{0}|\Iop{1}{\e}+\Iopd{1}{\e}|\ampnum{n+2}{0}})\,.
	\end{eqnarray}
	The appearance of the sum $\Iop{1}{\e}+\Iopd{1}{\e}$ indicates that only the real part of the insertion operator affects the description of the poles at the matrix element level, as expected. In the gluons-only case we can write the previous expression as 
	\begin{equation}
		\poles\left(\me{n+2}{1}\right)=\poles\left[\sum_{(i_g,j_g)}\braket{\ampnum{n+2}{0}|\T_{i_g}\cdot \T_{j_g}|\ampnum{n+2}{0}}\,2\text{Re}\left(\ourIop{1}{i_g j_g}{\e,\mu_r^2}\right)\right]\,,
	\end{equation}
	where we moved the colour sandwich outside the real part, since, as argued in section \ref{sec:LCandSLC} it is a real quantity. At the cross section level we have
	\begin{eqnarray}\label{Vpoles}
		\poles\left(\sigpart{V}{gg,NLO}\right)&=&\coeffVNLO\int\dphi{n}(p_3,\dots,p_{n+2};p_1,p_2)\,\jet{n}{n}{\lb p \rb_n}\nonumber\\
		&&\times\poles\left[\sum_{(i_g,j_g)}\braket{\ampnum{n+2}{0}|\T_{i_g}\cdot \T_{j_g}|\ampnum{n+2}{0}}2\,\text{Re}\left(\ourIop{1}{i_g j_g}{\e,\mu_r^2}\right)\right].
	\end{eqnarray}
	In~\eqref{Vpoles}, $\dphi{n}$ denotes the standard $(2\to n)$-particle phase space and $\jet{n}{m}{\lb p \rb_m}$ is the jet algorithm which selects $n$ jets from $m$ final-state parton momenta. The factor $\coeffVNLO$ is given by
	\begin{equation}
		\coeffVNLO=\coeff\coeffLO,
	\end{equation}
	where $\coeffLO$ contains the overall factors appropriate for the LO process, such as the strong coupling, symmetry factors and factors coming from the spin- and colour-average over the initial-state partons.
	
	\subsection{NLO mass factorization}\label{sec:MFNLO}
	
	For processes involving initial-state partons, it is necessary to treat the singularities arising when a final-state parton becomes collinear to the initial-state ones. These singularities are removed defining physical parton density functions which are obtained from the bare ones by means of mass factorization counterterms. At NLO this counterterm is given by
	\begin{equation}\label{MF}
		\dd\sigpart{MF}{ab,NLO}=-\coeff\sum_{c,d}\int\dr{x_1}\dr{x_2}\Gamma^{(1)}_{ab;cd}(x_1,x_2)\,\dd\sigpart{}{cd,LO},
	\end{equation}
	where $x_1$ and $x_2$ represent the momentum fractions transferred to the hard process and $\Gammaone{ab;cd}{x_1,x_2}$ denotes the NLO mass factorization kernel:
	\begin{equation}
		\Gammaone{ab;cd}{x_1,x_2}=\Gammaone{ca,\text{full}}{x_1}\delta_{db}\delta(1-x_2)+\Gammaone{db,\text{full}}{x_1}\delta_{ca}\delta(1-x_1).
	\end{equation}
	The $\Gammaone{ca,\text{full}}{x_i}$ contain the LO Altarelli-Parisi splitting kernels $p^0_{ca}(x_i)$~\cite{Altarelli:1977zs} and can be organized into several layers corresponding to different colour factors (the subscript `full' precisely indicates this). In principle, equation~\eqref{MF} has both identity preserving and identity changing contributions, respectively when $(c,d)=(a,b)$ and $(c,d)\ne(a,b)$. In the gluons-only case, we have $(c,d)=(a,b)=(g,g)$, so only gluon-to-gluon splitting kernels contribute:
	\begin{equation}
		\Gammaone{gg;gg}{x_1,x_2}=\Gammaone{gg,\text{full}}{x_1}\delta(1-x_2)+\Gammaone{gg,\text{full}}{x_1}\delta(1-x_1),
	\end{equation}
	where we only extract the gluon component of the splitting kernels at $N_f=0$:
	\begin{equation}
		\Gammaone{gg,\text{full}}{x_i}=N_c\,\Gammaone{gg}{x_i}=-\dfrac{1}{\e}N_c\,p^{0}_{gg}(x_i),
	\end{equation}
	with~\cite{Altarelli:1977zs}:
	\begin{equation}
		p^{0}_{gg}(x)=b_0\delta(1-x)+2\left(\dfrac{1}{1-x}\right)_{+}+\dfrac{2}{x}-2x^2+2x-4.
	\end{equation}
	It is possible to express the identity preserving mass factorization kernels in colour space, in analogy to the formalism used in the previous section to represent the singularity structure of the one-loop amplitudes. We define:
	\begin{equation}
		\BGammaone{gg;gg}{x_1,x_2}=\BGammaone{gg,\text{full}}{x_1}\delta(1-x_2)+\BGammaone{gg,\text{full}}{x_2}\delta(1-x_1),
	\end{equation}
	where 
	\begin{equation}
		\BGammaone{gg,\text{full}}{x_i}=-\Gammaone{gg,\text{full}}{x_i}\dfrac{1}{C_A}\sum_{j\ne i}\T_i\cdot\T_j,\quad i=1,2.
	\end{equation}
	This colour operator is proportional to the identity in colour space when it acts on a colour singlet vector, due to colour conservation~\eqref{colour_conservation} and \eqref{id_dip_2}, therefore
	\begin{equation}
		\BGammaone{gg;gg}{x_1,x_2}\ket{\ampnum{n+2}{\ell}}=\Gammaone{gg;gg}{x_1,x_2}\ket{\ampnum{n+2}{\ell}},
	\end{equation}
	which restores the original result, namely that the mass factorization counterterm factorizes onto the full corresponding LO matrix element. We can then rewrite~\eqref{MF} in the gluons-only case as
	\begin{eqnarray}\label{MFv2}
		\dd\sigpart{MF}{gg,NLO}&=&-\coeffVNLO\int\dr{x_1}\dr{x_2}\int\dphi{n}(p_3,\dots,p_{n+2};x_1 p_1,x_2 p_2)\,\jet{n}{n}{\lb p \rb_n}\nonumber\\
		&&\times\braket{\ampnum{n+2}{0}|\BGammaone{gg;gg}{x_1,x_2}|\ampnum{n+2}{0}}\,.
	\end{eqnarray}
	
	\subsection{NLO virtual subtraction term}\label{sec:NLOV}
	
	
	The virtual subtraction term at NLO $\sigpart{T}{gg,NLO}$ has to reproduce the explicit poles of the virtual matrix element as well as include the mass factorization contribution. Traditionally this was achieved summing the mass factorization counterterm with the real subtraction term after analytical integration over the phase space of an unresolved parton, as indicated by~\eqref{sigTNLO}. In this approach we exploit a combination of equations~\eqref{Vpoles} and~\eqref{MFv2} to directly construct $\sigpart{T}{gg,NLO}$. To do so, we define a NLO singularity dipole operator in colour space for an $(n+2)$-parton process:
	\begin{eqnarray}\label{J21}
		\Jcol{1}(\e)&&=\sum_{(i,j)\geq 3}(\T_i\cdot\T_j)\,\J{1}(i_g,j_g)+\sum_{i\ne 1,2}(\T_1\cdot\T_i)\,\J{1}(1_g,i_g)\nonumber\\
		&&\,+\sum_{i\ne 1,2}(\T_2\cdot\T_i)\,\J{1}(2_g,i_g)+(\T_1\cdot\T_2)\,\J{1}(1_g,2_g)\,.
	\end{eqnarray}
	The first sum runs over all pairs of gluons in the final state, the second and the third sums include all pairs with an initial-state gluon (respectively $1_g$ and $2_g$) and a final-state one and the last term addresses the configuration where both gluons are in the initial state. The scalar functions $\J{1}(i,j)$ are colour stripped one-loop integrated dipoles~\cite{Currie:2013vh,Currie:2013dwa}, given by a combination of integrated three-parton tree-level antenna functions and NLO mass factorization kernels. The explicit expressions of the gluon-gluon integrated dipoles for final-final (FF), initial-final (IF) and initial-initial (II) configurations are the following:
	\begin{eqnarray}
		\J{1}(i_g,j_g)&=&\dfrac{1}{3}\XFFint{F}{3}{0}(s_{ij}),\nonumber\\
		\J{1}(1_g,j_g)&=&\dfrac{1}{2}\XIFint{F}{3}{0}{g}(s_{1j})-\dfrac{1}{2}\Gammaone{gg}{x_1},\nonumber\\
		\J{1}(1_g,2_g)&=&\XIIint{F}{3}{0}{gg}(s_{12})-\dfrac{1}{2}\Gammaone{gg}{x_1}\deltatwo-\dfrac{1}{2}\Gammaone{gg}{x_2}\deltaone,		
	\end{eqnarray}
	where $\delta_{i}=\delta(1-x_i)$. The structure of the integrated dipoles is chosen in such a way that the singularities carried by the mass factorization kernels cancel with poles in the integrated initial-final and initial-initial antenna functions associated with initial-state collinear divergences. The remaining $\e$-poles exactly match the ones of the virtual matrix element, once the operator in~\eqref{J21} is evaluated on the corresponding Born-level amplitude in colour space. In particular, at one loop the following relation holds:
	\begin{equation}\label{J21relation}
		\poles\left[\J{1}(i_g,j_g)\right]=\poles\left[\text{Re}\left(\ourIop{1}{i_g j_g}{\e,\mu_r^2}\right)\right],
	\end{equation}
	where the integrated dipole on the left-hand side can be in the FF, IF or II configuration. Therefore, exploiting the dipole operator defined in~\eqref{J21}, it is possible to express the NLO virtual subtraction term as
	\begin{eqnarray}\label{sigT}
		\dd\sigpart{T}{gg,NLO}&=&\coeffVNLO\int\dr{x_1}\dr{x_2}\dphi{n}(p_3,\dots,p_{n+2};x_1 p_1,x_2 p_2)\jet{n}{n}{\lb p \rb_n}\nonumber\\
		&&\times 2\braket{\ampnum{n+2}{0}|\Jcol{1}(\e)|\ampnum{n+2}{0}}\,.
	\end{eqnarray}
	This virtual subtraction term describes in a general way the singularity structure of an \mbox{$(n+2)$-gluon} one-loop matrix element and includes the mass factorization contribution. Once this is subtracted from $\dd\sigpart{V}{gg,NLO}$, an $\e$-finite quantity is obtained, which can be integrated via Monte Carlo techniques. The explicit expression for the subtraction term in~\eqref{sigT} is obtained computing the colour sandwiches $\braket{\ampnum{n+2}{0}|\T_i\cdot\T_j|\ampnum{n+2}{0}}$ and dressing them with the associated colour stripped integrated dipoles, as indicated by the structure of~\eqref{J21}. We remark that~\eqref{sigT} is a completely general result in the case of gluon scattering: it is valid for any number of external legs $n$ and retains the full $N_c$ dependence.
	
	\subsection{NLO real subtraction term}\label{sec:NLOR}
	
	In the traditional approach, the NLO real subtraction term, describing the divergent behaviour of the real emission matrix elements in the IR limits, is constructed combining unintegrated NLO antenna functions with reduced matrix elements. In the colourful antenna approach, as explained in section~\ref{sec:NLOV}, the starting point is the virtual subtraction term, while the real subtraction term is obtained in an automated way from it. The key concept is the cancellation of the IR singularities between real and virtual corrections. Apart from the mass factorization counterterm, each term in $\dd\sigpart{T}{gg,NLO}$ must have an unintegrated counterpart in $\dd\sigpart{S}{gg,NLO}$. Therefore, it is possible to define a correspondence between integrated and unintegrated structures appearing at the virtual and real level respectively:
	\begin{equation}\label{insertion}
		\mathcal{X}_{3}^{0}(s_{ij})\,\anum{n+2}{0}(\sigma,\sigma',\{.,i,.,j,.\})\,\leftrightarrow\,X_{3}^{0}(i,k,j)\anum{n+2}{0}(\sigma,\sigma',\{.,\wt{ik},.,\wt{kj},.\}),
	\end{equation}
	where $\mathcal{X}_{3}^{0}(s_{ij})$ is the integrated tree-level antenna function obtained integrating $X_{3}^{0}(i,k,j)$ over the phase space of the unresolved parton $k$. The notation $\widetilde{ik}$ and $\widetilde{kj}$ indicates a suitable momentum mapping $\lb p_i,p_k,p_j\rb\to\{ p_{\wt{ik}},p_{\wt{kj}}\}$~\cite{Kosower:2002su} from the $(n+1)$-particle phase space to the $n$-particle phase space, obtained after the integration over the phase space of the unresolved gluon $k$. Due to this correspondence, once the virtual subtraction term is obtained, the structure of the real subtraction term can be completely determined by inserting an unresolved gluon between each pair of hard radiators appearing in the integrated dipoles. This involves a transition from an integrated NLO antenna to an unintegrated one and from a genuine LO colour interference to a reduced one where the $n$-particle momenta are meant to be obtained from a $(n+1)$-particle phase space through a suitable mapping. The right-hand-side of~\eqref{insertion} reproduces the divergent behaviour of the real interference $\anum{n+3}{0}(\sigma,\sigma';\{.,i,.,k,.,j,.\})$ when gluon $k$ is unresolved between the hard pair $(i,j)$. 
	
	The procedure to obtain $\dd\sigpart{S}{NLO,gg}$ from $\dd\sigpart{T}{NLO,gg}$ can be summarized as follows:
	
	\begin{itemize}
		\item remove the splitting kernels from the expression of the integrated dipoles $\J{1}$ in $\dd\sigpart{T}{NLO,gg}$;
		
		\item replace each integrated gluon-gluon $\XFFint{F}{3}{0}$ with its unintegrated version inserting an unresolved gluon:
		\begin{equation}\label{f30replace}
			\begin{aligned}
				\text{FF:  }&\quad\XFFint{F}{3}{0}(s_{ij})&\to&\quad 3\,f_{3}^{0}(i,k,j),\\
				\text{IF:  }&\quad\XIFint{F}{3}{0}{g}(s_{1i})&\to&\quad 2\,f_{3,g}^{0}(1,k,i),\\
				\text{II:  }&\quad\XIIint{F}{3}{0}{gg}(s_{12})&\to&\quad F_{3,gg}^{0}(1,k,2),
			\end{aligned}
		\end{equation}
		where the numerical coefficients are symmetry factors related to the antenna phase space;
		
		\item suitably replace the momenta in the colour interferences, according to the accompanying integrated antenna:
		\begin{equation}\label{ampreplace}
			\anum{n+2}{0}(\sigma,\sigma',\{.,i,.,j,.\})\to\anum{n+2}{0}(\sigma,\sigma',\{.,\wt{ik},.,\wt{kj},.\});
		\end{equation}
		
		\item apply the same momenta relabelling to the jet function:
		\begin{equation}\label{jetreplace}
			\jet{n}{m}{\dots,p_i,\dots,p_j,\dots}\to \jet{n}{m}{\dots,p_{\wt{ik}},\dots,p_{\wt{kj}},\dots};
		\end{equation}
		
		\item the obtained expression is now a function of $n+3$ momenta, so the phase space has to be adjusted accordingly:
		\begin{equation}
			\dphi{n}(p_3,\dots,p_{n+2};x_1 p_1,x_2 p_2)\to \dphi{n+1}(p_3,\dots,p_{n+3};x_1 p_1,x_2 p_2)
		\end{equation}
		and a sum over suitable permutations of these momenta is needed to have the full real subtraction term;
		
		\item the overall factor used at the virtual level needs to be replaced by the appropriate one for the real correction:
		\begin{equation}
			\coeffVNLO\to\coeffRNLO=s_{R}\left(4\pi\alpha_s\right)\coeffLO,
		\end{equation}
		where $s_{R}$ compensates the different final state symmetry factor for the real radiation and for a process with $n$ final-state gluons at the Born level we have:
		\begin{equation}
			s_{R}=\dfrac{n!}{(n+1)!}=\dfrac{1}{n+1}.
		\end{equation}	
	\end{itemize}
	
	With this procedure for the derivation of the real subtraction term, the cancellation of IR singularities between the real and virtual correction is trivially guaranteed. What is left to check is that the constructed $\dd\sigpart{S}{NLO,gg}$ correctly reproduces the divergent behaviour of the real correction matrix element in each unresolved limit once and only once. If one inserts an unresolved gluon between each pair in the $(n+2)$-gluon set and correctly sums over the relevant permutations of the resulting set of $n+3$ partons (with $n+1$ in the final state), it is clear that all possible unresolved limits at the real level are taken into account. The double-counting of any unresolved configuration is avoided since the unintegrated antenna functions in~\eqref{f30replace} are explicitly constructed to only reproduce the divergence arising when gluon $k$ is unresolved between the two hard radiators $i$ and $j$. In other words, each unintegrated $f_{3}^{0}$ addresses one and only one configuration of unresolved gluon and hard radiators pair.

	We introduce the following notation for the procedure that we have just illustrated to convert integrated quantities into their unintegrated counterparts:
	\begin{equation}\label{unint}
		\dsigSNLO{gg}=-\ins{\dsigTNLO{gg}}.
	\end{equation}
	The action of the $\ins{\cdot}$ operator on its argument is comprehensive of the steps discussed above and can be systematically formulated as follows:
	
	\begin{enumerate}
		\item Removal of the splitting kernels from the integrated dipoles;
		\item Transition from integrated three-parton antenna functions to unintegrated ones, as indicated by~\eqref{f30replace};
		\item Momenta relabeling within colour interferences and jet functions according to the accompanying antenna function;
		\item Sum over permutations of the $n+3$ momenta to cover all possible IR limits;
		\item Dressing of the obtained expression with the appropriate phase space and overall coefficient factor.
	\end{enumerate}
	The minus sign in~\eqref{unint} is explained in~\eqref{sigTNLO} and is needed to make the virtual subtraction term cancel with the integrated real subtract term once the full result is computed. This set of operations is sufficient at NLO, while it needs to be extended to be applied at NNLO. Nevertheless, as we show in section~\ref{sec:subNNLO}, the majority of the NNLO subtraction terms can be generated through the application of $\ins{\cdot}$ to integrated quantities, namely through the very same insertion used at NLO. The only exception is represented by configurations that require a simultaneous double insertion of unresolved gluons.
	
	A clear parallelism can be identified between the colourful antenna approach at NLO and the Catani-Seymour dipole formalism \cite{Catani:1996jh}. In both schemes, the real subtraction term is obtained as a combination of dipole contributions exploiting the general factorization properties of QCD in soft and collinear limits. However, in the Catani-Seymour dipole formalism, as well as in other subtraction schemes both at NLO and NNLO, the principal effort precisely lies in the construction of the subtraction term for the real emission. The integrated version of such subtraction term eventually cancels the $\e$-poles in virtual correction. On the opposite, the peculiarity of the described colourful antenna method consists in prioritizing the construction of the virtual subtraction term, from which the real subtraction term is then systematically inferred. As in the traditional subtraction schemes the key requirement is the analytical integrability of the real subtraction term over the phase space of the unresolved radiation, here it is crucial that the virtual subtraction term is expressed in a language suitable for the \textit{unintegration} procedure. The correspondence between integrated antenna functions, collected in \eqref{J21}, and their unintegrated versions guarantees this property.
	
	On a practical standpoint, both the computation of colour operators insertions like $\braket{\ampnum{n+2}{0}|\Jcol{1}(\e)|\ampnum{n+2}{0}}$ and the application of $\ins{\cdot}$ can be performed with any symbolic algebra program such as FORM, \textit{Mathematica} or \textit{Maple}. Once the symbolic expressions for the subtraction terms are generated, they can be implemented in a numerical code which performs the Monte Carlo integration.  
	

	\section{Colourful antenna subtraction at NNLO}\label{sec:subNNLO}
	
	The NNLO QCD correction to an $n$-jet cross section is given by:
	\begin{eqnarray}\label{NNLOcs}
		\dsigNNLO{ab}&=&\int_n\left(\dsigVV{ab}+\dsigMFNNLO{ab}{2}\right)\nonumber\\
		&+&\int_{n+1}\left(\dsigRV{ab}+\dsigMFNNLO{ab}{1}\right) + \int_{n+2}\dsigRR{ab},
	\end{eqnarray}
	where $\dsigVV{ab}$ represents the double virtual correction, $\dsigRV{ab}$ the real virtual correction and $\dsigRR{ab}$ the double real correction. The mass factorization counterterm is split into two terms associated with $n$- and $(n+1)$-particle final states, respectively $\dsigMFNNLO{ab}{2}$ and $\dsigMFNNLO{ab}{1}$.
	
	As for the NLO case, the quantity in~\eqref{NNLOcs} cannot be directly computed with numerical methods. The singular behaviour of both the double real and real virtual corrections in the IR limits must be subtracted and the explicit poles in the double virtual and real virtual matrix elements need to be properly removed. To achieve this, the NNLO cross section is rewritten in the context of antenna subtraction as~\cite{Currie:2013vh}:
	\begin{eqnarray}\label{NNLOcssub}
		\dsigNNLO{ab}&=&\int_n\left[\dsigVV{ab}-\dsigUNNLO{ab}\right]\nonumber\\
		&+&\int_{n+1}\left[\dsigRV{ab}-\dsigTNNLO{ab}\right]\nonumber\\
		&+&\int_{n+2}\left[\dsigRR{ab}-\dsigSNNLO{ab}\right],
	\end{eqnarray}
	where the subtracted quantities are the double virtual, the real virtual and the double real subtraction term. These contributions have the following form:
	\begin{eqnarray}\label{subtermsNNLO}
		\dsigSNNLO{ab}&=&\dsigSNNLOspe{ab}{1}+\dsigSNNLOspe{ab}{2}\,,\nonumber\\
		\dsigTNNLO{ab}&=&\dsigVSNNLO{ab}-\int_1 \dsigSNNLOspe{ab}{1}-\dsigMFNNLO{ab}{1}\,,\nonumber\\
		\dsigUNNLO{ab}&=&-\int_1 \dsigVSNNLO{ab}-\int_2 \dsigSNNLOspe{ab}{2}-\dsigMFNNLO{ab}{2}\,.
	\end{eqnarray}
	The double real subtraction term has been decomposed into two contributions which contain single and double unresolved IR limits. In the real virtual subtraction term, $\dsigVSNNLO{ab}$ cancels the implicit singular behaviour of the real virtual correction in the soft and collinear limits, while the remaining contributions remove the explicit $\e$-poles. 
	
	Analogously to NLO, in the traditional antenna subtraction approach the IR limits~\cite{Campbell:1997hg,Catani:1999ss,Bern:1999ry,Catani:2000vq}
	 of the double real and real virtual matrix elements are studied first to construct $\dsigSNNLO{ab}$ and $\dsigVSNNLO{ab}$. These terms are then integrated and combined with the mass factorization contributions to obtain the full infrastructure in~\eqref{subtermsNNLO}. 
	
	The application of the colourful antenna subtraction method at NNLO begins by addressing the double virtual correction in colour space. The singularity structure of two-loop amplitudes can be predicted in a general way and therefore the construction of the double virtual subtraction term can be achieved without the need of studying the real emission corrections first. The predictability of the poles of two-loop amplitudes in colour space is neither limited by the number of partons in the process nor conceptually more involved beyond leading colour. For this reason, if one aims at a general application of antenna subtraction at NNLO, starting from the construction of $\dsigUNNLO{ab}$ in colour space is significantly more convenient. Once the double virtual subtraction term is constructed, the real virtual and double real subtraction terms are generated by the insertion of unresolved partons, exploiting the relations between integrated and unintegrated blocks in the subtraction terms in~\eqref{subtermsNNLO} and removing any spurious singularity.
	
\begin{figure}[t]
		\centering
		\includegraphics[width=\linewidth,keepaspectratio]{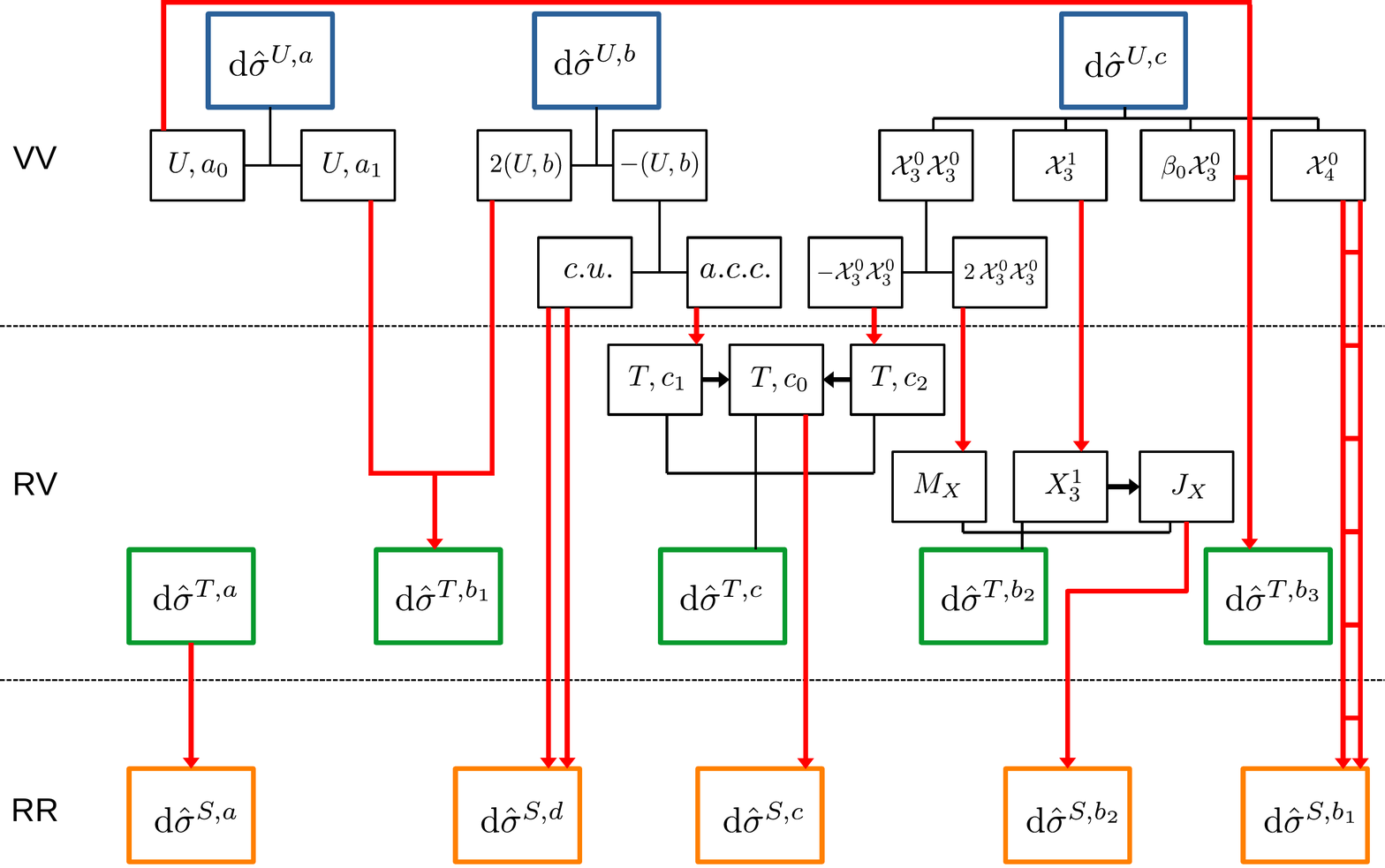}
		\caption{Structure of colourful antenna subtraction at NNLO. Descendant red arrows represent the transition from an integrated quantity to its unintegrated counterpart via single insertion (single arrow), double iterated insertion (two disjoint arrows) or double simultaneous insertion (two connected arrows) of unresolved gluons. The definitions of each component are listed in Table \ref{tab:equations}.}\label{fig:colant_scheme}
	\end{figure}

	In the rest of this section we show how this can be achieved in the case of gluon scattering. To support the explanation, we summarize the procedure in Figure~\ref{fig:colant_scheme}. Single descendant red arrows represent the transition from an integrated quantity to its unintegrated counterpart by means of the insertion of an unresolved gluon. Two disjoint red arrows indicate the iterated insertion of two unresolved gluons, while two connected red arrows indicate the simultaneous insertion of two unresolved gluons. 
	
	\begin{table}[t]
		\begin{minipage}{.5\linewidth}
			\centering
			\setlength{\tabcolsep}{5pt}
			\begin{tabular}{cc|cc}
				\hline
				\multirow{2}{*}{$U,a$} & \multirow{2}{*}{eq. \eqref{sigUa}}
				& $U,a_0$ & eq. \eqref{sigUa0} \\\cline{3-4}
				&& $U,a_1$ & eq. \eqref{sigUa1} \\\cline{3-4}
				\hline
				\multirow{2}{*}{$U,b$}  & \multirow{2}{*}{eq. \eqref{sigUb}}
				& $U,b,c.u.$ & eq. \eqref{sigUbcu} \\\cline{3-4}
				&& $U,b,a.c.c.$ & eq. \eqref{sigUbacc} \\\cline{3-4}
				\hline	
				$U,c$ & eq. \eqref{sigUc} & & \\\cline{3-4}
				\hline
				\hline
				$S,a$ & eq. \eqref{sigSa} &&\\
				$S,b_1$ & eq. \eqref{sigSb1} && \\
				$S,b_2$ & eq. \eqref{sigSb2} && \\
				$S,c$ & eq. \eqref{sigSc} && \\
				$S,d$ & eq. \eqref{sigSd} && \\
				\hline
			\end{tabular}
		\end{minipage}%
		\quad
		\begin{minipage}{.5\linewidth}
			\vspace{-0.8cm}
			\centering
			\setlength{\tabcolsep}{5pt}
			\begin{tabular}{cc|cc}
				\hline
				$T,a$ & eq. \eqref{sigTa} & & \\
				$T,b_1$ & eq. \eqref{sigTb1} & & \\\cline{3-4}
				\multirow{3}{*}{$T,b_2$} & \multirow{3}{*}{eq. \eqref{sigTb2}}
				& $T,b_2,X_3^1$ & eq. \eqref{sigTb2X31} \\\cline{3-4}
				&& $T,b_2,J_X$ & section \ref{sec:dsigTb} \\\cline{3-4}
				&& $T,b_2,M_X$ & eq. \eqref{sigTb2MX} \\\cline{3-4}
				$T,b_3$ & eq. \eqref{sigTb3} & & \\\cline{3-4}
				\multirow{3}{*}{$T,c$} & \multirow{3}{*}{eq. \eqref{sigTc}}
				& $T,c_1$ & eq. \eqref{sigTc1} \\\cline{3-4}
				&& $T,c_2$ & eq. \eqref{sigTc2} \\\cline{3-4}
				&& $T,c_0$ & section \ref{sec:dsigTc}\\\cline{3-4}
				\hline
			\end{tabular}
		\end{minipage} 
		\caption{Definitions of each term appearing in Figure \ref{fig:colant_scheme}.}\label{tab:equations}
	\end{table}
	To facilitate the navigation through the paper, we list in Table \ref{tab:equations} the location of the definition of each term appearing in Figure \ref{fig:colant_scheme}.
	
	\subsection{IR singularity structure at two loops}
	
	The singularity structure of renormalized two-loop amplitudes in QCD is known~\cite{Catani:1998bh} and can be described in colour space by:
	\begin{equation}\label{2l-sing}
		\ketamp{n+2}{2}=\Iop{1}{\epsilon,\mu_r^2}\ketamp{n+2}{1}+\Iop{2}{\epsilon,\mu_r^2}\ketamp{n+2}{0}+\ketampdep{n+2}{2,\text{fin}}{\mu_r^2},
	\end{equation}
	where, as before, $\ketampdep{n+2}{2,\text{fin}}{\mu_r^2}$ is a finite remainder. The two-loop Catani IR insertion operator has the following expression~\cite{Catani:1998bh}:
	\begin{eqnarray}\label{I2}
		\Iop{2}{\epsilon,\mu_r^2}&=&-\dfrac{\beta_0}{\e}\Iop{1}{\epsilon,\mu_r^2;\pset}-\dfrac{1}{2}\Iop{1}{\epsilon,\mu_r^2}\Iop{1}{\epsilon,\mu_r^2}\nonumber\\
		&&+e^{-\e\gamma_E}\dfrac{\Gamma(1-2\e)}{\Gamma(1-\e)}\left(\dfrac{\beta_0}{\e}+K\right)\Iop{1}{2\epsilon,\mu_r^2}\nonumber\\
		&&+\Hop{2}{\epsilon,\mu_r^2}\,.
	\end{eqnarray}
	In the gluons-only case with $N_f=0$ we have:
	\begin{eqnarray}
		\beta_0&=&b_0 N_c=\dfrac{11}{6}N_c,\\
		K&=&k_0 N_c=\left(\dfrac{67}{18}-\dfrac{\pi^2}{6}\right)N_c.
	\end{eqnarray}
	
	The colour structure of~\eqref{I2} is more involved than the colour charge dipole structure of~\eqref{I1}. As anticipated, products of two colour charge dipoles appear. The last line of~\eqref{I2} contains the hard radiation function $\Hop{2}{\epsilon,\mu_r^2;\pset}$~\cite{Catani:1998bh,Bern:2003ck,Becher:2009qa}, which can be decomposed in the following manner:
	\begin{equation}\label{H2}
		\Hop{2}{\epsilon,\mu_r^2}=\sum_{i}C_i\ourHop{2}{i}{\e}\mathbf{Id}+\check{\bm{H}}^{(2)}(\epsilon,\mu_r^2),
	\end{equation}
	where the sum runs over the $n+2$ external partons and $C_i$ are the Casimir coefficients. The first term in~\eqref{H2} is proportional to the identity in colour space, while the second term has a non-trivial colour structure, which can not be in general expressed in terms of colour charge dipoles. However, the second term vanishes when sandwiched between tree-level states~\cite{Bern:2003ck,Becher:2009qa}:
	\begin{equation}\label{H2vanish}
		\braamp{n+2}{0}\check{\bm{H}}^{(2)}(\epsilon,\mu_r^2)\ketamp{n+2}{0}=0.
	\end{equation}
	For the purpose of describing the IR singularity structure of the two-loop squared matrix elements, the hard radiation function $\Hop{2}{\epsilon,\mu_r^2}$ needs to be evaluated on tree-level states and therefore it is possible to neglect $\check{\bm{H}}^{(2)}(\epsilon,\mu_r^2)$ in its decomposition. As we did for the splitting kernels in section~\ref{sec:MFNLO}, we can express a colour operator proportional to the identity as a sum of colour charge dipoles using colour conservation. We can then rewrite:
	\begin{eqnarray}\label{H2todipole}
		\sum_{i}C_i\ourHop{2}{i}{\e}\mathbf{Id}&=&-\sum_{i}\ourHop{2}{i}{\e}\sum_{j\ne i}\T_i\cdot\T_j\nonumber\\
		&=&-\sum_{(i,j)}\ourHop{2}{ij}{\e}\T_i\cdot\T_j,
	\end{eqnarray}
	where, as usual, the sum runs over pairs of partons and $\ourHopnodep{2}{ij}=\ourHopnodep{2}{i}+\ourHopnodep{2}{j}$. In the gluon-only case we only need $\ourHop{2}{g}{\e}$ for $N_f=0$: 
		\begin{eqnarray}
		\ourHop{2}{g}{\epsilon}= \dfrac{e^{\e\gamma_E}}{4\Gamma(1-\e)}\dfrac{N_c}{\e}\left[\dfrac{5}{12}+\dfrac{11}{144}\pi^2+\dfrac{\zeta_3}{2}\right]\,.
	\end{eqnarray}
	Using~\eqref{I1v2} and~\eqref{H2todipole} and neglecting $\check{\bm{H}}^{(2)}(\epsilon,\mu_r^2)$ we can rearrange equation~\eqref{I2} as:
	\begin{eqnarray}\label{I2v2}
		\Iop{2}{\epsilon,\mu_r^2}&=&-\dfrac{\beta_0}{\e}\sum_{(i,j)}\ourIop{1}{ij}{\e,\mu_r^2}\T_i\cdot\T_j\nonumber\\
		&&-\dfrac{1}{2}\sum_{(i,j)}\sum_{(k,l)}\ourIop{1}{ij}{\e,\mu_r^2}\ourIop{1}{kl}{\e,\mu_r^2}(\T_i\cdot\T_j)(\T_k\cdot\T_l)\nonumber\\
		&&+\sum_{(i,j)}\ourIop{2}{ij}{\e,\mu_r^2}\T_i\cdot\T_j,
	\end{eqnarray}
	where 
	\begin{equation}
		\ourIop{2}{ij}{\e,\mu_r^2}=e^{-\e\gamma_E}\dfrac{\Gamma(1-2\e)}{\Gamma(1-\e)}\left(\dfrac{\beta_0}{\e}+K\right)\ourIop{1}{ij}{2\e,\mu_r^2}-\ourHop{2}{ij}{\epsilon}\,.
	\end{equation}
	
	We can now use~\eqref{1l-sing},~\eqref{2l-sing} and~\eqref{I2} to express the singularity structure of a two-loop matrix element:
	\begin{eqnarray}\label{M2poles}
		\poles\left(\me{n}{2}\right)&=&\poles\left(\braket{\ampnum{n}{2}|\ampnum{n}{0}}+\braket{\ampnum{n}{0}|\ampnum{n}{2}}+\braket{\ampnum{n}{1}|\ampnum{n}{1}}\right)\nonumber\\
		&=&\poles\Big\lbrace\Big.\braket{\ampnum{n}{1}|\Iop{1}{\e}+\Iopd{1}{\e}|\ampnum{n}{0}}+\braket{\ampnum{n}{0}|\Iop{1}{\e}+\Iopd{1}{\e}|\ampnum{n}{1}}\nonumber\\
		&&-\dfrac{1}{2}\braket{\ampnum{n}{0}|\left(\Iop{1}{\e}+\Iopd{1}{\e}\right)\left(\Iop{1}{\e}+\Iopd{1}{\e}\right)|\ampnum{n}{0}}\nonumber\\
		&&-\betaoe\braket{\ampnum{n}{0}|\Iop{1}{\e}+\Iopd{1}{\e}|\ampnum{n}{0}}\nonumber\\
		&&+\Itwoepsiloncoeff\braket{\ampnum{n}{0}|\Iop{1}{2\e}+\Iopd{1}{2\e}|\ampnum{n}{0}}\nonumber\\
		&&+\braket{\ampnum{n}{0}|\Hop{2}{\e}+\Hopd{2}{\e}|\ampnum{n}{0}}\Big.\Big\rbrace\,.
	\end{eqnarray}
	We see again that only the real part of the insertion operators is needed to describe the singularity structure. Using~\eqref{I1v2} and~\eqref{I2v2} it is possible to recast equation~\eqref{M2poles} as:
	\begin{eqnarray}\label{M2poles2}
		\poles\left(\me{n}{2}\right)&=&\poles\Big\lbrace\Big.\sum_{(i,j)}2\text{Re}\left[\ourIop{1}{ij}{\e,\mu_r^2}\right]\left[\braket{\ampnum{n}{1}|\T_i\cdot\T_j|\ampnum{n}{0}}+\braket{\ampnum{n}{0}|\T_i\cdot\T_j|\ampnum{n}{1}}\right]\nonumber\\
		&&-\dfrac{1}{2}\sum_{(i,j)}\sum_{(k,l)}2\text{Re}\left[\ourIop{1}{ij}{\e,\mu_r^2}\right]2\text{Re}\left[\ourIop{1}{lk}{\e,\mu_r^2}\right]\braket{\ampnum{n}{0}|(\T_i\cdot\T_j)(\T_k\cdot\T_l)|\ampnum{n}{0}}\nonumber\\
		&&-\betaoe\sum_{(i,j)}2\text{Re}\left[\ourIop{1}{ij}{\e,\mu_r^2}\right]\braket{\ampnum{n}{0}|\T_i\cdot\T_j|\ampnum{n}{0}}\nonumber\\
		&&+\sum_{(i,j)}2\text{Re}\left[\ourIop{2}{ij}{\e,\mu_r^2}\right]\braket{\ampnum{n}{0}|\T_i\cdot\T_j|\ampnum{n}{0}}\Big.\Big\rbrace.
	\end{eqnarray}
	The poles of the double virtual cross section for gluons-only processes are therefore given by:
	\begin{eqnarray}\label{VVpoles}
	\lefteqn{\poles\left(\sigpart{VV}{gg,NNLO}\right)=\coeffVVNNLO\int\dphi{n}(p_3,\dots,p_{n+2};p_1,p_2)\,\jet{n}{n}{\lb p \rb_n}}\nonumber\\
		&\times&\poles\Big\lbrace\Big.\sum_{(i_g,j_g)}2\text{Re}\left[\ourIop{1}{i_gj_g}{\e,\mu_r^2}\right]\left[\braket{\ampnum{n}{1}|\T_{i_g}\cdot\T_{j_g}|\ampnum{n}{0}}+\braket{\ampnum{n}{0}|\T_{i_g}\cdot\T_{j_g}|\ampnum{n}{1}}\right]\nonumber\\
		&&-\dfrac{1}{2}\sum_{(i_g,j_g)}\sum_{(k_g,l_g)}2\text{Re}\left[\ourIop{1}{i_gj_g}{\e,\mu_r^2}\right]2\text{Re}\left[\ourIop{1}{l_gk_g}{\e,\mu_r^2}\right]\braket{\ampnum{n}{0}|(\T_{i_g}\cdot\T_{j_g})(\T_{k_g}\cdot\T_{l_g})|\ampnum{n}{0}}\nonumber\\
		&&-\betaoe\sum_{(i,j)}2\text{Re}\left[\ourIop{1}{i_gj_g}{\e,\mu_r^2}\right]\braket{\ampnum{n}{0}|\T_{i_g}\cdot\T_{j_g}|\ampnum{n}{0}}\nonumber\\
		&&+\sum_{(i,j)}2\text{Re}\left[\ourIop{2}{i_gj_g}{\e,\mu_r^2}\right]\braket{\ampnum{n}{0}|\T_{i_g}\cdot\T_{j_g}|\ampnum{n}{0}}\Big.\Big\rbrace,
	\end{eqnarray}
	where 
	\begin{equation}
		\coeffVVNNLO=\coeff^2\coeffLO\,.
	\end{equation}
	
	\subsection{NNLO mass factorization}\label{sec:MFNNLO}
	
	At NNLO, as indicated in~\eqref{NNLOcssub} and~\eqref{subtermsNNLO}, we have two different contributions to the mass factorization counterterm: the double virtual and the real virtual mass factorization terms. The decomposition is meant to separate contributions that are defined on the $n$-particle phase space from terms that are defined on the $(n+1)$-particle phase space.
	
	\subsubsection{Double virtual mass factorization term}
	
	We first address the double virtual mass factorization term at NNLO. It can be written as~\cite{Currie:2013vh}:
	\begin{eqnarray}\label{MFVV}
		\dd\sigpart{MF,2}{ab,NNLO}&=&-\int\dr{x_1}\dr{x_2}\sum_{c,d}\Bigg\lbrace\Bigg.\coeff\left[\Gammaone{ab;cd}{x_1,x_2}\left(\dd\sigpart{V}{cd,NLO}-\dd\sigpart{T}{cd,NLO}\right)\right]\nonumber\\
		&&\hspace{0.5cm}+\coeff^2\Big[\Big.\Gammatwo{ab;cd}{x_1,x_2}-\dfrac{\beta_0}{\e}\Gammaone{ab;cd}{x_1,x_2}\nonumber\\
		&&\hspace{3cm}+\dfrac{1}{2}\sum_{\alpha,\beta}\left[\Gamma^{(1)}_{ab;\alpha\beta}\otimes\Gamma^{(1)}_{\alpha\beta;cd}\right](x_1,x_2)\Big.\Big]\dd\sigpart{}{cd,LO}\Bigg.\Bigg\rbrace.
	\end{eqnarray}
	The reduced two-loop mass factorization kernel is defined as~\cite{Currie:2013vh}:
	\begin{equation}
		\Gammatwo{ab;cd}{x_1,x_2}=\Gammatwo{ca,\text{full}}{x_1}\delta_{db}\delta(1-x_2)+\Gammatwo{db,\text{full}}{x_2}\delta_{ca}\delta(1-x_1),
	\end{equation}
	where the $\Gammatwo{ca,\text{full}}{x_1}$ can be written in terms of LO and NLO Altarelli-Parisi spitting kernels~\cite{Altarelli:1977zs,Curci:1980uw,Furmanski:1980cm} and have their own colour decomposition. In the gluons-only case we simply need:
	\begin{equation}
		\Gammatwo{gg;gg}{x_1,x_2}=\Gammatwo{gg,\text{full}}{x_1}\delta(1-x_2)+\Gammatwo{gg,\text{full}}{x_2}\delta(1-x_1),
	\end{equation}
	where we discard any contribution in the two-loop splitting kernel except for the gluon-to-gluon one for $N_f=0$. Thus we have:
	\begin{equation}
		\Gammatwo{gg,\text{full}}{x_i}=N_c^2\,\Gammatwo{gg}{x_i}=-\dfrac{1}{2\e}\left(N_c^2\,p^1_{gg}(x_i)+\dfrac{\beta_0\,N_c}{\e}p^0_{gg}(x_i)\right),
	\end{equation}
	where $p^1_{gg}(x_i)$ is the Altarelli-Parisi NLO splitting kernel given by~\cite{Furmanski:1980cm}:
	\begin{eqnarray}
		p^1_{gg}(x)=&&\left[\dfrac{67}{9}-4\ln(x)\ln(1-x)+\ln^2(x)-\dfrac{\pi^2}{3}\right]\left[\left(\dfrac{1}{1-x}\right)_{+}+\dfrac{1}{x}-2+x(1-x)\right]\nonumber\\
		&&+2S_2(x)\left[\dfrac{1}{1+x}-\dfrac{1}{x}-2-x(1+x)\right]\nonumber\\
		&&+\dfrac{27}{2}(1-x)+\dfrac{67}{9}\left(x^2-\dfrac{1}{x}\right)-\left(\dfrac{25}{3}-\dfrac{11}{3}x+\dfrac{44}{3}x^2\right)\ln(x)\nonumber\\
		&&+4(1+x)\ln^2(x)+\left(\dfrac{8}{3}+3\zeta_3\right)\delta(1-x),
	\end{eqnarray}
	with
	\begin{equation}
		S_2(x)=-2\text{Li}_{2}(-x)+\dfrac{1}{2}\ln^2(x)-2\ln(x)\ln(1+x)-\dfrac{\pi^2}{6}.
	\end{equation}
	As we did in section~\ref{sec:MFNLO}, it is possible to express the two-loop identity-preserving mass factorization kernel in colour space as:
	\begin{equation}
		\BGammatwo{gg;gg}{x_1,x_2}=\BGammatwo{gg,\text{full}}{x_1}\delta(1-x_2)+\BGammatwo{gg,\text{full}}{x_2}\delta(1-x_1),
	\end{equation}
	where 
	\begin{equation}
		\BGammatwo{gg,\text{full}}{x_i}=-\Gammatwo{gg,\text{full}}{x_i}\dfrac{1}{C_A}\sum_{j\ne i}\T_i\cdot\T_j,\quad i=1,2.
	\end{equation}
	We can then write the expression for the required double virtual mass factorization counterterm as:
	\begin{eqnarray}\label{MFVVgg}
		\dd\sigpart{MF,2}{gg,NNLO}&=&-\coeffVVNNLO\int\dr{x_1}\dr{x_2}\dphi{n}(p_3,\dots,p_{n+2};x_1p_1,x_2p_2)\jet{n}{n}{\pset}\nonumber\\
		&&\Big\lbrace\Big.\braket{\ampnum{n+2}{0}|\BGammaone{gg;gg}{x_1,x_2}|\ampnum{n+2}{1}}+\braket{\ampnum{n+2}{1}|\BGammaone{gg;gg}{x_1,x_2}|\ampnum{n+2}{0}}\nonumber\\
		&&\hspace{2mm}-2\braket{\ampnum{n+2}{0}|\left[\BGammaonenodep{gg;gg}\otimes\Jcol{1}(\e)\right](x_1,x_2)|\ampnum{n+2}{0}}
		\nonumber\\
		&&\hspace{2mm}
		+\dfrac{1}{2}\braket{\ampnum{n+2}{0}|\left[\BGammaonenodep{gg;gg}\otimes\BGammaonenodep{gg;gg}\right](x_1,x_2)|\ampnum{n+2}{0}}\nonumber\\
		&&\hspace{2mm}-\dfrac{\beta_0}{\e}\braket{\ampnum{n+2}{0}|\BGammaone{gg;gg}{x_1,x_2}|\ampnum{n+2}{0}}
		\nonumber\\
		&&\hspace{2mm}+\braket{\ampnum{n+2}{0}|\BGammatwo{gg;gg}{x_1,x_2}|\ampnum{n+2}{0}}\Big.\Big\rbrace,
	\end{eqnarray}
	where we used the expression of $\dd\sigpart{T}{cd,NLO}$ given in \eqref{sigT}.
	
	\subsubsection{Real virtual mass factorization term}
	
	The real virtual mass factorization term is given by~\cite{Currie:2013vh}:
	\begin{equation}\label{MFRV}
		\dd\sigpart{MF,1}{ab,NNLO}=-\coeff\sum_{c,d}\int\dr{x_1}\dr{x_2}\Gammaone{ab;cd}{x_1,x_2}\left(\dd\sigpart{R}{cd,NLO}-\dd\sigpart{S}{cd,NLO}\right).
	\end{equation}
	For later convenience, it can be split into two contributions:
	\begin{equation}
		\dd\sigpart{MF,1}{ab,NNLO}=\dd\sigpart{MF,1,a}{ab,NNLO}+\dd\sigpart{MF,1,b}{ab,NNLO},
	\end{equation}
	with 
	\begin{eqnarray}
		\dd\sigpart{MF,1,a}{ab,NNLO}&=&-\coeff\sum_{c,d}\int\dr{x_1}\dr{x_2}\Gammaone{ab;cd}{x_1,x_2}\dd\sigpart{R}{cd,NLO},\\
		\dd\sigpart{MF,1,b}{ab,NNLO}&=&\coeff\sum_{c,d}\int\dr{x_1}\dr{x_2}\Gammaone{ab;cd}{x_1,x_2}\dd\sigpart{S}{cd,NLO}.
	\end{eqnarray}
	In the gluons-only case, in analogy with~\eqref{MFv2}, $\dd\sigpart{MF,1,a}{ab,NNLO}$ can be rewritten as:
	\begin{eqnarray}\label{MF1a}
		\dd\sigpart{MF,1,a}{gg,NNLO}&=&-\coeffRVNNLO\int\dr{x_1}\dr{x_2}\int\dphi{n}(p_3,\dots,p_{n+2};x_1 p_1,x_2 p_2)\,\jet{n}{n+1}{\lb p \rb_{n+1}}\nonumber\\
		&&\times\braket{\ampnum{n+3}{0}|\BGammaone{gg;gg}{x_1,x_2}|\ampnum{n+3}{0}},
	\end{eqnarray}
	where $\ket{\ampnum{n+3}{0}}$ indicates the real correction to the $(n+2)$-gluon Born process and $\coeffRVNNLO$ is the appropriate overall coefficient at the real virtual level:
	\begin{equation}\label{coeffRV}
		\coeffRVNNLO=s_{RV}(4\pi\alpha_s)\coeff\coeffLO,\quad\text{with}\quad s_{RV}=s_{R}.
	\end{equation}
	Similarly, $\dd\sigpart{MF,1,b}{ab,NNLO}$ can be rewritten as:
	\begin{equation}\label{MF1b}
		\dd\sigpart{MF,1,b}{gg,NNLO}=\coeff\int\dr{x_1}\dr{x_2}\Gammaone{gg;gg}{x_1,x_2}\dd\sigpart{S}{gg,NLO}.
	\end{equation}
	We notice that the terms appearing in this mass factorization contribution have a structure like $\Gamma^{(1)}_{gg}X_{3}^{0}\anum{n+2}{0}$. As we show in section \ref{sec:dsigTb}, this term is used to reconstruct one-loop integrated dipoles that are needed in the real virtual subtraction term to remove the explicit poles of one-loop reduced matrix elements.

	\subsection{NNLO double virtual subtraction term}
	
	The double virtual subtraction term at NNLO $\dd\sigpart{U}{gg,NNLO}$, reproduces the explicit poles of the two-loop matrix element and contains the double virtual mass factorization counterterm. In the following we see how to construct $\dd\sigpart{U}{gg,NNLO}$ in a general way relying on the results of the previous sections, in particular equations~\eqref{VVpoles} and~\eqref{MFVVgg}. In analogy with~\eqref{J21}, we define a two-loop insertion operator in colour space:
	\begin{eqnarray}\label{J22}
		\Jcol{2}(\e)&&=N_c\sum_{(i,j)\geq 3}(\T_i\cdot\T_j)\,\J{2}(i_g,j_g)+N_c\sum_{i\ne 1,2}(\T_1\cdot\T_i)\,\J{2}(\hat{1}_g,i_g)\nonumber\\
		&&\,+N_c\sum_{i\ne 1,2}(\T_2\cdot\T_i)\,\J{2}(\hat{2}_g,i_g)+N_c\,(\T_1\cdot\T_2)\,\J{2}(\hat{1}_g,\hat{2}_g)\,.
	\end{eqnarray}
	The two-loop colour stripped integrated dipoles $\J{2}$ have a more involved structure with respect to their one-loop counterparts. The expressions of the gluon-gluon $\J{2}$ are given by~\cite{Currie:2013vh,Currie:2013dwa}:
	\begin{eqnarray}\label{J22expression}
		\J{2}(i_g,j_g)&=&\dfrac{1}{4}\XFFint{F}{4}{0}+\dfrac{1}{3}\XFFint{F}{3}{1}+\dfrac{1}{3}\boe\QQs{ij}\XFFint{F}{3}{0}-\dfrac{1}{9}\left[\XFFint{F}{3}{0}\otimes\XFFint{F}{3}{0}\right],\nonumber\\
		\J{2}(\hat{1}_g,j_g)&=&\dfrac{1}{2}\XIFint{F}{4}{0}{g}+\dfrac{1}{2}\XIFint{F}{3}{1}{g}+\dfrac{1}{2}\boe\QQs{1j}\XIFint{F}{3}{0}{g}-\dfrac{1}{4}\left[\XIFint{F}{3}{0}{g}\otimes\XIFint{F}{3}{0}{g}\right]-\dfrac{1}{2}\Gammatwo{gg}{x_1}\deltatwo,\nonumber\\
		\J{2}(\hat{1}_g,\hat{2}_g)&=&\XIIint{F}{4}{0,adj}{gg}+\dfrac{1}{2}\XIIint{F}{4}{0,n.adj}{gg}+\XIIint{F}{3}{1}{gg}+\boe\QQs{12}\XIIint{F}{3}{0}{gg}-\left[\XIIint{F}{3}{0}{gg}\otimes\XIIint{F}{3}{0}{gg}\right]\nonumber\\
		&&\hspace{5.8cm}-\dfrac{1}{2}\Gammatwo{gg}{x_1}\deltatwo-\dfrac{1}{2}\Gammatwo{gg}{x_2}\deltaone,
	\end{eqnarray}
	where we omitted the dependence on the scale $s_{ij}$ in the integrated antennae. At NNLO, as expected, we see the appearance of integrated four-parton antennae, integrated three-parton one-loop antennae~and a convolution of two three-parton integrated antennae, as well as two-loop mass factorization kernels for initial-final and initial-initial configurations. In analogy with~\eqref{J21relation}, we can relate the pole structure of~\eqref{J22expression} to the  insertion operators in~\eqref{I2v2}:
	\begin{equation}
		\poles\left[N_c\,\J{2}(i_g,j_g)-\dfrac{\beta_0}{\e}\J{1}(i_g,j_g)\right]=\poles\left[\text{Re}\left(\ourIop{2}{gg}{\e,\mu_r^2}-\dfrac{\beta_0}{\e}\ourIop{1}{gg}{\e,\mu_r^2}\right)\right].
	\end{equation}
	Thus, using \eqref{J21} and \eqref{J22} we can construct an expression for the double virtual subtraction term in colour space:
	\begin{eqnarray}\label{sigU}
		\dd\sigpart{U}{gg,NNLO}&=&\coeffVVNNLO\int\dr{x_1}\dr{x_2}\dphi{n}(p_3,\dots,p_{n+2};x_1p_1,x_2p_2)\jet{n}{n}{\pset}\nonumber\\
		&&\times2\Big\lbrace\Big.\braket{\ampnum{n+2}{0}|\Jcol{1}(\e)|\ampnum{n+2}{1}}+\braket{\ampnum{n+2}{1}|\Jcol{1}(\e)|\ampnum{n+2}{0}}\nonumber\\
		&&\hspace{2mm}-\braket{\ampnum{n+2}{0}|\Jcol{1}(\e)\otimes\Jcol{1}(\e)|\ampnum{n+2}{0}}\nonumber\\
		&&\hspace{2mm}-\dfrac{\beta_0}{\e}\braket{\ampnum{n+2}{0}|\Jcol{1}(\e)|\ampnum{n+2}{0}}\nonumber\\
		&&\hspace{2mm}+\braket{\ampnum{n+2}{0}|\Jcol{2}(\e)|\ampnum{n+2}{0}}\Big.\Big\rbrace.
	\end{eqnarray}
	The poles of the mass factorization kernels cancel against initial-state collinear poles in the integrated IF and II antennae and the remaining singularities in~\eqref{sigU} exactly cancel the poles coming from the $(n+2)$-gluon two-loop matrix element in \eqref{VVpoles}.
	
	According to the usual decomposition of the double virtual subtraction term~\cite{Currie:2013vh}, we conveniently split~\eqref{sigU} into the following contributions:
	\begin{eqnarray}\label{sigUa}
		\dd\sigpart{U,a}{gg,NNLO}&=&\coeffVVNNLO\int\dr{x_1}\dr{x_2}\dphi{n}(p_3,\dots,p_{n+2};x_1p_1,x_2p_2)\jet{n}{n}{\pset}\nonumber\\
		&&\times2\,\Big\lbrace\Big.\braket{\ampnum{n+2}{0}|\Jcol{1}(\e)|\ampnum{n+2}{1}}+\braket{\ampnum{n+2}{1}|\Jcol{1}(\e)|\ampnum{n+2}{0}}\nonumber\\
		&&\hspace{1cm}-\dfrac{\beta_0}{\e}\braket{\ampnum{n+2}{0}|\Jcol{1}(\e)|\ampnum{n+2}{0}}\Big.\Big\rbrace,
	\end{eqnarray}
	\begin{eqnarray}\label{sigUb}
		\dd\sigpart{U,b}{gg,NNLO}&=&\coeffVVNNLO\int\dr{x_1}\dr{x_2}\dphi{n}(p_3,\dots,p_{n+2};x_1p_1,x_2p_2)\jet{n}{n}{\pset}\nonumber\\
		&&\times2\,\Big\lbrace\Big.-\braket{\ampnum{n+2}{0}|\Jcol{1}(\e)\otimes\Jcol{1}(\e)|\ampnum{n+2}{0}}\Big.\Big\rbrace,
	\end{eqnarray}
	\begin{eqnarray}\label{sigUc}
		\dd\sigpart{U,c}{gg,NNLO}&=&\coeffVVNNLO\int\dr{x_1}\dr{x_2}\dphi{n}(p_3,\dots,p_{n+2};x_1p_1,x_2p_2)\jet{n}{n}{\pset}\nonumber\\
		&&\times2\,\braket{\ampnum{n+2}{0}|\Jcol{2}(\e)|\ampnum{n+2}{0}}.
	\end{eqnarray}
	We further decompose $\dsigUNNLOshort{a}$ into	
	\begin{eqnarray}\label{sigUa0}
		\dd\sigpart{U,a_0}{gg,NNLO}&=&\coeffVVNNLO\int\dr{x_1}\dr{x_2}\dphi{n}(p_3,\dots,p_{n+2};x_1p_1,x_2p_2)\jet{n}{n}{\pset}\nonumber\\
		&&\times2\,\Big\lbrace\Big.-\dfrac{\beta_0}{\e}\braket{\ampnum{n+2}{0}|\Jcol{1}(\e)|\ampnum{n+2}{0}}\Big.\Big\rbrace;
	\end{eqnarray}
	\begin{eqnarray}\label{sigUa1}
		\dd\sigpart{U,a_1}{gg,NNLO}&=&\coeffVVNNLO\int\dr{x_1}\dr{x_2}\dphi{n}(p_3,\dots,p_{n+2};x_1p_1,x_2p_2)\jet{n}{n}{\pset}\nonumber\\
		&&\times2\,\Big\lbrace\Big.\braket{\ampnum{n+2}{0}|\Jcol{1}(\e)|\ampnum{n+2}{1}}+\braket{\ampnum{n+2}{1}|\Jcol{1}(\e)|\ampnum{n+2}{0}}\Big.\Big\rbrace,
	\end{eqnarray}
	where we separated the contribution of the one-loop amplitude from the $\beta_0$ term, which only contains tree-level amplitudes. We also label the contributions in $\dd\sigpart{U,c}{gg,NNLO}$ according to the different terms in the two-loop integrated dipoles~\eqref{J22}: $\dd\sigpart{U,c,\mathcal{X}_{4}^{0}}{gg,NNLO}$, $\dd\sigpart{U,c,\mathcal{X}_{3}^{1}}{gg,NNLO}$, $\dd\sigpart{U,c,{\mathcal{X}_{3}^{0}\otimes\mathcal{X}_{3}^{0}}}{gg,NNLO}$ and $\dd\sigpart{U,c,b_0}{gg,NNLO}$.
	
	\subsection{NNLO real virtual subtraction term}
	
	At NLO, once the virtual subtraction term is obtained, it is straightforward to systematically construct the real subtraction term. At NNLO, the structure of the subtraction is significantly more involved, due to the presence of two additional layers besides the double virtual correction: real virtual and double real. The real virtual subtraction term $\dd\sigpart{T}{gg,NNLO}$ must cancel the explicit $\e$-poles in the real virtual matrix element, as well as remove the divergent behaviour in single unresolved IR limits. The double real subtraction term $\dd\sigpart{S}{gg,NNLO}$ is needed to remove the single and double unresolved divergences in the double real correction. Due to the complexity of the NNLO IR structure, the systematic generation of the real virtual and double real subtraction terms from the double virtual one is non-trivial. 
	
	In the following, we illustrate how the real virtual subtraction term can be generated for the gluons-only case. Once again, the correspondence between unintegrated real emission subtraction terms and integrated virtual subtraction term is the key. The deduction of $\dd\sigpart{T}{gg,NNLO}$ from $\dd\sigpart{U}{gg,NNLO}$ is articulated in two main steps: 
	
	\begin{itemize}
		\item integrated terms are translated from $\dd\sigpart{U}{gg,NNLO}$ to $\dd\sigpart{T}{gg,NNLO}$ inserting an unresolved gluon between a pair of hard radiators, via the application of $\ins{\cdot}$, in complete analogy to what is done at NLO;
		\item suitable terms are generated to remove spurious $\e$-poles.
	\end{itemize}
	Any additional contribution which is added at the real virtual level and does not have a direct correspondence to terms in $\dd\sigpart{U}{gg,NNLO}$ will eventually generate corresponding terms at the double real level after the insertion of a second unresolved gluon. We note that not the entirety of $\dd\sigpart{U}{gg,NNLO}$ will be translated to $\dd\sigpart{T}{gg,NNLO}$, some terms have to undergo a double insertion of unresolved gluons and directly move from $\dd\sigpart{U}{gg,NNLO}$ to $\dd\sigpart{S}{gg,NNLO}$, as explained in section~\ref{sec:RR}.
	
	We recall the usual decomposition of $\dd\sigpart{T}{gg,NNLO}$ in the context of antenna subtraction~\cite{Currie:2013vh}:
	\begin{equation}
		\dd\sigpart{T}{gg,NNLO}=\dd\sigpart{T,a}{}+\dd\sigpart{T,b}{}+\dd\sigpart{T,c}{},
	\end{equation}
	where we dropped the subscript `$gg,NNLO$' in the right-hand-side. The meaning of this decomposition is the following:
	
	\begin{itemize}
		\item $\dd\sigpart{T,a}{}$ reproduces the explicit poles of the real virtual matrix element;
		\item $\dd\sigpart{T,b}{}$ describes the divergent behaviour of the real virtual matrix element in single unresolved limits;
		\item $\dd\sigpart{T,c}{}$  removes the overlap in the
		single unresolved behaviour between the two terms above \cite{Currie:2013vh}.
	\end{itemize}
	
	\subsubsection{\text{d}\boldmath{$\sigma^{T,a}$}}
	
	This part of the real virtual subtraction term is needed to remove the explicit poles of the $(n+3)$-particle one-loop matrix element. Moreover, it contains the mass factorization contribution $\dd\sigpart{MF,1,a}{gg,NNLO}$ given in~\eqref{MF1a}. The construction of $\sigma^{T,a}$ is completely analogous to the one adopted for the NLO virtual subtraction term in section~\ref{sec:NLOV}, with the difference that here we have an additional particle. It is clear that this term is not generated by any integrated contribution at the double virtual level and so, as we show in section~\ref{sec:dsigSa} below, its unintegrated counterpart will be added to the the double real subtraction term. For the gluons-only case we have:
	\begin{eqnarray}\label{sigTa}
		\dd\sigpart{T,a}{}&=&\coeffRVNNLO\int\dr{x_1}\dr{x_2}\dphi(p_3,\dots,p_{n+2};x_1 p_1,x_2 p_2)\jet{n+1}{n}{\lb p \rb_{n+1}}\nonumber\\
		&&\times 2\braket{\ampnum{n+3}{0}|\Jcol{1}(\e)|\ampnum{n+3}{0}}.
	\end{eqnarray}
	
	\subsubsection{\text{d}\boldmath{$\sigma^{T,b}$}}\label{sec:dsigTb}
	
	This block of the subtraction term reproduces the divergent behaviour of the real virtual matrix element in single unresolved IR limits. The subtraction of infrared divergences from a one loop matrix element is more involved than the one required at tree-level. In particular, along with tree-level antennae and reduced one-loop matrix elements (tree $\times$ loop), suitable combinations of three-particle one-loop antenna functions and tree-level reduced matrix elements (loop $\times$ tree) have to be used. Both these structures are present, in their integrated form, in the double virtual subtraction term, since they describe the emission of soft and collinear gluons from the one-loop amplitude. It is possible to systematically generate these terms at the real virtual level starting from $\dsigUNNLO{gg}$, as we show in the following. According to~\cite{Currie:2013vh}, we introduce a suitable decomposition of $\dsigTNNLOshort{b}$:
	\begin{equation}
		\dsigTNNLOshort{b}=\dsigTNNLOshort{b_1}+\dsigTNNLOshort{b_2}+\dsigTNNLOshort{b_3}, 
	\end{equation}
	where the elements on the right-hand-side respectively contain (tree $\times$ loop) contributions, (loop $\times$ tree) contributions and suitable terms needed to ensure the correct renormalization of loop quantities in the real virtual subtraction term.
	
	We first focus on $\dsigTNNLOshort{b_1}$. The respective integrated counterparts appear in the double virtual subtraction term as a combination of a one-loop integrated dipole and a one-loop reduced matrix element. In particular, these contributions are given by $\dsigUNNLOshort{a_1}$. The procedure of inserting an unresolved gluon is the same as the one depicted in section~\ref{sec:NLOR}, with the tree-level amplitudes replaced by the one-loop ones. The result has the following form:
	\begin{equation}\label{treexloop_1}
		\ins{\dsigUNNLOshort{a_1}}\sim \sum_{ijk} \,f_{3}^{0}(i,k,j)\,\anum{n+2,c}{1}(\sigma,\sigma';\{.,\widetilde{ik},.,\widetilde{kj},.\}).
	\end{equation}
	The resulting contribution partially takes care of the divergent IR limits of the real virtual matrix element. However, the one-loop interferences $\anum{n+2,c}{1}$ contain explicit $\e$-poles which must be removed to ensure the finiteness of the real virtual subtraction term. In fact, it is possible to systematically generate suitable terms that cancel these poles, exploiting once again the predictability of the singularity structure of one-loop amplitudes. The following relation holds:
	\begin{eqnarray}\label{UaUbrel1}
	\lefteqn{\poles\lb2\sum_{(i,j)}\left(\braket{\ampnum{n+2}{0}|\left(\T_{i}\cdot\T_{j}\right)|\ampnum{n+2}{1}}+\braket{\ampnum{n+2}{1}|\left(\T_{i}\cdot\T_{j}\right)|\ampnum{n+2}{0}}\right)\rb}\nonumber\\
		&=&\poles\lb4\sum_{(i,j)}\text{Re}\left(\braket{\ampnum{n+2}{0}|\left(\T_{i}\cdot\T_{j}\right)|\ampnum{n+2}{1}}\right)\rb\nonumber\\
		&=&\poles\lb4\sum_{(i,j)}\text{Re}\left(\braket{\ampnum{n+2}{0}|\left(\T_{i}\cdot\T_{j}\right)\left(\sum_{(k,l)}\left(\T_{k}\cdot\T_{l}\right)\ourIop{1}{kl}{\e,\mu_r^2}\right)|\ampnum{n+2}{0}}\right)\rb\nonumber\\
		&=&-2\,\poles\lb-2\sum_{(i,j)}\sum_{(k,l)}\text{Re}\left[\ourIop{1}{kl}{\e,\mu_r^2}\right]\braket{\ampnum{n+2}{0}|\left(\T_{i}\cdot\T_{j}\right)\left(\T_{k}\cdot\T_{l}\right)|\ampnum{n+2}{0}}\rb,
	\end{eqnarray}
	which is analogous to state that the $\e^{-4}$ and $\e^{-3}$ poles in the first and second lines of~\eqref{M2poles2} are related by the same factor $-2$ we see in the last line here. Equation~\eqref{UaUbrel1} indicates that the pole structure of $\ins{\dsigUNNLOshort{a_1}}$ can be obtained in a general way by applying the unresolved parton insertion operator in $\dsigUNNLOshort{b}$. In particular we have:
	\begin{equation}\label{UaUbrel2}
		\poles\lb\ins{\dsigUNNLOshort{a_1}}\rb=\poles\lb-2\,\ins{\dsigUNNLOshort{b}}-\dd\sigpart{MF,1,b}{}\rb. 
	\end{equation}
	Equation~\eqref{UaUbrel2} requires some comments. First of all, the $\ins{\cdot}$ operator removes the splitting kernels from both the $\Jcol{1}$ present in $\dsigUNNLOshort{b}$ and $\dd\sigpart{MF,1,b}{}$, given by~\eqref{MF1b}, exactly reconstructs the one-loop integrated dipole which is not affected by the insertion of the unresolved parton. Secondly, step $3$ of the list in section~\ref{sec:NLOR} needs to be extended: the momentum relabelling here affects not only the colour interferences and the jet function, but also the integrated dipole which is not converted into unintegrated antenna functions. This applies in general at NNLO, namely the transition to higher multiplicities via momenta relabelling occurs within any function of the external momenta accompanying the integrated dipole which is converted into unintegrated antennae. The choice of which integrated dipole to convert into an unintegrated antenna function should be done in such a way any pair of hard radiators is addressed once and only once. In practise this can be easily achieved performing the decomposition:
	\begin{equation}
		\J{1}(i,j)\J{1}(l,k)=\dfrac{1}{2}\J{1}(i,j)\J{1}(l,k)+\dfrac{1}{2}\J{1}(k,l)\J{1}(i,j)
	\end{equation}
	and then fixing the first dipole in each term of the symmetrized expression on the right-hand-side to be the one converted into unintegrated antennae. Therefore, the structure of the right-hand-side of \eqref{UaUbrel2} is given by:
	\begin{equation}\label{Ubstructure}
		-2\,\ins{\dsigUNNLOshort{b}}-\dd\sigpart{MF,1,b}{}\sim-2\sum_{ijk,lm}f_{3}^{0}(i,k,j)\J{1}(l,m)\anum{n+2}{0}(\sigma,\sigma';\{.,\widetilde{ik},.,\widetilde{kj},.\}),
	\end{equation}
	where $l$ and $m$ can represent either $\wt{ik}$ and $\wt{kj}$ or any other parton not belonging to the $(i,j)$ dipole.	We can thus write down the first contribution to the real virtual subtraction term as:
	\begin{equation}\label{sigTb1}
		\dsigTNNLOshort{b_1}=-\ins{\dsigUNNLOshort{a_1}}-2\,\ins{\dsigUNNLOshort{b}}-\dd\sigpart{MF,1,b}{gg,NNLO}, 
	\end{equation}
	which is free of $\e$-poles. We observe that, to compensate $\dsigUNNLOshort{b}$ in the double virtual subtraction term, we should add back a factor $+\ins{\dsigUNNLOshort{b}}$. This contribution is indeed added later in the generation of the real virtual and double real subtraction terms.

	We consider now $\dsigTNNLOshort{b_2}$, namely (loop $\times$ tree) contributions. The core part of this term is given by unintegrated three-parton one-loop antenna functions $X_{3}^{1}$ combined with tree-level reduced matrix elements. The integrated counterpart of these terms is contained in $\dsigUNNLOshort{c,\mathcal{X}_{3}^{1}}$. Once again, we can insert an unresolved gluon and obtain from $\dsigUNNLOshort{c,\mathcal{X}_{3}^{1}}$ the contribution needed at the real virtual level. As it happened for $\dsigTNNLOshort{b_1}$, if one simply considers $\ins{\dsigUNNLOshort{c,\mathcal{X}_{3}^{1}}}$, spurious $\e$-poles coming from the one-loop antennae would remain in the real virtual subtraction term. Therefore, these singularities need to be removed to obtain a finite subtraction term. This can be done systematically since the singularity structure of the unintegrated three-parton one-loop antenna functions is known and can be expressed by means of one-loop integrated dipoles and three-parton tree-level antennae~\mbox{\cite{Gehrmann-DeRidder:2005btv,Currie:2013vh}}. The construction of the required blocks in the specific case of a $f_{3}^{1}$ antenna, the only one needed for gluon scattering, is achieved through the following replacement:
	\begin{equation}\label{f31decomp}
		f_{3}^{1}(i,k,j)\to f_{3}^{1}(i,k,j) + \sum_{(l,m)=1}^{3}\J{1}(l,m)f_{3}^{0}(i,k,j) - 2\J{1}(\widetilde{ik},\widetilde{kj})f_{3}^{0}(i,k,j),
	\end{equation}
	where the sum in the second term runs over the $3$ pairs of colour-connected gluons in the $f_{3}^{1}(i,k,j)$ antenna. The expression obtained after the replacement in~\eqref{f31decomp} is free of poles and can be used to construct the missing part of the real virtual subtraction term. To use a similar notation to~\cite{Currie:2013vh}, we label $\dsigTNNLOshort{b_2,X_3^1}$, $\dsigTNNLOshort{b_2,J_X}$ and $\dsigTNNLOshort{b_2,M_X}$ the three blocks coming from the three components in~\eqref{f31decomp}. As we have already stated
	\begin{equation}\label{sigTb2X31}
		\dsigTNNLOshort{b_2,X_3^1}=-\ins{\dsigUNNLOshort{c,\mathcal{X}_{3}^{1}}}\sim\sum_{ijk} f_{3}^{1}(i,k,j)\anum{n+2}{0}(\sigma,\sigma';\{.,\widetilde{ik},.,\widetilde{kj},.\}).
	\end{equation}
	The two remaining blocks are treated differently: $\dsigTNNLOshort{b_2,M_X}$ comes from $\dsigUNNLOshort{c,\mathcal{X}_{3}^{0}\otimes\mathcal{X}_{3}^{0}}$ after the insertion of an unresolved gluon, while $\dsigTNNLOshort{b_2,J_X}$ is a genuinely new contribution added at the real virtual level, which needs to be compensated by its unintegrated counterpart at the double real level. This can be noticed looking at the arguments of the integrated dipoles appearing in the two blocks. $\dsigTNNLOshort{b_2,M_X}$ depends on mapped momenta, which come from the insertion of an extra unresolved gluon in the $n$-particle phase space, where the double virtual subtraction term lives. On the other hand, the integrated dipoles in $\dsigTNNLOshort{b_2,J_X}$ depend on $(n+1)$-particle phase space momenta, which are not accessible at the double virtual level. It is trivial to verify that the mass factorization kernels in the integrated dipoles in $\dsigTNNLOshort{b_2,J_X}$ and $\dsigTNNLOshort{b_2,M_X}$ exactly cancel for any configuration of gluons $i$, $j$ and $k$. For bookkeeping purposes we label this mass factorization $\dd\sigpart{MF,1,b_2}{}$.
	
	For $\dsigTNNLOshort{b_2,M_X}$ one obtains the relation: 
	\begin{equation}\label{sigTb2MX}
		\dsigTNNLOshort{b_2,M_X}=-2\ins{\dsigUNNLOshort{c,\mathcal{X}_{3}^{0}\otimes\mathcal{X}_{3}^{0}}}-\dd\sigpart{MF,1,b_2}{}.
	\end{equation}
	The structure of $\dsigTNNLOshort{b_2,M_X}$ is indeed the required one, according to \eqref{f31decomp}:
	\begin{equation}\label{Mxstructure}
		\dsigTNNLOshort{b_2,M_X}\sim-2\sum_{ijk} f_{3}^{0}(i,k,j)\J{1}(\widetilde{ik},\widetilde{kj})\anum{n+2}{0}(\sigma,\sigma';\{.,\widetilde{ik},.,\widetilde{kj},.\}).
	\end{equation}
	The leftover $+\ins{\dsigUNNLOshort{c,\mathcal{X}_{3}^{0}\otimes\mathcal{X}_{3}^{0}}}$ is used in $\dsigTNNLOshort{c}$, as we show in the next section. In the gluons-only case, the two integrated antennae are identical (see \eqref{J22expression}), so the choice of which one has to be translated into its unintegrated form is irrelevant. In general, the same symmetrization procedure employed for $\dsigUNNLOshort{b}$ can be used here as well. In conclusion, the (loop $\times$ tree) block is given by:
	\begin{equation}\label{sigTb2}
		\dsigTNNLOshort{b_2}=-\ins{\dsigUNNLOshort{c,\mathcal{X}_{3}^{1}}}+\dsigTNNLOshort{b_2,J_X}-2\ins{\dsigUNNLOshort{c,\mathcal{X}_{3}^{0}\otimes\mathcal{X}_{3}^{0}}}-\dd\sigpart{MF,1,b_2}{},
	\end{equation}
	which is free of $\e$-poles. We remark again that $\dd\sigpart{MF,1,b_2}{}$ is compensated by an identical contribution in $\dsigTNNLOshort{b_2,J_X}$ and no new mass factorization kernels are actually added in $\dsigTNNLOshort{b_2}$.
	
	The last contribution to $\dsigTNNLOshort{b}$ is required to fix the renormalization of one-loop quantities in $\dsigTNNLOshort{b_1}$ and $\dsigTNNLOshort{b_2}$. To fix the correct renormalization of both the one-loop matrix elements and the one-loop antenna functions it is sufficient to perform the following replacement~\cite{Gehrmann-DeRidder:2005btv}:
	\begin{equation}
		X_{3}^{1}(i,k,j)\to X_{3}^{1}(i,k,j) +\dfrac{b_0}{\e}X_{3}^{0}(i,k,j)\left(\left(\dfrac{\left|s_{ijk}\right|}{\mu_r^2}\right)^{-\e}-1\right). 
	\end{equation}
	$\dsigTNNLOshort{b_3}$ is entirely constructed with terms coming from the double virtual subtraction term:
	\begin{equation}\label{sigTb3}
		\dsigTNNLOshort{b_3}=-\ins{\dsigUNNLOshort{a_0}}-\ins{\dsigUNNLOshort{c,b_0}},
	\end{equation}
	as can be easily checked keeping track of the QCD $\beta$-function coefficient $b_0$.
	
	\subsubsection{\text{d}\boldmath{$\sigma^{T,c}$}}\label{sec:dsigTc}
	
	The last block of the real virtual subtraction term is $\dsigTNNLOshort{c}$. This block has two components, one which is better identified as the integrated version of part of the double real subtraction term and one which can be obtained as the unintegrated counterpart of contributions at the double virtual level. In the original antenna subtraction approach~\cite{Currie:2013vh}, $\dsigTNNLOshort{c}$ was constructed starting from the former contribution and adding the latter subsequently to remove spurious $\e$-poles and ensuring the finiteness of the full real-virtual contribution. In particular, $\dsigSNNLOshort{c}$, the block containing almost colour-connected contributions~\cite{NigelGlover:2010kwr} and large angle soft terms~\cite{Gehrmann-DeRidder:2007foh,Weinzierl:2008iv} at the double real level, is integrated over the phase space of an unresolved parton and used as part of $\dsigTNNLOshort{c}$. To remove the singularities present in these results, two blocks denoted as $\dsigTNNLOshort{c_1}$ and $\dsigTNNLOshort{c_2}$ in~\cite{Currie:2013vh} are generated. The integrated counterparts of these two blocks are then used respectively in $\dsigUNNLOshort{b}$ and $\dsigUNNLOshort{c}$.
	
	The procedure we present in the following to construct $\dsigTNNLOshort{c}$ proceeds in reverse with respect to the one summarized above. We first deduce $\dsigTNNLOshort{c_1}$ and $\dsigTNNLOshort{c_2}$ from the double virtual subtraction term and then we infer the complete structure of the combination of integrated almost colour-connected terms and large angle soft terms. Again, the guiding principle is the cancellation of unwanted $\e$-poles, which, combined with the knowledge of the general structure of $\dsigSNNLOshort{c}$, allows for a systematic generation of $\dsigTNNLOshort{c}$. 
	
	The first block $\dsigTNNLOshort{c_1}$ comes from the leftover $+\ins{\dsigUNNLOshort{b}}$ that we did not use in section~\ref{sec:dsigTb}. However, we can identify in $\dsigUNNLOshort{b}$ two types of structures: almost colour-connected contributions and colour-unconnected contributions~\cite{NigelGlover:2010kwr,Currie:2013vh}. These latter terms do not need to be included at the real virtual level and can actually be moved from the double virtual to the double real subtraction term directly. The reason is that the integration over the phase space of two unresolved colour-unconnected partons can be performed independently. In our top-down approach (see Figure \ref{fig:colant_scheme}), this means that two unresolved partons can be iteratively inserted to produce a contribution to the double real subtraction term, as we show in section~\ref{sec:dsigSd}. The colour-unconnected contribution in $\dsigUNNLOshort{b}$, which we label as $\dsigUNNLOshort{b,c.u.}$ can be easily identified since, in the convolution of two one-loop integrated dipoles, the two partons that form the first dipole are both different from either of the two partons appearing as arguments of the second. Namely:
	\begin{equation}\label{sigUbcu}
		\dsigUNNLOshort{b,c.u.}\sim \J{1}(i,j)\otimes\J{1}(k,l),\quad\text{with } i\ne k,l\,\,\text{and}\,\,j\ne k,l.
	\end{equation}
	For the purpose of constructing $\dsigTNNLOshort{c_1}$, we can then decompose $\dsigUNNLOshort{b}$ into
	\begin{equation}\label{sigUbacc}
		\dsigUNNLOshort{b}=\dsigUNNLOshort{b,a.c.c.}+\dsigUNNLOshort{b,c.u.},
	\end{equation}
	where $\dsigUNNLOshort{b,a.c.c.}$ contains almost colour-connected contributions and is identified removing from $\dsigUNNLOshort{b}$ contributions with a structure like \eqref{sigUbcu}. $\dsigTNNLOshort{c_1}$ is then obtained as
	\begin{equation}\label{sigTc1}
		\dsigTNNLOshort{c_1}= +\ins{\dsigUNNLOshort{b,a.c.c.}}+\dd\sigpart{MF,1,c_1}{}.
	\end{equation}
	The structure of this term is the same as the one in \eqref{Ubstructure}, where the contribution with $l$ and $m$ both different from $\wt{ik}$ or $\wt{kj}$ have been removed.
	To generate $\dsigTNNLOshort{c_2}$, we directly use the leftover $+\ins{\dsigUNNLOshort{c,\mathcal{X}_{3}^{0}\otimes\mathcal{X}_{3}^{0}}}$:
	\begin{equation}\label{sigTc2}
		\dsigTNNLOshort{c_2}= +\ins{\dsigUNNLOshort{c,\mathcal{X}_{3}^{0}\otimes\mathcal{X}_{3}^{0}}}+\dd\sigpart{MF,1,c_2}{},
	\end{equation}
	which has a structure analogous to \eqref{Mxstructure}. In the previous expressions, the two contributions $\dd\sigpart{MF,1,c_1}{}$ and $\dd\sigpart{MF,1,c_2}{}$ contain the mass factorization kernels needed to reconstruct integrated dipoles, however they will be immediately subtracted back again and are included explicitly here only for consistency of notation. As we said, the sum $\dsigTNNLOshort{c_1}+\dsigTNNLOshort{c_2}$ exhibits explicit singularities that need to be cancelled adding a suitable block which represents the integrated counterpart of almost colour-connected contributions and large angle soft terms in the double real subtraction term. The systematic generation of this contribution is a crucial step in the colourful antenna subtraction method and in the gluons-only case can be achieved through the following replacements in the combination $\dsigTNNLOshort{c_1}+\dsigTNNLOshort{c_2}$:
	\begin{eqnarray}\label{replaceTc}
		f_{3}^{0}(i,k,j)\J{1}(\widetilde{ik},\widetilde{kj})&\to& f_{3}^{0}(i,k,j)\Big[\Big.\J{1}(\widetilde{ik},\widetilde{kj})-\J{1}(i,j)\nonumber\\
		&&\hspace{1.3cm}+\mathcal{S}(s_{ij},s_{i'j'},1)-\mathcal{S}(s_{(ik)(kj)},s_{i'j'},x_{(ik)(kj),i'j'})\Big.\Big],\nonumber\\
		f_{3}^{0}(i,k,j)\J{1}(\widetilde{ik},a)&\to& f_{3}^{0}(i,k,j)\Big[\Big.\J{1}(\widetilde{ik},a)-\J{1}(i,a)\nonumber\\
		&&\hspace{1.3cm}+\mathcal{S}(s_{ia},s_{i'j'},x_{ia,ij})-\mathcal{S}(s_{(ik)a},s_{i'j'},x_{(ik)a,i'j'})\Big.\Big],\nonumber\\
		f_{3}^{0}(i,k,j)\J{1}(b,\widetilde{kj})&\to& f_{3}^{0}(i,k,j)\Big[\Big.\J{1}(b,\widetilde{kj})-\J{1}(b,j)\nonumber\\
		&&\hspace{1.3cm}+\mathcal{S}(s_{bj},s_{i'j'},x_{bj,ij})-\mathcal{S}(s_{b(kj)},s_{i'j'},x_{b(kj),i'j'})\Big.\Big],
	\end{eqnarray}
	where $\mathcal{S}$ denotes the integrated large angle soft terms~\cite{Gehrmann-DeRidder:2007foh,Gehrmann-DeRidder:2011jwo}. These replacements cover all possible structures appearing in $\dsigTNNLOshort{c_1}+\dsigTNNLOshort{c_2}$. The expressions on the right-hand-side of~\eqref{replaceTc} are free of poles, since the residual singularities in the difference of one-loop integrated dipoles are removed by the integrated large angle soft terms and this cancellation is in fact the guiding principle through which~\eqref{replaceTc} is constructed. Moreover, as we anticipated, the mass factorization kernels needed for the newly introduced integrated dipoles exactly cancel the ones present in $\dd\sigpart{MF,1,c_1}{}$ and $\dd\sigpart{MF,1,c_2}{}$, in such a way that the resulting expression is actually free of mass factorization kernels. Partons $i'$ and $j'$ can be chosen arbitrarily since there is a priori no singular behaviour associated with them. They appear as reference momenta in a phase space mapping for a soft gluon radiation between other momenta $i$ and $j$ that can but do not have to be different from $i'$ and $j'$ (all arrangements of $i, j$ and $i',j'$ are discussed in Section~\ref{sec:dsigSc} below). In particular, for process involving four or more partons at the Born level, $i'$ and $j'$ can be chosen to be two final-state partons~\cite{Gehrmann-DeRidder:2011jwo}. This implies that the integrated large angle soft terms appearing in~\eqref{replaceTc} are obtained through the integration of soft eikonal factors over a FF configuration, which requires a simpler momentum mapping with respect to IF or II configurations. The identification of $i'$ and $j'$ must be inherited properly to the double real level, where the unintegrated counterpart of the large angle soft terms in \eqref{replaceTc} is needed to compensate for the oversubtraction of large angle soft gluon radiation \cite{Gehrmann-DeRidder:2007foh,Weinzierl:2008iv,Weinzierl:2009nz}. We label $\dsigTNNLOshort{c_0}$ the new block generated through~\eqref{replaceTc} and we can finally construct $\dsigTNNLOshort{c}$ as:
	\begin{equation}\label{sigTc}
		\dsigTNNLOshort{c}=\dsigTNNLOshort{c_0}+\dsigTNNLOshort{c_1}+\dsigTNNLOshort{c_2}.
	\end{equation} 
	
	This last step completes the construction of the real virtual subtraction term. We notice that almost the entire double virtual subtraction term has been converted into its unintegrated counterpart and has been used at the real virtual level, with the only exceptions being $\dsigUNNLOshort{b,cu}$ and $\dsigUNNLOshort{c,\mathcal{X}_{4}^{0}}$, which are directly converted to the double real subtraction term via the insertion of two unresolved gluons. Moreover, new components had to be added at the real virtual level to cancel unwanted singularities and need a corresponding counterpart in the double real subtraction term. These components are $\dsigTNNLOshort{a}$, $\dsigTNNLOshort{b_2,J_X}$ and $\dsigTNNLOshort{c_0}$.

	\subsection{NNLO double real subtraction term}\label{sec:RR}
	
	The last ingredient for an NNLO calculation is the double real subtraction term $\dsigSNNLO{gg}$, which removes the divergent behaviour of the double real matrix element in single and double unresolved limits. In the colourful antenna subtraction approach, the generation of $\dsigSNNLO{gg}$ is performed at the end, with the significant advantage of avoiding to deal with the involved IR structure of the matrix elements arising from the large number of partons. Once the double virtual and the real virtual subtraction terms are available, it is indeed straightforward to complete the subtraction procedure with the missing blocks needed to cancel the unmatched contributions in those two layers. The double real subtraction term is constructed inserting a second unresolved gluon in contributions coming from $\dsigTNNLO{gg}$ and two unresolved gluons in terms coming from $\dsigUNNLO{gg}$. The only non-trivial step, as we show later in this section, is related to these latter contributions, in particular to the insertion of an unresolved gluon pair in the integrated four-parton antennae $\mathcal{X}_{4}^{0}$. The appropriate overall coefficient to be used to dress the double real subtraction term is given by:
	\begin{equation}
		\coeffRRNNLO=s_{RR}(4\pi\alpha_s)^2\coeffLO,
	\end{equation}
	where, in the gluons-only case we have
	\begin{equation}
		s_{RR}=\dfrac{n!}{(n+2)!}=\dfrac{1}{(n+1)(n+2)}.
	\end{equation}
	
	We recall the usual decomposition of $\dsigSNNLO{gg}$~\cite{Currie:2013vh}:
	\begin{equation}
		\dsigSNNLO{gg}=\dsigSNNLOshort{a}+\dsigSNNLOshort{b}+\dsigSNNLOshort{c}+\dsigSNNLOshort{d}.
	\end{equation}
	The first term $\dsigSNNLOshort{a}$ removes single unresolved limits and it is analogous to an NLO real subtraction term for an $(n+3)$-particle Born process. The remaining terms respectively reproduce the divergent behaviour of the double real correction in colour-connected, almost colour-connected and colour-unconnected configurations \cite{NigelGlover:2010kwr,Currie:2013vh}. Moreover, $\dsigSNNLOshort{c}$ also contains the large angle soft terms. In the following, we describe how to systematically generate each contribution.
	
	\subsubsection{\text{d}\boldmath{$\sigma^{S,a}$}}\label{sec:dsigSa}
	
	This part of the subtraction term can be straightforwardly generated from $\dsigTNNLOshort{a}$, since it can be seen as its corresponding real NLO subtraction term. Indeed, the following relation holds:
	\begin{equation}
		\dsigTNNLOshort{a}=-\int_1\dsigSNNLOshort{a}-\dd\hat{\sigma}^{MF,1,a},
	\end{equation}
	which reflects equation~\eqref{sigTNLO} and therefore, following what is done in section~\ref{sec:NLOR} for the NLO real subtraction term, we can write:
	\begin{equation}\label{sigSa}
		\dsigSNNLOshort{a}=-\ins{\dsigTNNLOshort{a}}.
	\end{equation}
	
	\subsubsection{\text{d}\boldmath{$\sigma^{S,b}$}}\label{sec:dsigSb}
	
	In the colour-connected configuration, the two unresolved gluons are emitted between the same pair of hard radiators. According to~\cite{Currie:2013vh}, we further decompose $\dsigSNNLOshort{b}$ into two contributions:
	\begin{equation}
		\dsigSNNLOshort{b}=\dsigSNNLOshort{b_1}+\dsigSNNLOshort{b_2},
	\end{equation}
	where $\dsigSNNLOshort{b_1}$ contains four-parton antennae $X_{4}^{0}$, while $\dsigSNNLOshort{b_2}$ contains suitable convolutions of two three-parton antennae $X_{3}^{0}\otimes X_{3}^{0}$ needed to remove single unresolved limits from the $X_{4}^{0}$. The generation of $\dsigSNNLOshort{b_2}$ is straightforward, since its integrated counterpart is exactly $\dsigTNNLOshort{b_2,J_X}$ and so:
	\begin{equation}\label{sigSb2}
		\dsigSNNLOshort{b_2}=-\ins{\dsigTNNLOshort{b_2,J_X}}, 
	\end{equation}
	where the momenta relabelling due to the insertion of an unresolved parton must occur within the unintegrated antenna functions which appear in $\dsigTNNLOshort{b_2,J_X}$ too. On the other hand, $\dsigSNNLOshort{b_1}$ comes from the insertion of two unresolved gluons in $\dsigUNNLOshort{c,\mathcal{X}_{4}^{0}}$. We therefore define a new transformation which acts on the integrated four-parton antennae as:
	\begin{equation}\label{double_insertion}
		\mathcal{X}_{4}^{0}(s_{ij})\anum{n+2}{0}(\sigma,\sigma',\{.,i,.,j,.\})\leftrightarrow X_{4}^{0}(i,k,l,j)\anum{n+2}{0}(\sigma,\sigma',\{.,\widetilde{ikl},.,\widetilde{lkj},.\}),
	\end{equation}
	where partons $k$ and $l$ are unresolved between the pair of hard radiators $i$ and $j$. The considered momentum mapping maps the final state $(n+2)$-particle momenta to $n$-particle momenta, possibly together with a rescaling of the initial state momenta~\cite{Kosower:2002su}. 
	As it happens in the case of a single insertion, after the replacement described by~\eqref{double_insertion}, a suitable sum over the permutations of the external momenta is needed to construct the full contribution. The conversion of the $\mathcal{X}_{4}^{0}$ to their unintegrated counterparts is less trivial than the one needed for the $\mathcal{X}_{3}^{0}$, given the more involved structure of four-particle antenna functions. For gluon scattering the required replacements are given by:
	\begin{equation}\label{f40replace}
		\begin{aligned}
			\text{FF:  }&\quad\XFFint{F}{4}{0}(s_{ij})&\to&\quad 4\left[F_{4,a}^{0}(i,k,l,j) + F_{4,b}^{0}(i,k,l,j)\right],\\
			\text{IF:  }&\quad\XIFint{F}{4}{0}{g}(s_{1i})&\to&\quad F_{4}^{0}(1,k,l,i),\\
			\text{II:  }&\quad\XIIint{F}{4}{0,adj.}{gg}(s_{12})&\to&\quad F_{4}^{0}(1,k,l,2),\\
			&\quad\XIIint{F}{4}{0,n.adj.}{gg}(s_{12})&\to&\quad F_{4}^{0}(1,k,2,l),
		\end{aligned}
	\end{equation}
	where the sub-antennae $F_{4,a}^{0}$ and $F_{4,b}^{0}$ are defined in~\cite{NigelGlover:2010kwr}. We notice that the order of the unresolved gluons $k$ and $l$ in \eqref{f40replace} matters, since different orderings are associated to different colour connections within the unintegrated four-parton antennae. This must be properly taken into account when the sum over permutations of the external momenta is performed, in such a way that both the $(a,k,l,b)$ and $(a,l,k,b)$ orderings are considered. We introduce a new operator $\insdouble{\cdot}$ to indicate the simultaneous double insertion of two unresolved gluons, in the sense indicated by~\eqref{double_insertion}. The application of $\insdouble{\cdot}$ occurs through the following steps:
	
	\begin{enumerate}
		\item Removal of the splitting kernels from the integrated dipoles;
		\item Transition from integrated four-parton antenna functions to unintegrated ones, as indicated by~\eqref{f40replace};
		\item Momenta relabelling within colour interferences and jet functions according to the accompanying antenna function;
		\item Sum over permutations of the $n+4$ momenta to cover all possible IR limits;
		\item Dressing of the resulting expression with the appropriate phase space and overall coefficient. 
	\end{enumerate}
	We can then obtain $\dsigSNNLOshort{b_1}$ as
	\begin{equation}\label{sigSb1}
		\dsigSNNLOshort{b_1}=-\insdouble{\dsigUNNLOshort{c,\mathcal{X}_{4}^{0}}}.
	\end{equation}
	Since the integrated version of a four-particle antenna is obtained after analytic integration over the double unresolved antenna phase space, in principle the simultaneous insertion of a pair of unresolved gluons can not be equated to the iterated insertion of a single gluon. Nevertheless, as shown by the instructions above, on a practical standpoint the action of the $\insdouble{\cdot}$ operator is very similar to the application of $\ins{\cdot}$ with two unresolved partons, provided the appropriate momentum mapping is used and the correct sum over external momenta is performed. In Figure \ref{fig:colant_scheme}, we denote the action of $\insdouble{\cdot}$ with two connected descendant red arrows.
	
	We notice that the two contributions $\dsigSNNLOshort{b_1}$ and $\dsigSNNLOshort{b_2}$ are obtained here from a priori independent blocks, while a precise relation between these two terms should hold to ensure the removal of single unresolved limits from the four-particle antenna functions at the double real level. In fact, the structure of $\dsigSNNLOshort{b_2}$ directly descends from the one-loop three-parton antennae appearing in the two-loop integrated dipoles, which are in turn related to the four-parton antenna functions. Therefore, the relation among $\dsigSNNLOshort{b_1}$ and $\dsigSNNLOshort{b_2}$ is actually mirrored by the inner structure of the two-loop integrated dipoles and the interplay between the $\mathcal{X}_{4}^{0}$ and the $\mathcal{X}_{3}^{1}$, which gives the correct $\e$-poles at the double virtual level, manifest here in the form of the correct arrangement of $X_{4}^{0}$ and $X_{3}^{0}\otimes X_{3}^{0}$ contributions.
	
	\subsubsection{\text{d}\boldmath{$\sigma^{S,c}$}}\label{sec:dsigSc}
	
	In the almost colour-connected configuration, the unresolved gluons are emitted between two pairs of hard radiators which share one common hard radiator. The structure of the blocks needed to remove the divergences associated to these configurations is shared by the large angle soft terms too, which are thus naturally incorporated in $\dsigSNNLOshort{c}$~\cite{NigelGlover:2010kwr,Currie:2013vh}. The integrated counterpart of $\dsigSNNLOshort{c}$ is generated at the real virtual level and it is part of $\dsigTNNLOshort{c}$, as depicted in section~\ref{sec:dsigTc}. In particular we have:
	\begin{equation}\label{sigSc}
		\dsigSNNLOshort{c}=-\ins{\dsigTNNLOshort{c_0}}.
	\end{equation}
	The action of the $\ins{\cdot}$ operator on $\dsigTNNLOshort{c_0}$ actually requires the insertion of an unresolved parton within integrated large angle soft terms. Calling $k$ and $l$ respectively the previously inserted unresolved gluon and the newly inserted unresolved gluon at the double real level, the required replacements are:
	\begin{equation}\label{Softreplace}
		\begin{aligned}
			\mathcal{S}(s_{IJ},s_{I'J'},x)\quad&\to& S(I,l,J)&\quad\text{for}\quad(I,J)\ne(I',J'),\\
			\mathcal{S}(s_{I'J},s_{I'J'},x)\quad&\to& S(\wt{I'l},l,J)&\quad\text{for}\quad J\ne J',\\
			\mathcal{S}(s_{IJ'},s_{I'J'},x)\quad&\to& S(I,l,\wt{lJ'})&\quad\text{for}\quad I\ne I',\\
			\mathcal{S}(s_{I'J'},s_{I'J'},x)\quad&\to& S(\wt{I'l},l,\wt{lJ'})&,\\
		\end{aligned}
	\end{equation}
	where $S(a,b,c)$ is the eikonal factor:
	\begin{equation}
		S(a,b,c)=2\dfrac{s_{ac}}{s_{ab}s_{bc}}.
	\end{equation} 
	$I$ and $J$ represent any unmapped or mapped parton and $I'$, $J'$ indicate either $i'$, $j'$ or $\wt{i'k}$, $\wt{kj'}$. In this latter case, namely when parton $i'$ or $j'$ acts as a hard radiator for both partons $k$ and $l$, the correct order of the momentum mapping is $i'\to\wt{i'l}\to\wt{\wt{i'l}k}$ or $j'\to\wt{j'l}\to\wt{\wt{j'l}k}$, since the first parton which is integrated over at the double real level is $l$. The eikonal factors generated via \eqref{Softreplace} remove the remnant soft gluon divergent behaviour associated to colour-connected and almost colour-connected contributions at the double real level \cite{Gehrmann-DeRidder:2007foh,Weinzierl:2008iv,Weinzierl:2009nz,NigelGlover:2010kwr,Currie:2013vh}.
		
	\subsubsection{\text{d}\boldmath{$\sigma^{S,d}$}}\label{sec:dsigSd}
	
	In the colour-unconnected configuration, the two unresolved gluons are emitted between two distinct pairs of hard radiators. As we explained in section~\ref{sec:dsigTb}, these terms do not appear at the real virtual level but can be inherited directly from the double virtual subtraction term to the double real one. This is achieved inserting two unresolved gluons in $\dsigUNNLOshort{b,c.u.}$, one in each of the two one-loop integrated dipoles. Since the two pairs of hard radiators are distinct, the two insertions can be performed independently. Therefore $\dsigSNNLOshort{d}$ is actually generated through the iterated application of the $\ins{\cdot}$ operator:
	\begin{equation}\label{sigSd}
		\dsigSNNLOshort{d}=+\ins{\ins{\dsigUNNLOshort{b,c.u.}}}.
	\end{equation}
	In Figure \ref{fig:colant_scheme}, we indicate the iterated insertion of two unresolved gluons as two disjoint descendant red arrows, to differentiate it from the simultaneous double insertion discussed in section \ref{sec:dsigSb}.
	
	\section{Gluonic three-jet production at NNLO}\label{sec:ggggg}
	
	To assess the applicability of the colourful antenna subtraction scheme, we compute the NNLO QCD corrections to the 
	hadron-collider all-gluons three-jet production process:
	\begin{equation}\label{ggggg}
		pp(gg)\to ggg
	\end{equation}
	We set $N_f=0$ and we only consider gluons both as internal and external particles, as if no quarks were present in the theory (pure Yang-Mills). To be consistent with this choice, the renormalization of one- and two-loop amplitudes is performed setting \mbox{$N_f=0$} and the mass factorization counterterms only contain gluon-to-gluon splitting kernels. The NNLO correction in the gluons-only scenario is theoretically well defined and all IR divergences cancel in the sum of virtual and real corrections. We perform the calculation in full colour, with the only exception represented by the finite remainder of the two-loop five-gluon amplitude~\cite{Abreu:2021oya}, which is only available at leading colour. 
	
	From a phenomenological point of view, this calculation is part of the NNLO corrections to three-jet production (which were computed 
	recently~\cite{Czakon:2021mjy,Czakon:2014oma} using a residue subtraction technique~\cite{Czakon:2010td}, with the same leading-colour restriction on the 
	finite remainders of the virtual two-loop amplitudes). However, our result can not be considered a faithful approximation of the full calculation, due to the absence of quarks. Both the $N_f\ne 0$ contributions to gluonic scattering and the quark-induced channels have a large impact, both in terms of the final numerical result and of the scale variation behaviour of the theoretical predictions. Indeed, as we discuss in more detail in section~\ref{sec:scale_variation} below, the gluons-only picture exhibits pathological issues when one assesses the theoretical uncertainty by means of the usual renormalization and factorization scale variation analysis. This is not surprising, given the unphysical nature of the gluons-only scenario.
	
	For these reasons, our calculation should be considered as a proof of concept for the automated implementation of the colourful antenna subtraction formalism, and not yet as a precision phenomenology prediction.

	\subsection{Computational setup}
	
	The computation is performed within the \textsc{NNLOjet} framework. \textsc{NNLOjet} is a Monte Carlo event generator which implements the antenna subtraction method to compute NNLO QCD corrections to a series of processes. For gluonic three-jet production at NNLO, high multiplicity tree and loop gluonic amplitudes are needed. The computation relies on a mixture of analytical results and numerical implementations for the amplitudes. The tree-level helicity amplitudes for the scattering of five, six and seven gluons~\cite{Berends:1987cv,Dixon:2010ik}, corresponding to the LO, real and double real contributions, are incorporated in an analytical form in \textsc{NNLOjet}. This is also the case for the five-gluon one-loop helicity amplitudes~\cite{Bern:1993mq}, which represent the virtual correction. 
	
	The planar five-parton two-loop amplitudes have recently been computed~\cite{Abreu:2019odu} using a basis of pentagon functions~\cite{Chicherin:2017dob,Gehrmann:2018yef,Chicherin:2020oor}. We rely on a public C\texttt{++} code~\cite{Abreu:2021oya} which implements the aforementioned amplitudes and computes the renormalized infrared-finite remainder of the five-gluon two-loop matrix element in the leading colour approximation. According to~\cite{Abreu:2021oya}, for a given configuration of external momenta and helicities, the two-loop infrared-finite remainder is defined as~\cite{Catani:1998bh}:
	\begin{equation}\label{finrem}
		\ket{\mathcal{R}^{(2)}}=\ket{\mathcal{A}^{(2)}}-\Iop{1}{\e}\ket{\mathcal{A}^{(1)}}-\Iop{2}{\e}\ket{\mathcal{A}^{(0)}},
	\end{equation}
	where the $\mathcal{A}^{(\ell)}$ represent the renormalized two-loop ($\ell=2$), one-loop ($\ell=1$) and tree-level ($\ell=0$) amplitudes. The quantity in~\eqref{finrem} is evaluated in the leading colour approximation. For the $\e$-poles of the two-loop matrix element we use the full colour result and we checked their complete cancellation against the full-colour double virtual subtraction term. We notice that in~\eqref{finrem}, not only the singularities, but also finite contributions are subtracted from the two-loop amplitude. Our numerical implementation is suitably designed to take this into account and restore the correct finite result in the combination $\dsigVV{gg}-\dsigUNNLO{gg}$.
	
	The six-gluon one-loop matrix elements are computed with the \textsc{OpenLoops} generator~\cite{Cascioli:2011va,Buccioni:2017yxi,Buccioni:2019sur}. In particular we use a new version of~\textsc{OpenLoops}~\cite{otter:2020} which implements the original algorithm of~\cite{Cascioli:2011va} in combination with the helicity summation technique of~\cite{Buccioni:2017yxi} as well as a new tensor reduction algorithm~\cite{otter:2020,Chen:2021azt}. The latter is based on the reduction techniques of~\cite{Buccioni:2017yxi}, which are implemented at the level of tensor integrals in a way that yields improved numerical stability in the deep infrared regions together with a very significant speedup of quadruple-precision evaluations. Internally, \textsc{OpenLoops} uses double-precision scalar integrals that are provided by \textsc{Collier}~\cite{Denner:2014gla,Denner:2016kdg}, as well as quadruple-precision scalar integrals provided by \textsc{OneLoop}~\cite{vanHameren:2010cp}. As explained in the following sections, to validate the generated subtraction terms and to estimate the impact of the subleading colour contribution, we had to extract the leading colour part of the six-gluon one-loop amplitude. This was possible since \textsc{OpenLoops} allows for the computation of partial amplitudes, which can be suitably combined to construct the leading colour contribution, according to~\eqref{loop1lLC}.
	
	The subtraction terms needed for the NNLO calculation are constructed in a systematic way with the colourful antenna subtraction method, as described in sections~\ref{sec:subNLO} and~\ref{sec:subNNLO}. The subtraction terms only depend on five- and six-gluon tree-level amplitudes and five-gluon one-loop amplitudes, so they are implemented in a completely analytical fashion within \textsc{NNLOjet}. The real, double real and real virtual subtraction terms have been extensively tested against the corresponding matrix elements to check the pointwise cancellation of IR divergences in single and double unresolved limits, as discussed in section~\ref{sec:spike}. 
	
	The NNLO correction to this process is very challenging from the computational point of view. The stability over the whole phase space of the numerical implementations of double virtual and real virtual matrix elements must be ensured exploiting quadruple-precision arithmetic when the standard evaluation fails. Moreover, due to the high multiplicity, the numerical integration of real virtual and double real corrections and their respective subtraction terms requires a very substantial number of evaluations to reach a satisfactory precision.
		
	\subsection{Tests of the subtraction terms}\label{sec:spike}
	
	Before presenting the results for the NNLO corrections to differential three-jet cross sections, we first assess the pointwise convergence 
	in all single and double unresolved IR limits
	of the NNLO subtraction terms at double real and real virtual level that we generated using the colourful antenna method
	against the respective squared matrix elements. 
	We do so in a similar way to what is done in~\cite{NigelGlover:2010kwr}: we generate a sample of $10'000$ phase space points at $\sqrt{s}=13$ TeV close to a given infrared limit and we compute:
	\begin{equation}
		R_{\text{RV}}=\dfrac{\dsigRV{gg}}{\dsigTNLO{gg}}\quad\text{and}\quad R_{\text{RR}}=\dfrac{\dsigRR{gg}}{\dsigSNLO{gg}},
	\end{equation}
	for the real virtual and double real case respectively. We then bin the events as function of the following quantity:
	\begin{equation}\label{tvar}
		t_{i}=\log_{10} \left(\left|1-R_i\right|\right),\quad\text{with}\quad i=\text{RV},\text{ RR},
	\end{equation}
	which provides an estimate of the number of correct digits reproduced by the subtraction terms. We probe each unresolved limit through the variables $x$ and $y$, which parametrize the IR depth at which each limit is tested. The definition of $x$ and $y$ varies according to the considered configuration and is given in Table~\ref{tab:smallx}. The squared centre-of-mass energy is $s$, while the other invariants are defined as:
	\begin{equation}
		\begin{aligned}
			s_{i_1\dots i_m}&=(p_{i_1}+\dots+p_{i_m})^2\quad&\text{small when }i_1,\dots,i_m\text{ are collinear},\\
			s_{-i_1\dots i_m}&=\left(\sum_{j\geq 3,\,j\ne i_1\dots i_m}p_j\right)^2\quad&\text{close to }s\text{ when }i_1,\dots,i_m\text{ are soft}.
		\end{aligned}
	\end{equation}
	The smaller $x$ and $y$ are, the more enhanced the divergent behaviour of matrix elements is. For configurations that require both $x$ and $y$ we choose to fix $x=y$. 
	
	\begin{table}
		\centering
		\begin{tabular}{c|c|c|c|c}
			\hline
			Configuration & Soft & Collinear & $x$ & $y$\\
			\hline
			Single soft & $i$ & - & $(s-s_{-i})/s$ & - \\
			Single collinear & - & $i//j$ & $s_{ij}/s$ & - \\
			Double soft & $i,j$ & - & $(s-s_{-ij})/s$ & - \\
			Triple collinear & - & $i//j//k$ & $s_{ijk}/s$ & - \\
			Soft and collinear & $i$ & $j//k$ & $(s-s_{-i})/s$ & $s_{jk}/s$ \\
			Double collinear & - & $i//j,\,k//l$ & $s_{jk}/s$ & $s_{kl}/s$ \\
		\end{tabular}
		\caption{Variables $x$ and $y$ used to test the IR limits. The first two lines refer to single unresolved limits, the remaining ones to double unresolved limits.}\label{tab:smallx}
	\end{table}
	
	We perform the test independently for the leading colour (LC) and the subleading colour (SLC) part of the real virtual and double real matrix elements, to assess the correct behaviour of the subtraction terms in both cases. Indeed, as can be noticed in the plots in section~\ref{sec:results}, the subleading colour contribution for gluon scattering has a very small numerical impact on the full colour result. Therefore, the inspection of the $t_i$ distributions in full colour might give no insights on the correct behaviour of the subtraction at the subleading colour level. Since the systematic treatment of the subleading colour contribution is a major achievement of the new formalism described in this paper, we focus on it specifically. To remove angular correlations and achieve a proper subtraction in IR limits with collinear partons, a point-by-point angular average is considered, as described in detail in~\cite{NigelGlover:2010kwr}.
	
\begin{figure}
		\begin{center}
			\includegraphics[width=0.45\columnwidth]{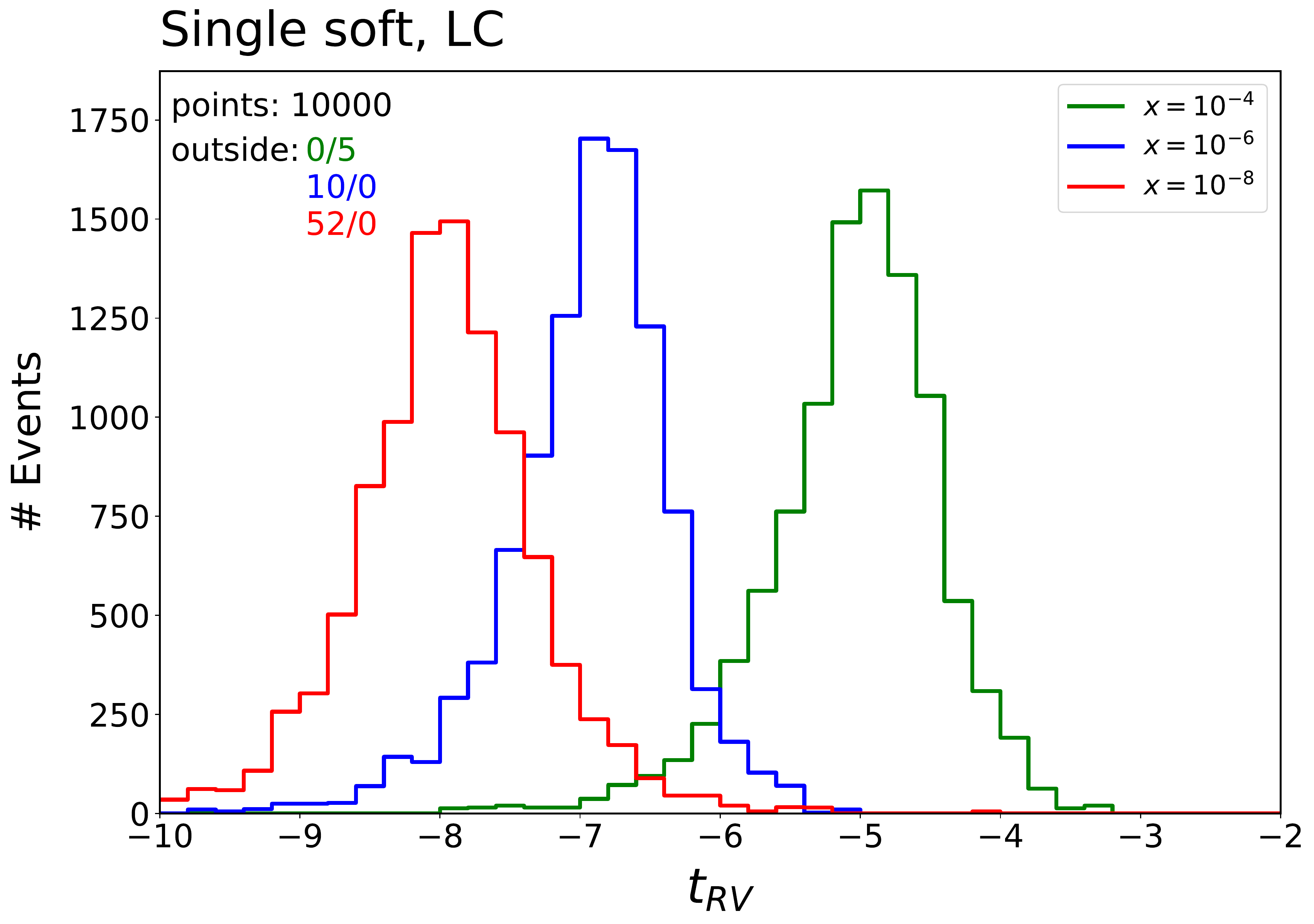}
			\hspace{1cm}
			\includegraphics[width=0.45\columnwidth]{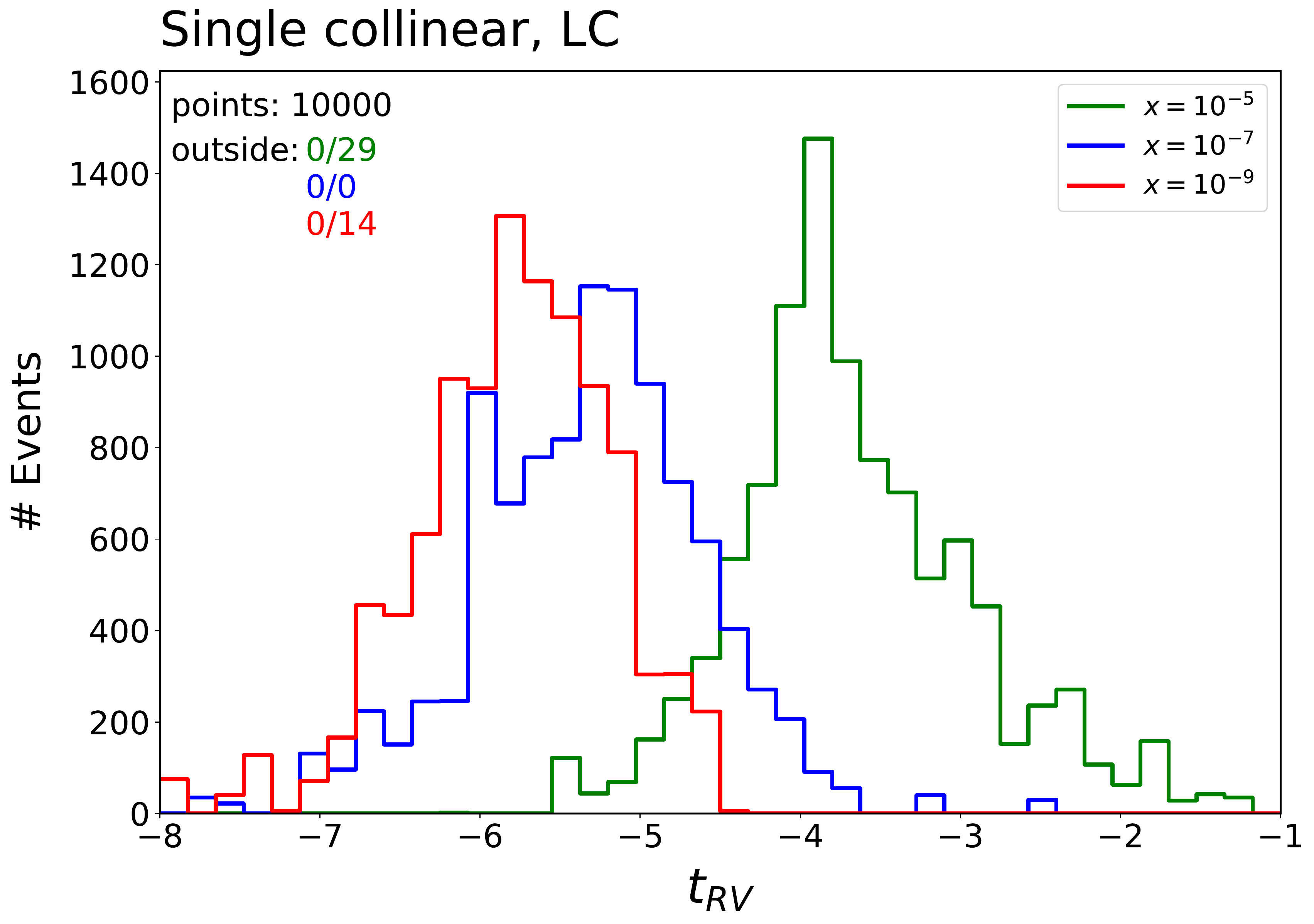}\\
			\includegraphics[width=0.45\columnwidth]{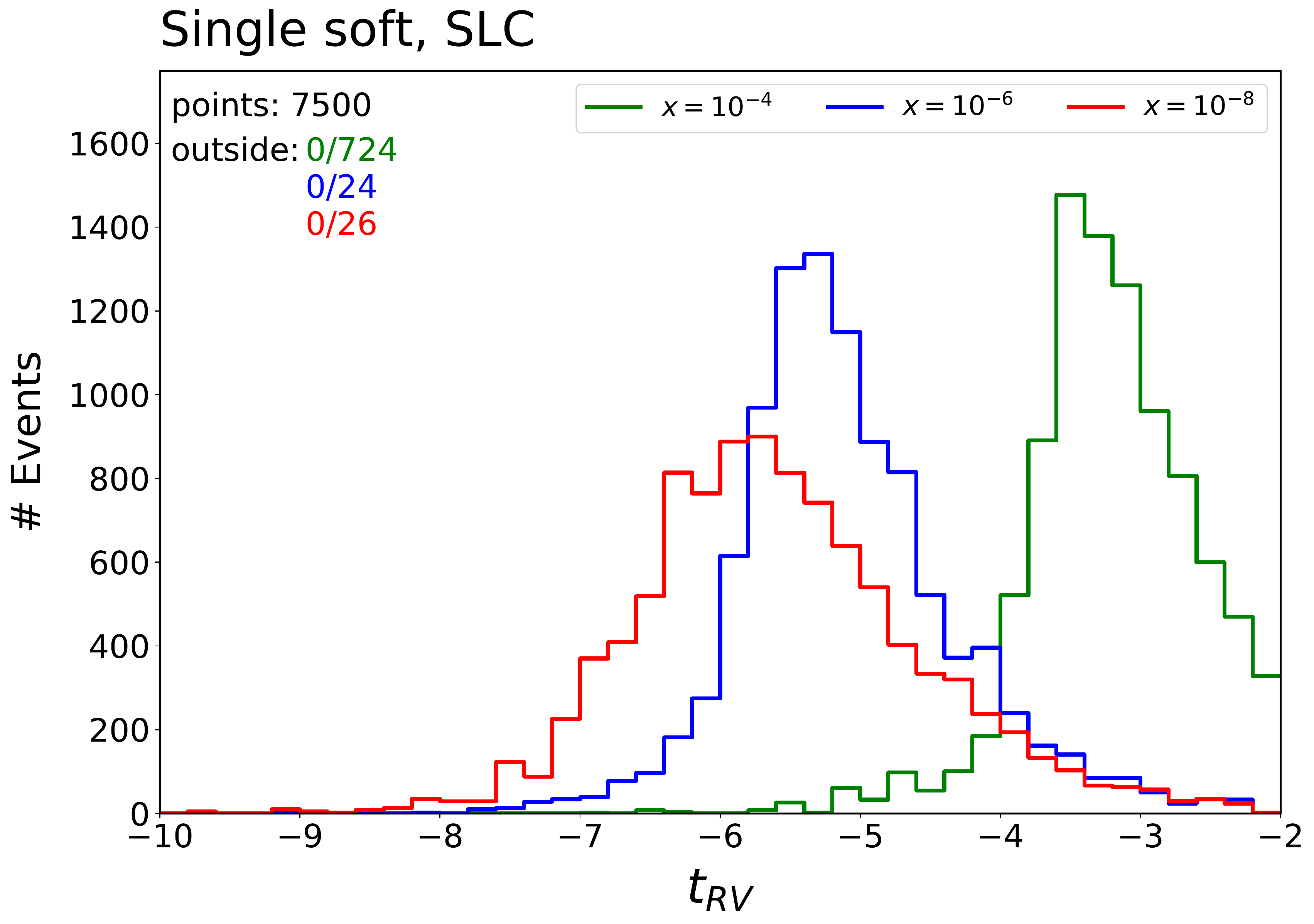}
			\hspace{1cm}
			\includegraphics[width=0.45\columnwidth]{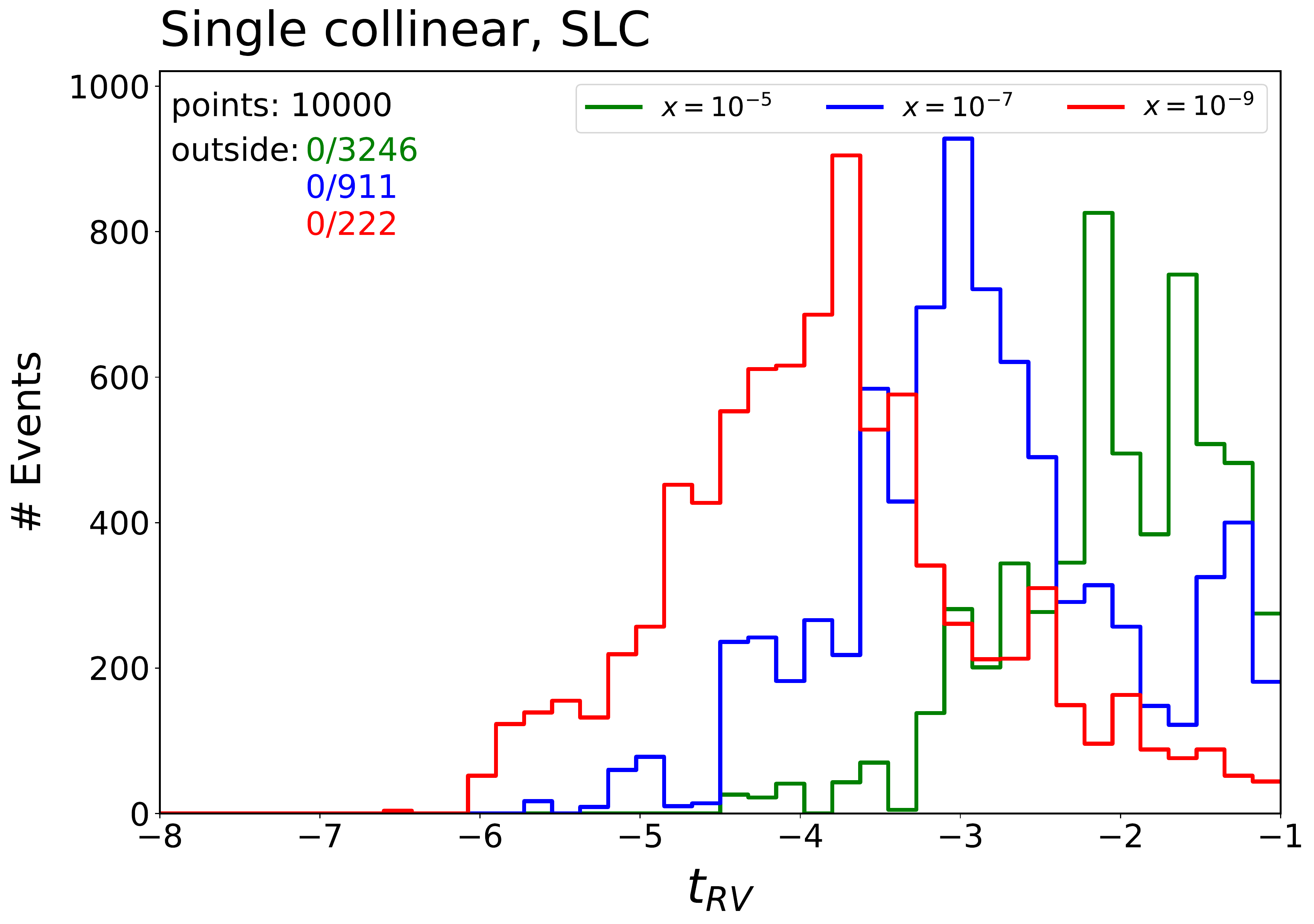}
		\end{center}
		\caption{Test of the real virtual subtraction term in single soft (left) and single collinear (right) limits for the leading colour (upper plots) and subleading colour (lower plots) contributions.}
		\label{fig:spikeRV}
	\end{figure}

	We start by addressing the single unresolved limits of the real virtual correction. There are two IR regions that we test: single soft emission and single collinear emission. The results are shown in Figure~\ref{fig:spikeRV}. The pairs of numbers reported in the plots under the label `outside' respectively indicate how many events fell on the left and on the right of the displayed range in $t_i$. 
	
We observe that the agreement between the squared matrix elements and the subtraction terms increases the deeper the IR regions are tested, with the single soft limit exhibiting more peaked distributions with respect to the single collinear one. Reasons for this are the more divergent behaviour of matrix elements and the exact locality of the subtraction in soft limits. For a given value of $x$, both in the soft and in the collinear limit, the subleading colour contribution is characterized by broader distributions, centred around higher values of $t_i$ with respect to the leading colour counterpart. This can be explained by an enhanced numerical noise for this contribution. 
The expressions for subtraction terms and squared matrix elements at subleading colour are considerably larger compared to the 
leading colour case, and each IR limit receives contributions from a substantial number of individual terms.  The numerical cancellations between these terms induce rounding errors in the final expressions for the squared matrix element and subtraction term, thereby leading to the observed 
deterioration of convergence. Nevertheless, the observed agreement between the subtraction terms and the matrix elements is largely satisfactory
both at leading and subleading colour. In general, we remark that the excellent performance of the \textsc{OpenLoops} implementation of the six-gluon one-loop amplitudes is crucial to probe the IR limits of the real virtual correction for such small values of $x$. 
	
	\begin{figure}
		\begin{center}
			\includegraphics[width=0.45\columnwidth]{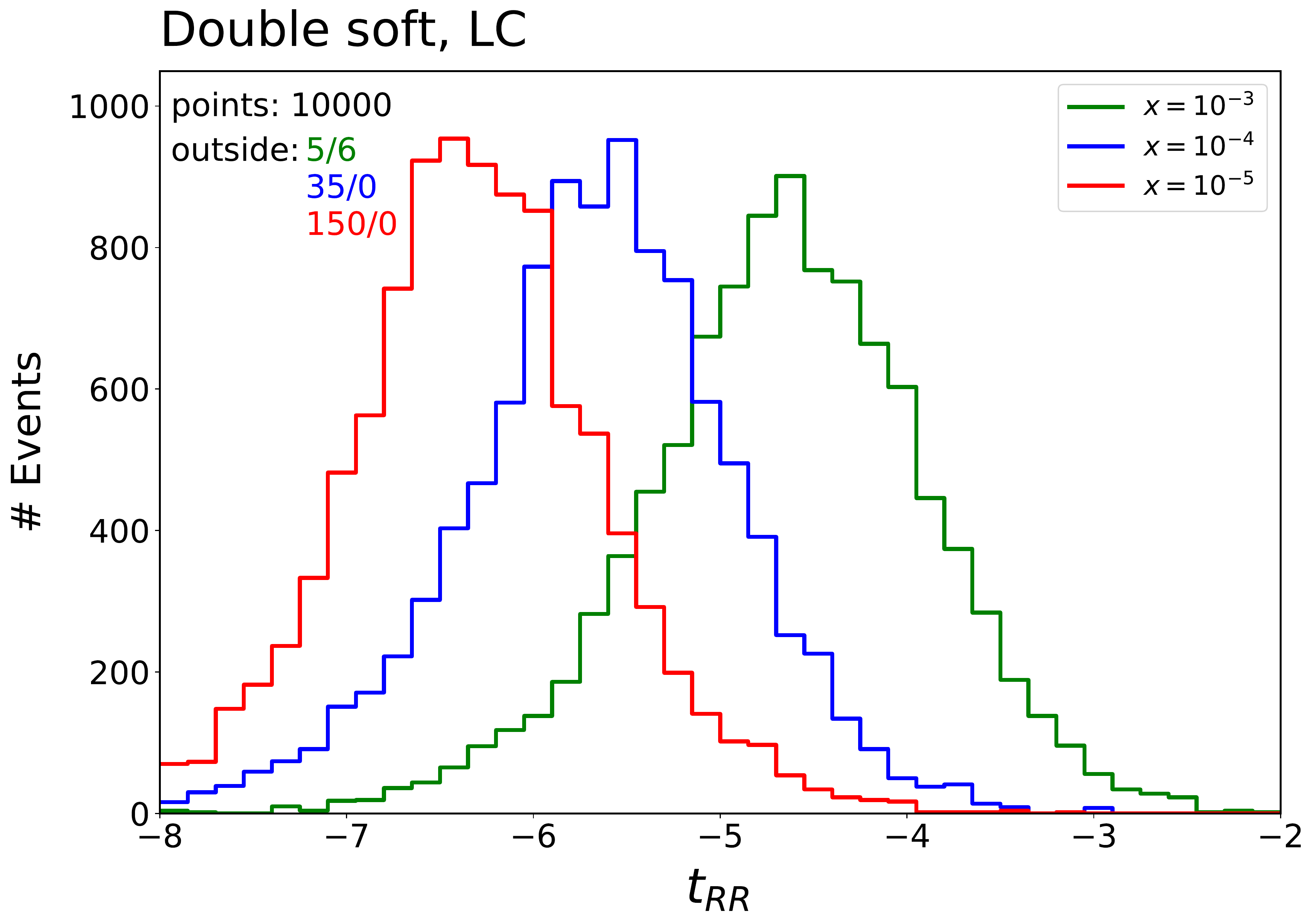}
			\hspace{1cm}
			\includegraphics[width=0.45\columnwidth]{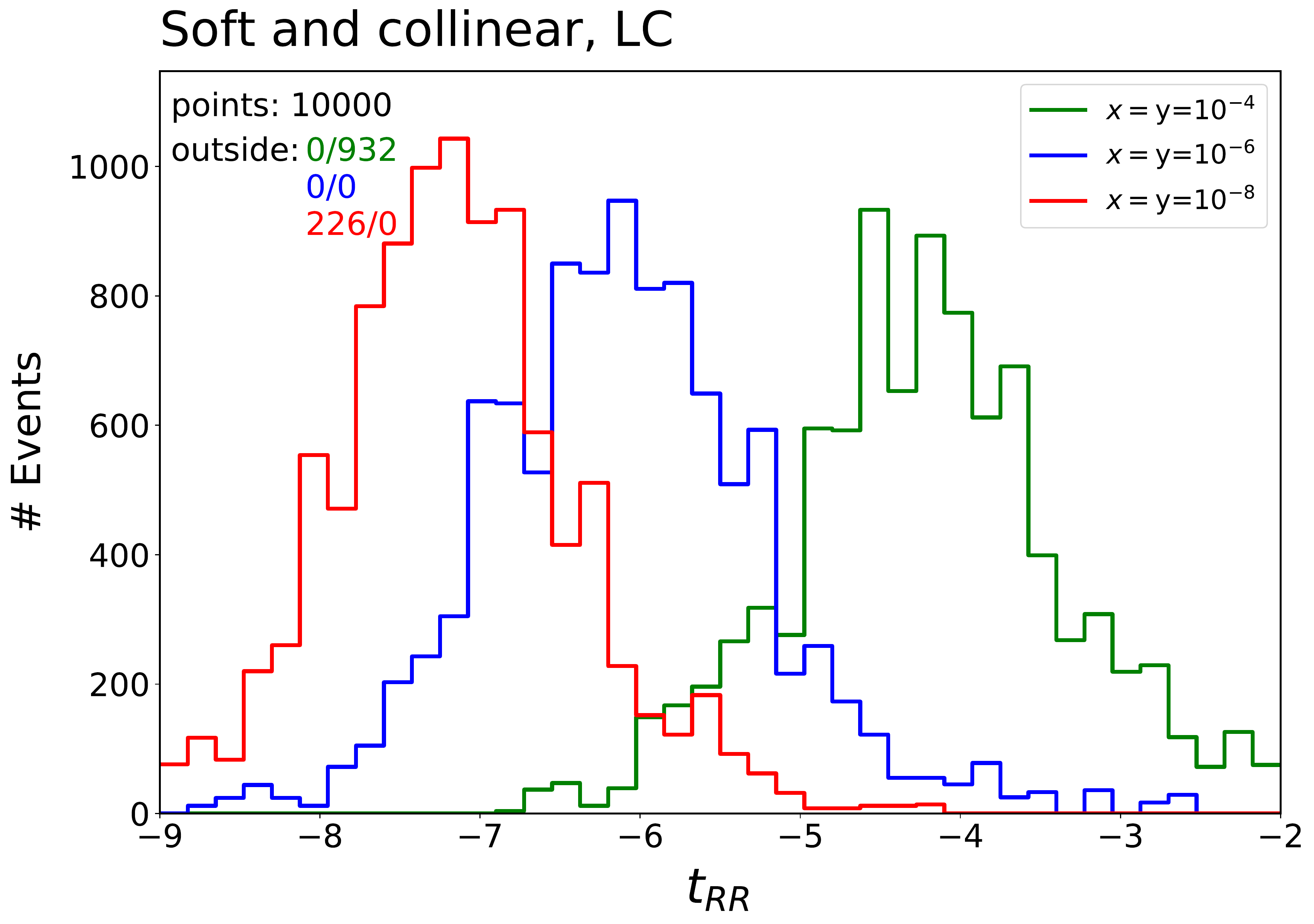}\\
			\includegraphics[width=0.45\columnwidth]{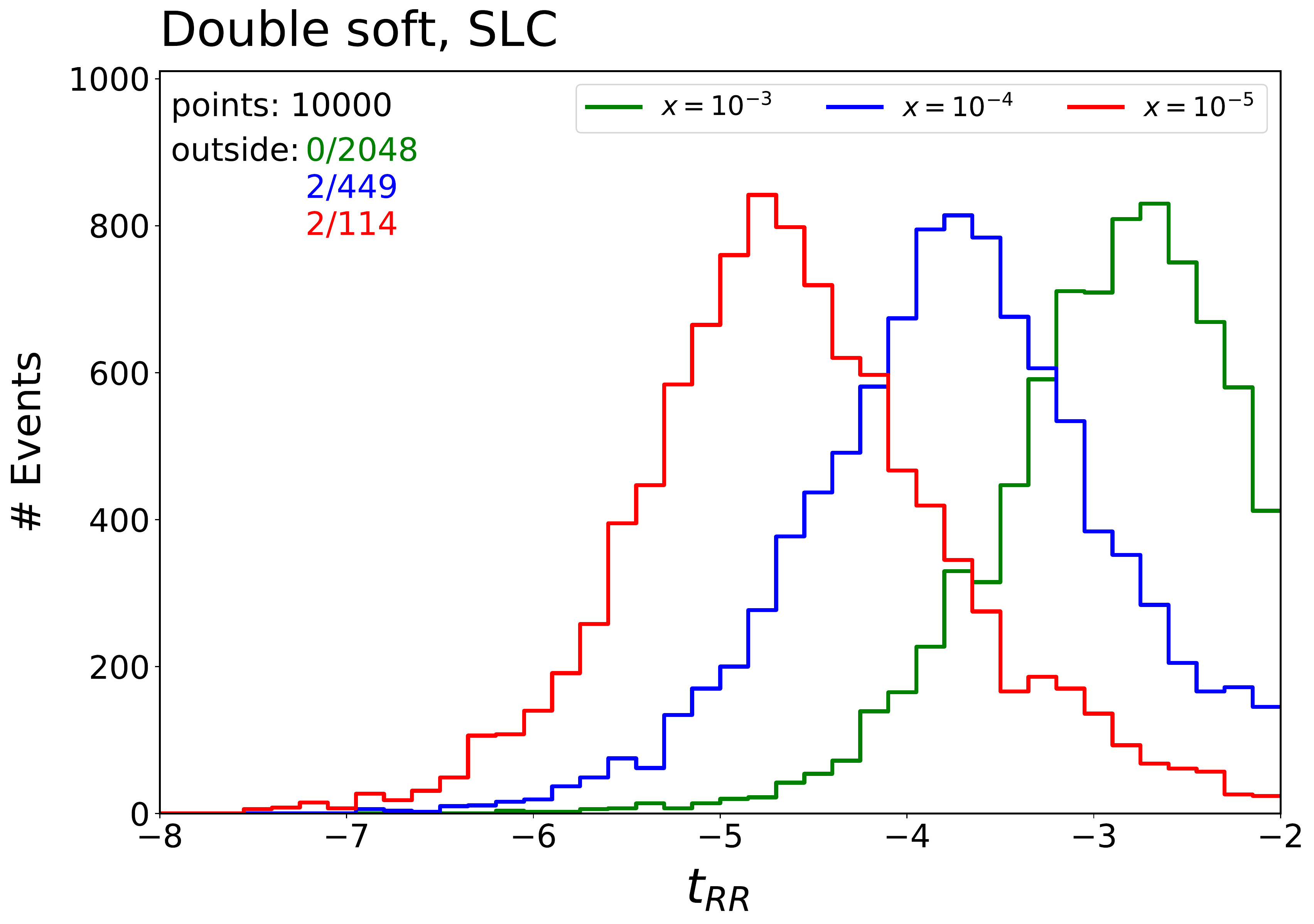}
			\hspace{1cm}
			\includegraphics[width=0.45\columnwidth]{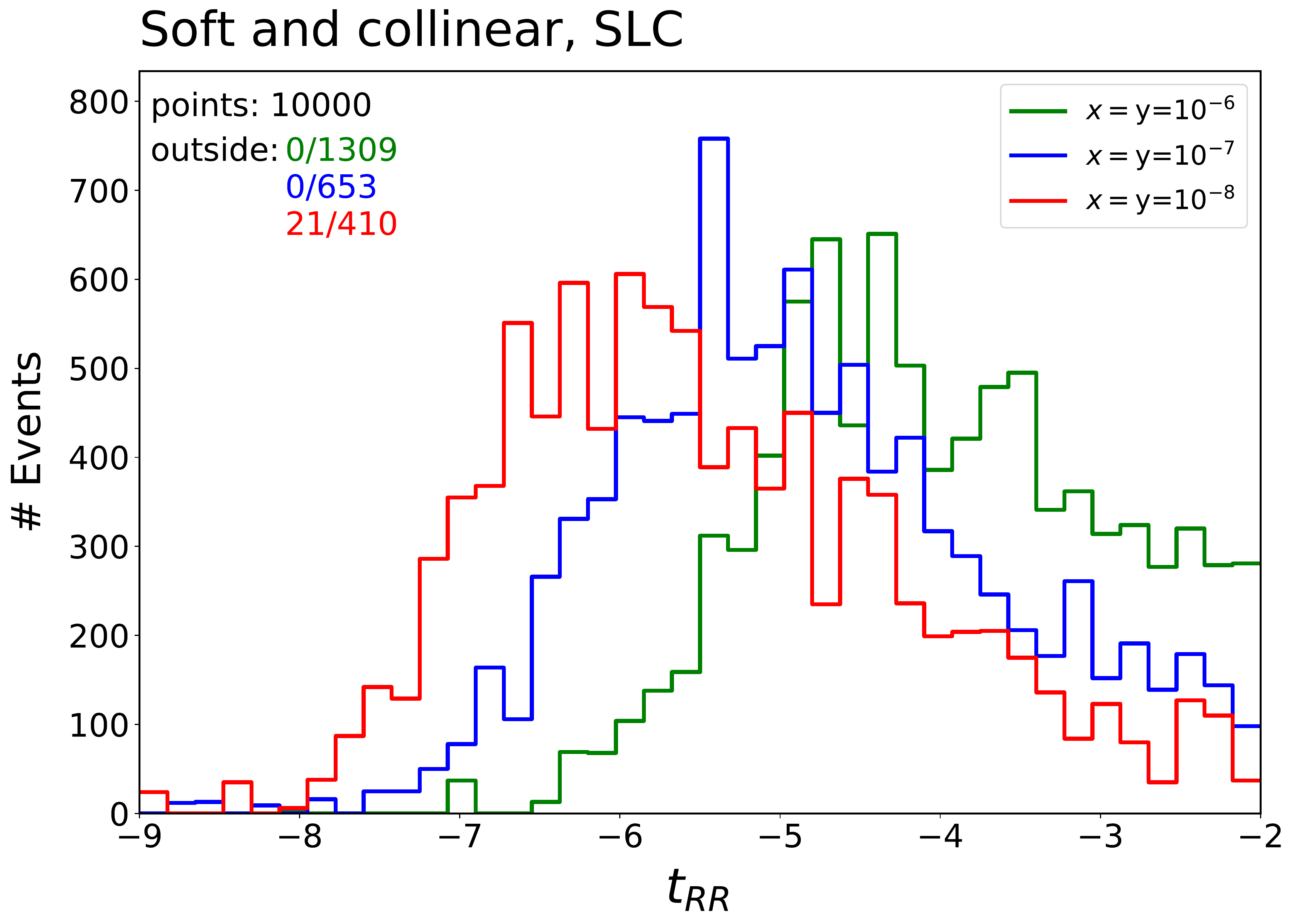}
		\end{center}
		\caption{Test of the double real subtraction term in double soft (left) and soft and collinear (right) limits for the leading colour (upper plots) and subleading colour (lower plots) contributions.}
		\label{fig:spikeRR_DS_SC}
	\end{figure}
\begin{figure}
		\begin{center}
			\includegraphics[width=0.45\columnwidth]{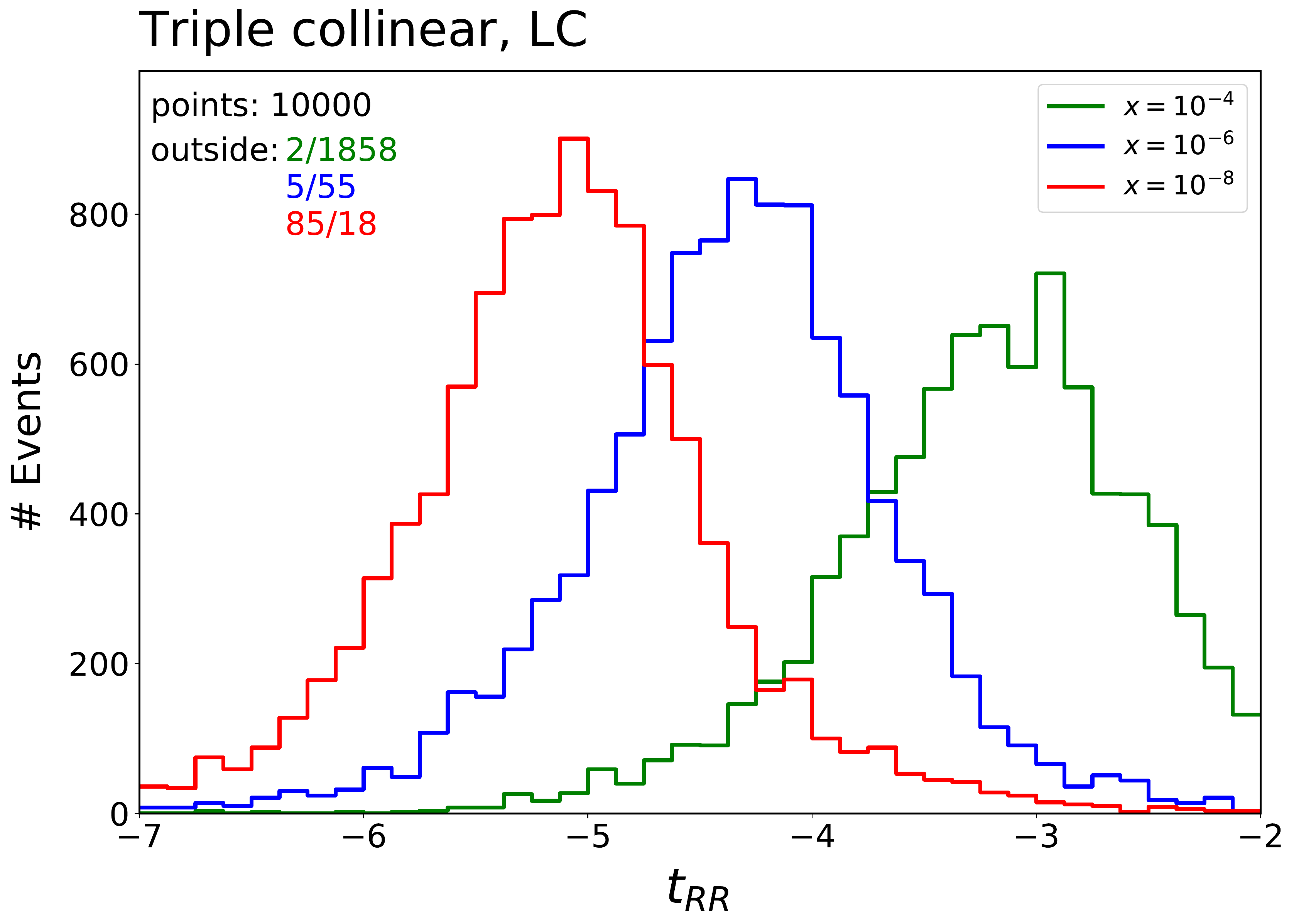}
			\hspace{1cm}
			\includegraphics[width=0.45\columnwidth]{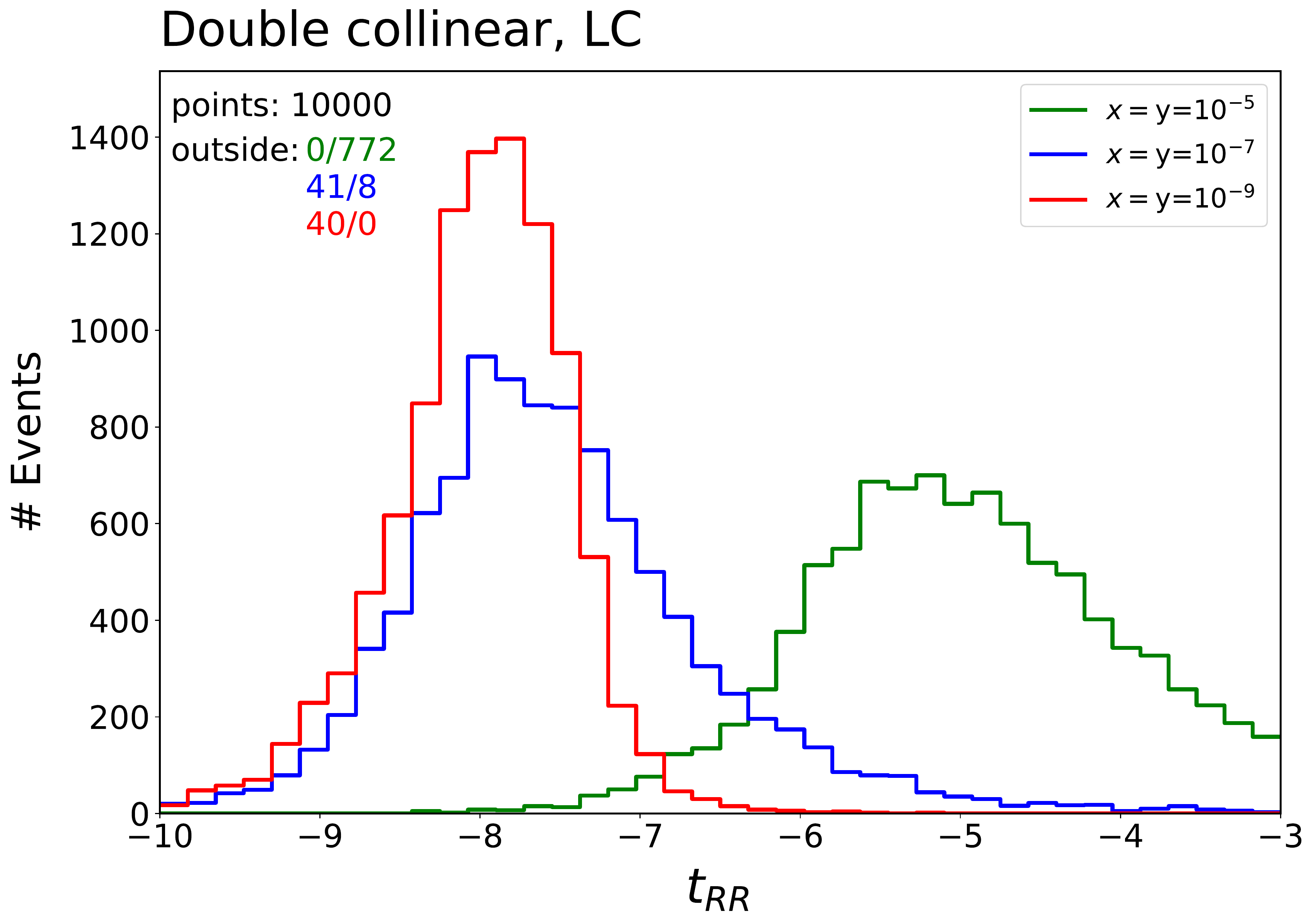}\\
		\end{center}
		\caption{Test of the double real subtraction term in triple collinear (left) and double collinear (right) limits at leading colour. The subleading colour contribution does not exhibit IR divergences in these limits.}
		\label{fig:spikeRR_TC_DC}
	\end{figure}
\begin{figure}
		\begin{center}
			\includegraphics[width=0.45\columnwidth]{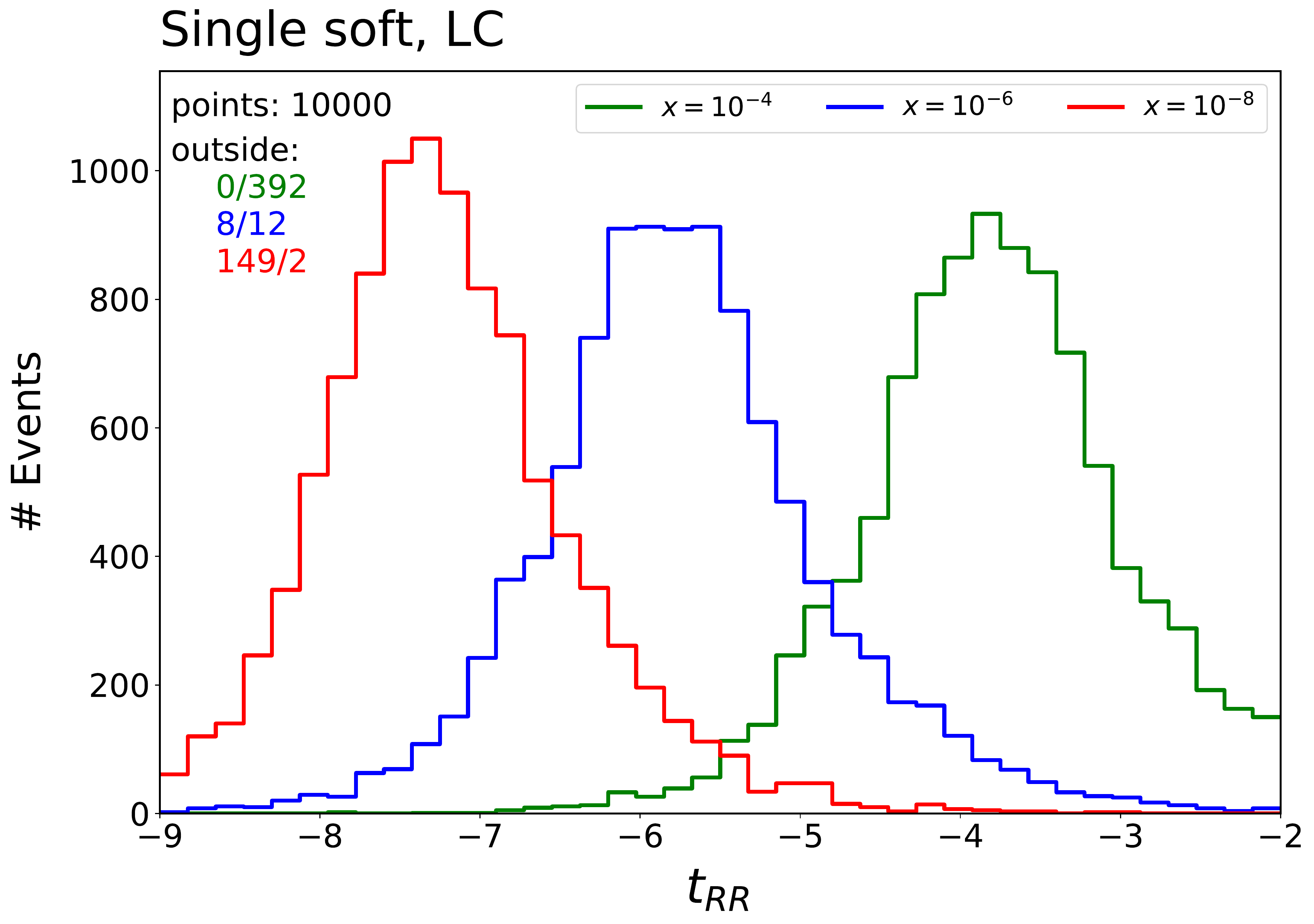}
			\hspace{1cm}
			\includegraphics[width=0.45\columnwidth]{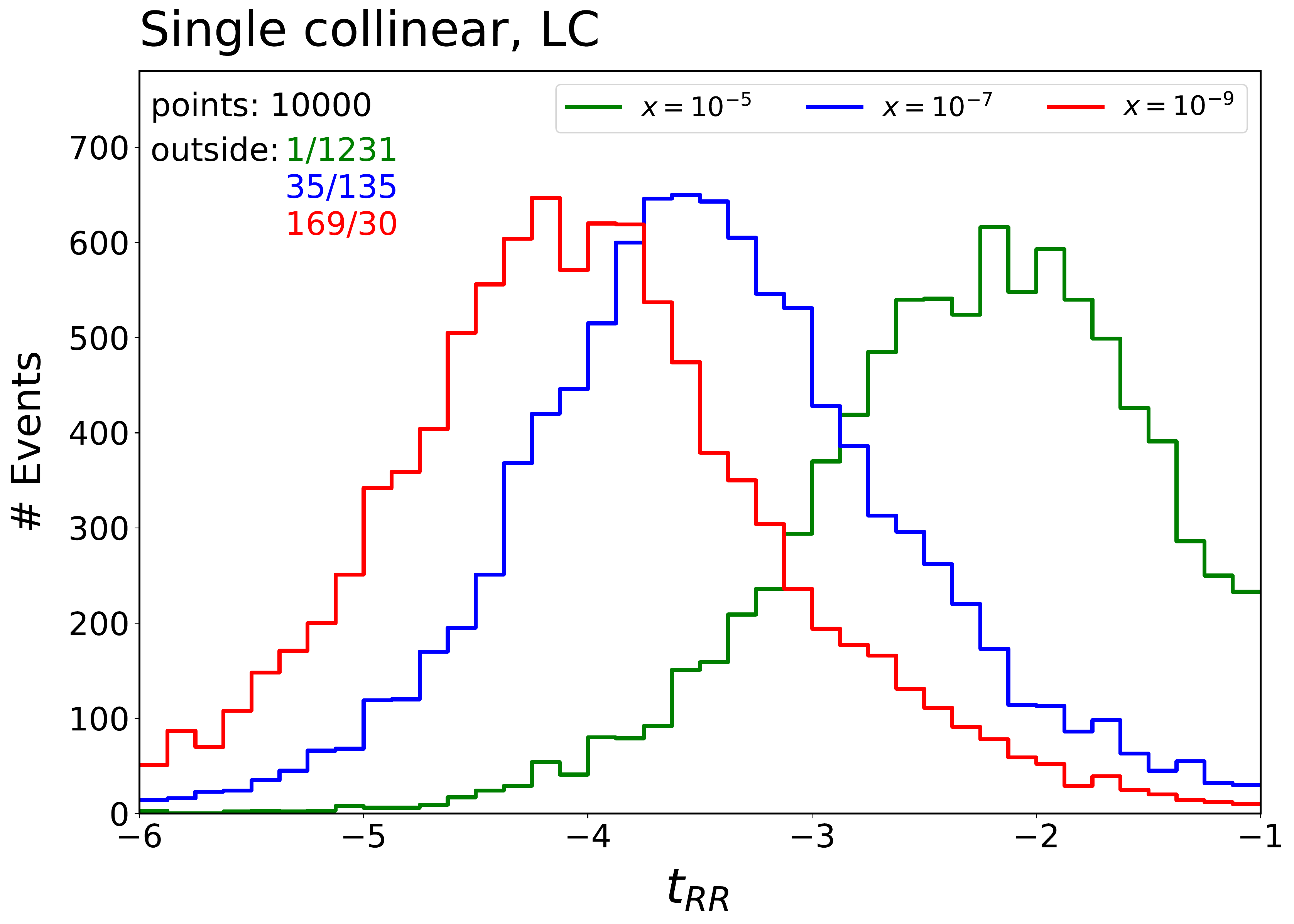}\\
			\includegraphics[width=0.45\columnwidth]{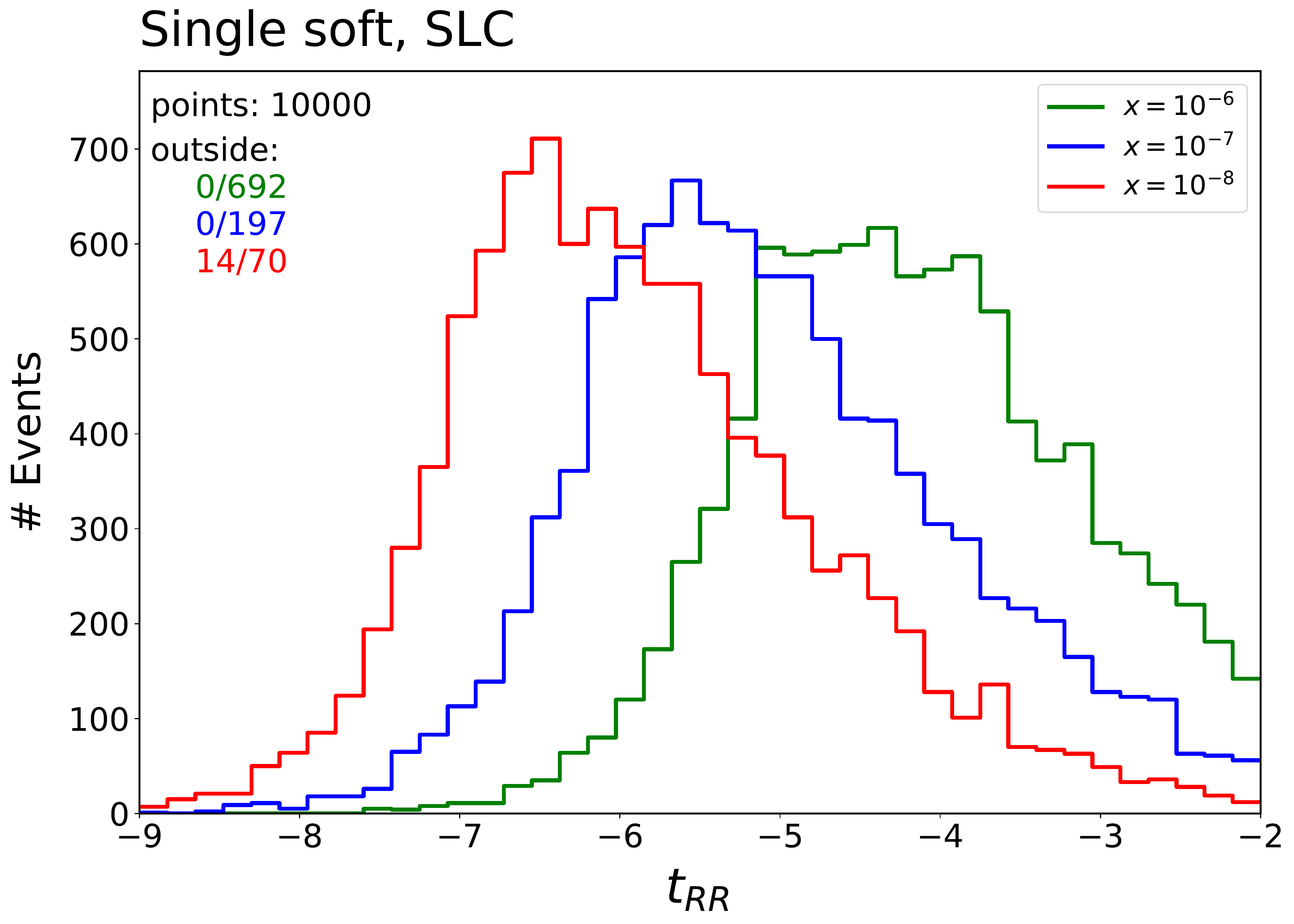}
			\hspace{1cm}
			\includegraphics[width=0.45\columnwidth]{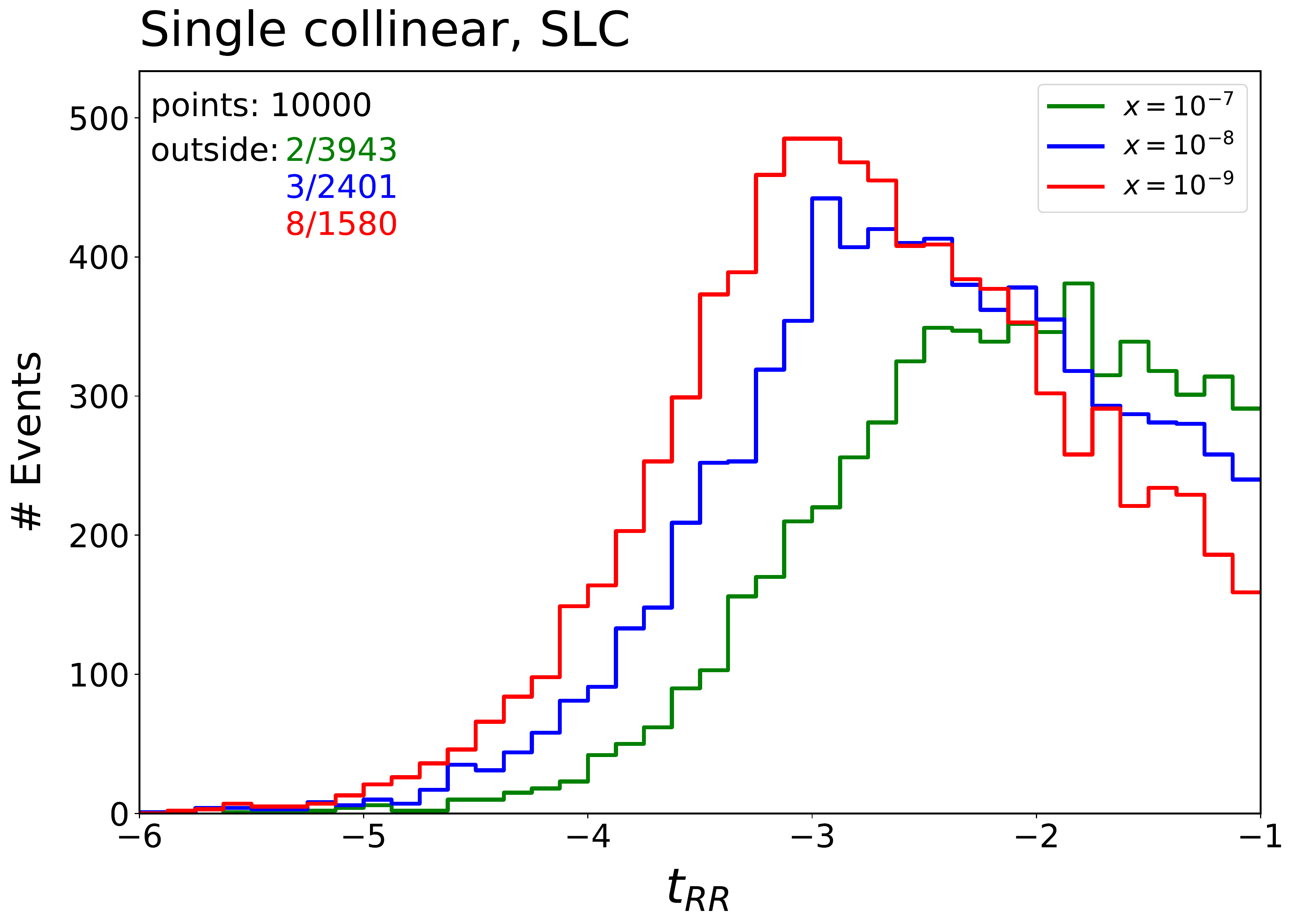}
		\end{center}
		\caption{Test of the double real subtraction term in single soft (left) and single collinear (right) limits for the leading colour (upper plots) and subleading colour (lower plots) contributions.}
		\label{fig:spikeRR_S_C}
	\end{figure}

		For the double real correction, we have both single and double unresolved limits. Concerning double unresolved limits we have double soft emission, soft and collinear emission, triple collinear emission and double collinear emission. The results for the double soft and soft and collinear limits are presented in Figure~\ref{fig:spikeRR_DS_SC}, while in Figure~\ref{fig:spikeRR_TC_DC} we report the results for triple collinear and double collinear limits. Finally we present the results for the single unresolved limits of the double real correction in Figure~\ref{fig:spikeRR_S_C}.

	In double and triple collinear limits, the subleading colour contribution does not exhibit a divergent behaviour. This is motivated by an argument analogous to the one adduced in~\cite{Currie:2013dwa} to explain why the six-gluon tree-level matrix element is not divergent in single, double and triple collinear limits at subleading colour. For the seven-gluon tree-level matrix element considered here, one can express the subleading colour contribution as a sum of incoherent interferences within which the two colour orderings share at most a single pair of adjacent gluons. This implies that the subleading colour part of the matrix element is indeed divergent in single collinear limits, but is finite in double and triple collinear limits.
	
	In general we observe that the subtraction terms correctly reproduce the divergent behaviour of the double real correction in both double and single unresolved configurations, with the quality of the agreement increasing with the infrared depth. Once again, the subleading colour contribution presents some numerical noise with respect to the leading colour part, for the same reasons explained in the real virtual case.

	\subsection{Results}\label{sec:results}
	
	In this section we present the results of our NNLO calculation. The considered centre-of-mass energy for the colliding protons is $13$ TeV and the applied kinematical cuts are the same used for the three-jet production NNLO calculation in~\cite{Czakon:2021mjy}. The cuts are as follows:
	
	\begin{itemize}
		\item minimal transverse momentum of a jet: $p_T(j)>60$ GeV;
		\item maximal jet rapidity: $|y(j)|<4.4$;
		\item minimal sum of the transverse momenta of the first two leading jets : $p_T(j_1)+p_T(j_2)>250$ GeV.
	\end{itemize}
	 Jets are reconstructed using the anti-$k_T$ algorithm~\cite{Cacciari:2008gp} with a radius $R=0.4$. We use the NNLO set of the NNPDF3.1 parton distribution functions~\cite{NNPDF:2017mvq} and we evaluate the PDFs using LHAPDF~\cite{Buckley:2014ana}. 
	The same PDF set is used for the predictions at LO, NLO and NNLO. LHAPDF is also used to evaluate the strong coupling constant, with $\alpha_s(m_Z)=0.118$. 
	 We notice here that these quantities are the only ones in the entire calculation which are evaluated in the full QCD theory, namely with the complete bosonic and fermionic degrees of freedom of QCD. As we explicitly show in section~\ref{sec:scale_variation}, this is the origin of an inconsistent estimation of the uncertainty on our theory prediction, which is assessed with a seven-point variation of the renormalization and factorization scales. The central value is chosen dynamically for each event and is given by
	the scalar sum of transverse momenta at parton level:
	\begin{equation}
		\mu_f=\mu_r=\hat{H}_{T}=\sum_{i\in \text{partons}} p_{T}(i).
	\end{equation}
	We consider distributions in the $H_T$ observable:
	\begin{equation}
		H_T=\sum_{j\in \text{jets}} p_{T}(j),
	\end{equation}
	in the three-jet invariant mass $m_{123}$, in individual jet transverse momenta and rapidities, jet-pair differences 
	in rapidity $|\Delta y|$ and azimuthal angle $|\Delta \Phi|$ as well as three-jet rapidity variables $y^\star_{123}$  and $|y_{{\rm max}}|$, defined as:
	\begin{eqnarray}
		y^*_{123}&=&|y_1-y_2|+|y_2-y_3|+|y_3-y_1|,\\
		|y_{{\rm max}}|&=&\max(|y_1|,|y_2|,|y_3|).
	\end{eqnarray}
	
		\begin{figure}
		\centering
		\includegraphics[width=0.42\textwidth]{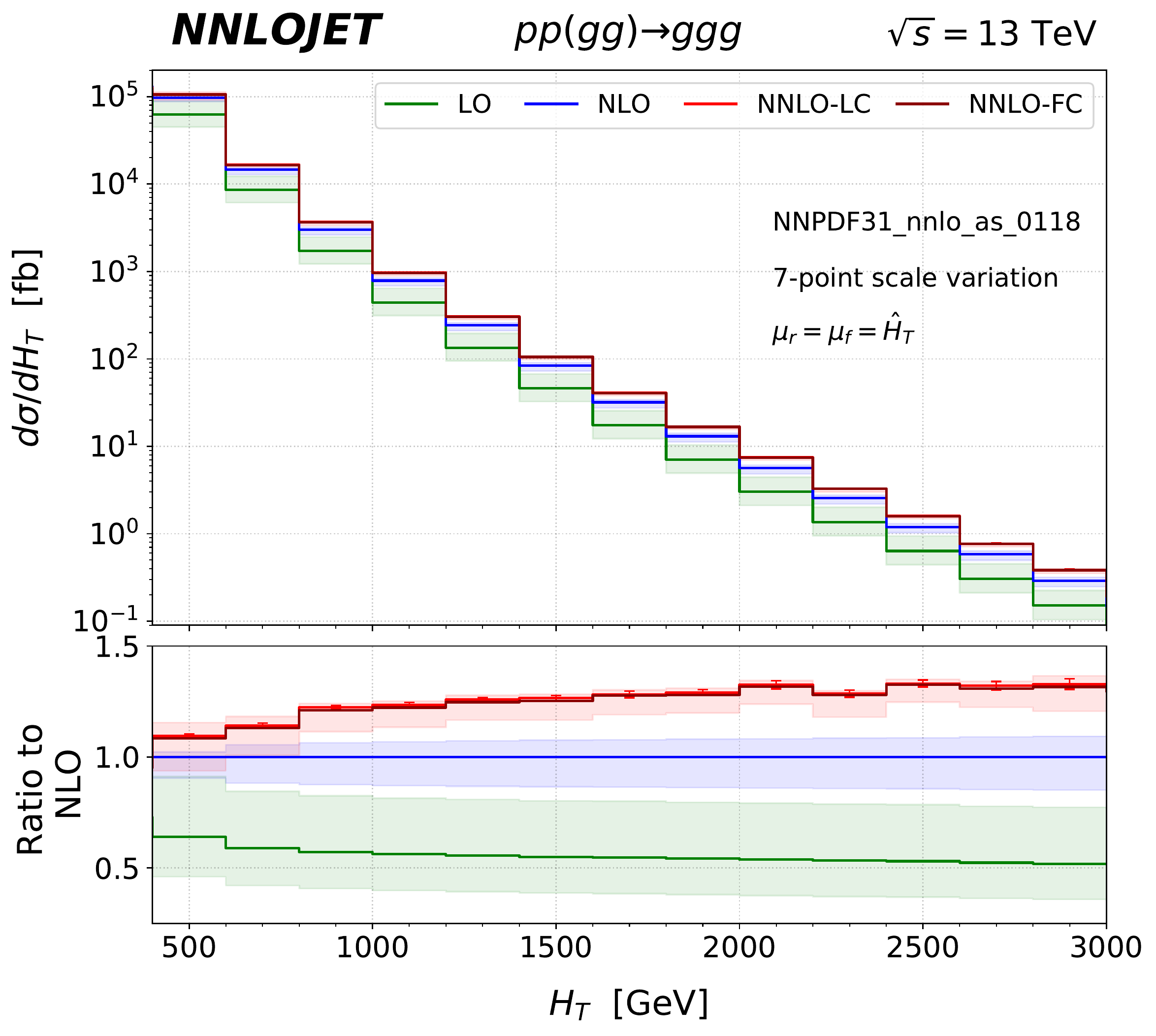}
		\includegraphics[width=0.42\textwidth]{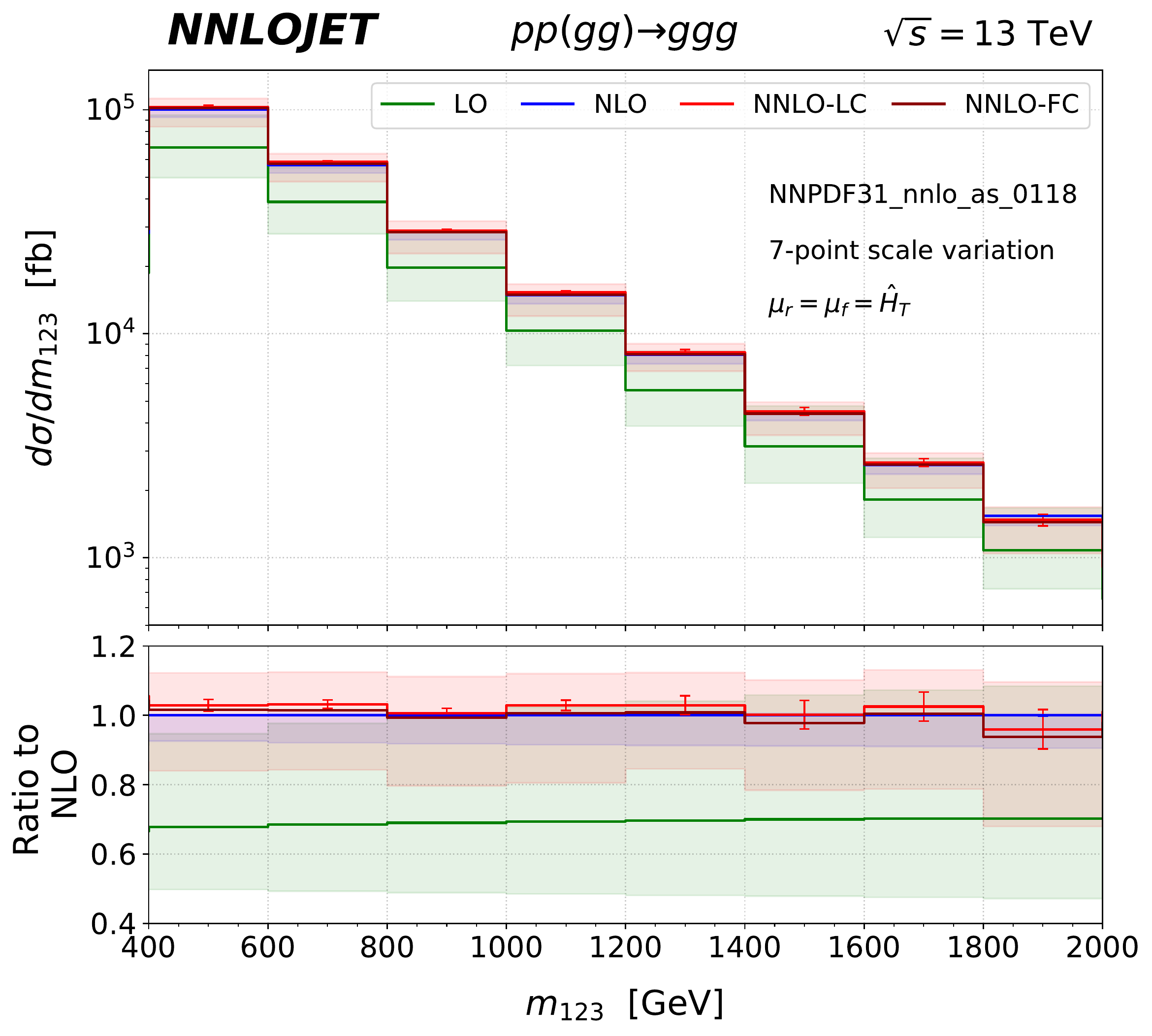}
\caption{Gluons-only three-jet production cross section differential in $H_{T}$ (left) and $m_{123}$ (right) up to NNLO. The NNLO-LC band 
corresponds to the leading-colour only contributions at NNLO, the NNLO-FC line includes subleading colour contributions except 
for the finite parts of the two-loop virtual corrections.}\label{fig:ht_m123}
	\end{figure}
The calculation is performed in full colour, however, the subleading colour contribution to this process at NNLO has an exiguous numerical impact and the inclusion of the missing subleading colour part of the finite reminder of the double virtual matrix element might significantly affect the value of the subleading colour correction. For this reason, in the plots presented here, we focus on the leading colour NNLO correction (NNLO-LC), which is a well defined and complete quantity. We then superimpose the full colour result without the finite two-loop remainder pieces 
	(NNLO-FC) to provide an estimate of the impact of the subleading colour contribution. The Monte Carlo integration error is reported for the NNLO correction at leading colour. A consistent effort on the computational side would be required to significantly reduce the statistical uncertainty on the presented results. Since this computation serves as an assessment of the colourful antenna subtraction method at NNLO more than a high-precision phenomenological study, we did not invest additional resources in the reduction of the integration error. 

	\begin{figure}
		\centering
		\includegraphics[width=0.32\textwidth]{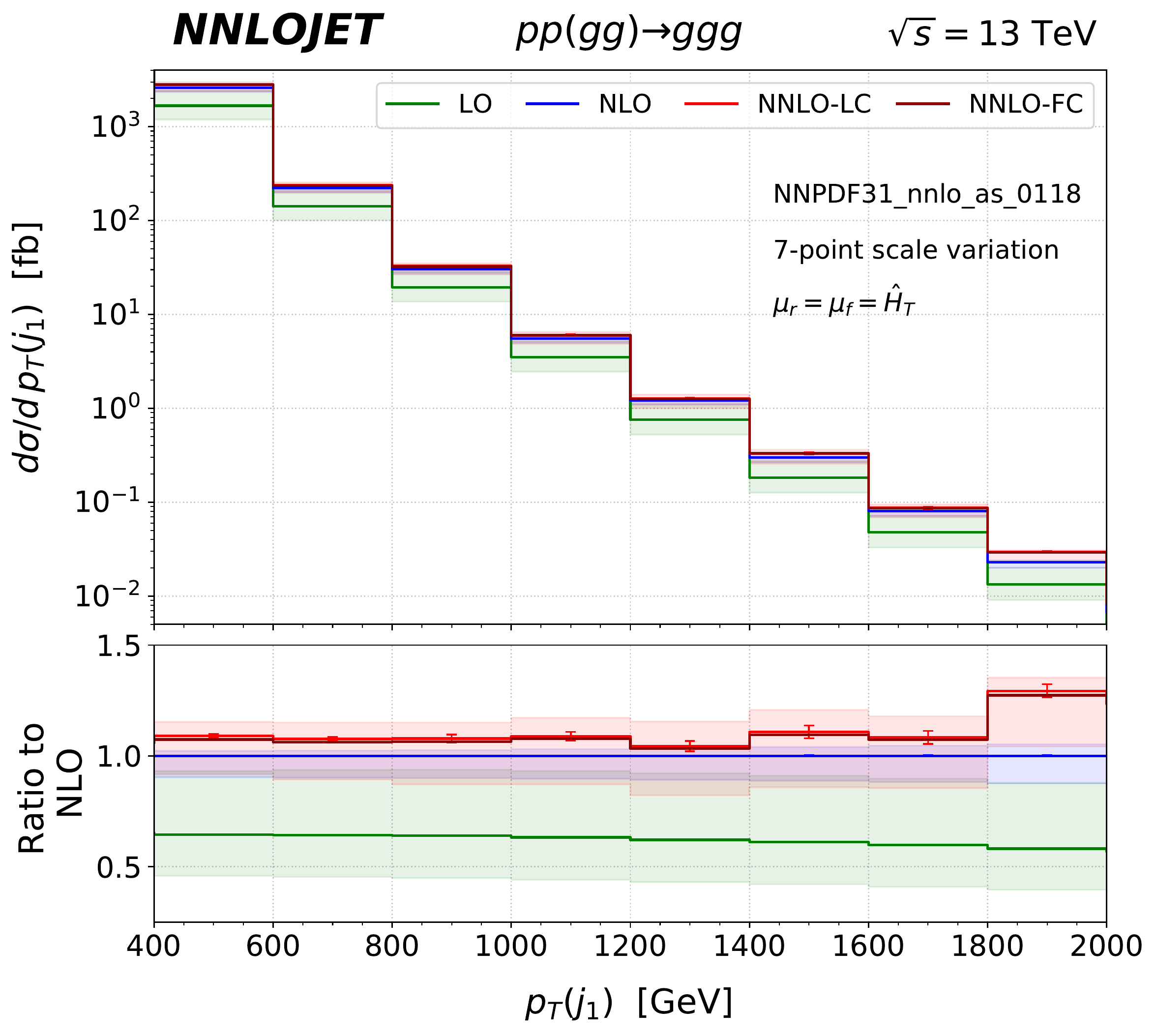}
		\includegraphics[width=0.32\textwidth]{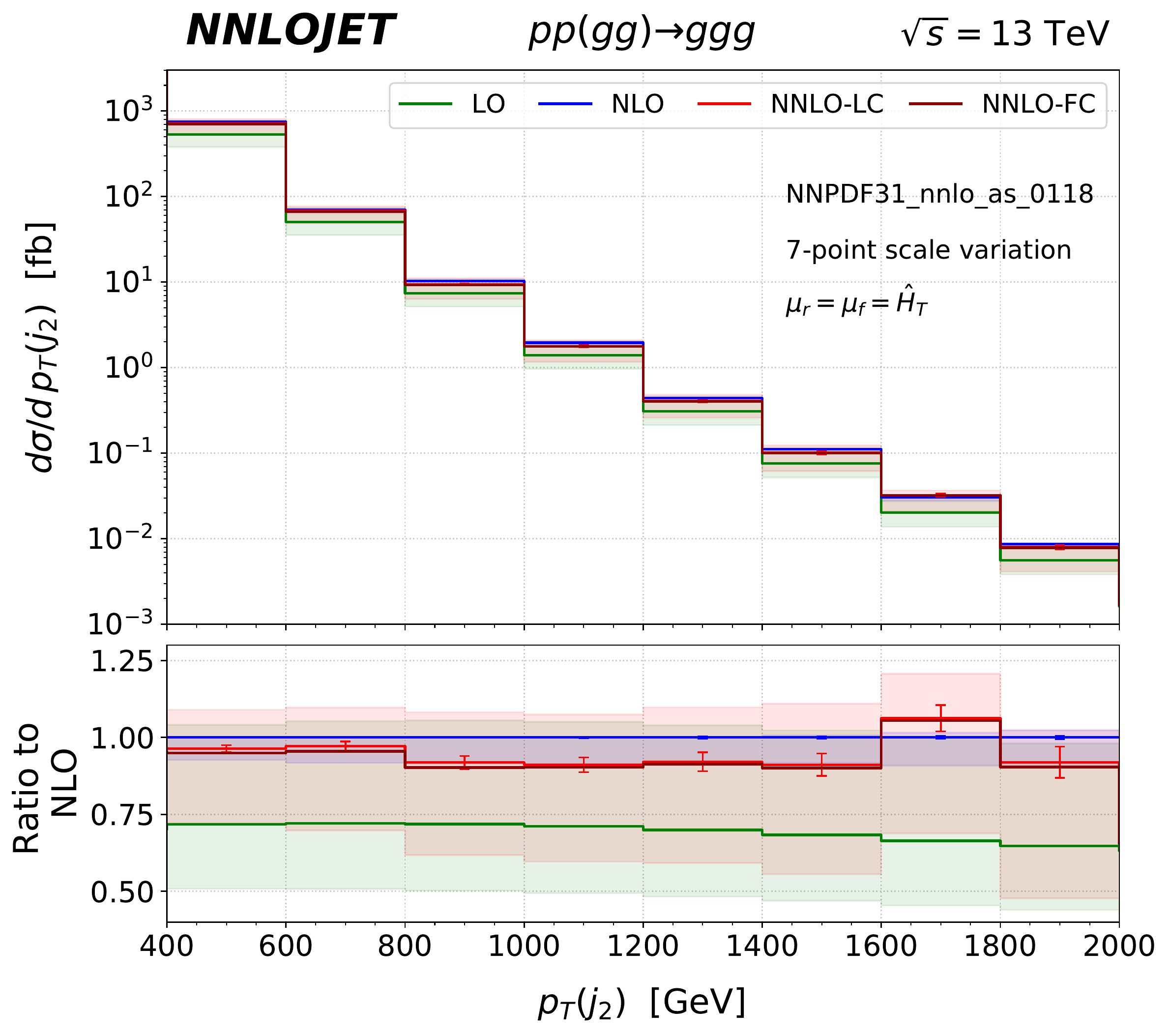}
		\includegraphics[width=0.32\textwidth]{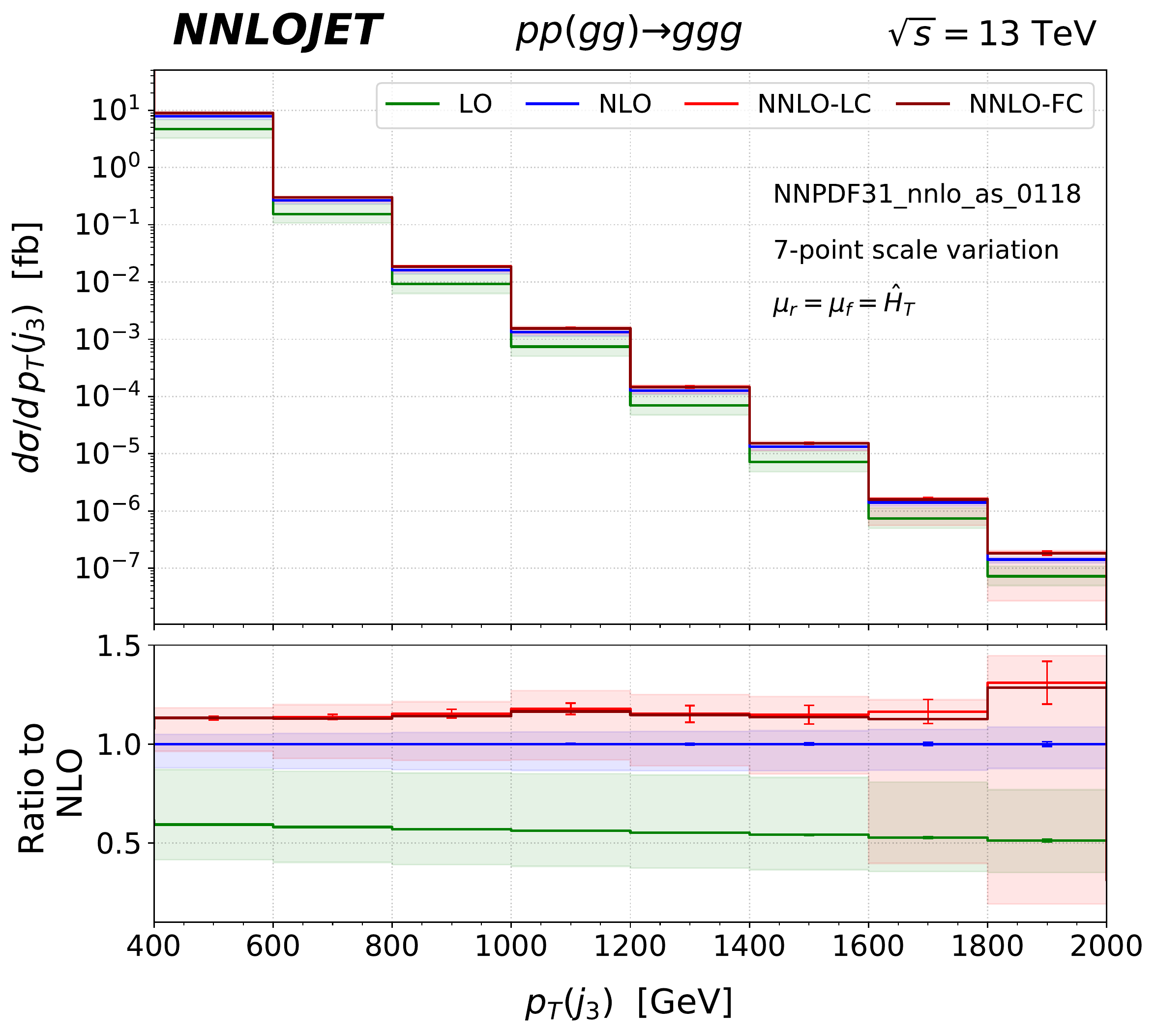}\\
		\includegraphics[width=0.32\textwidth]{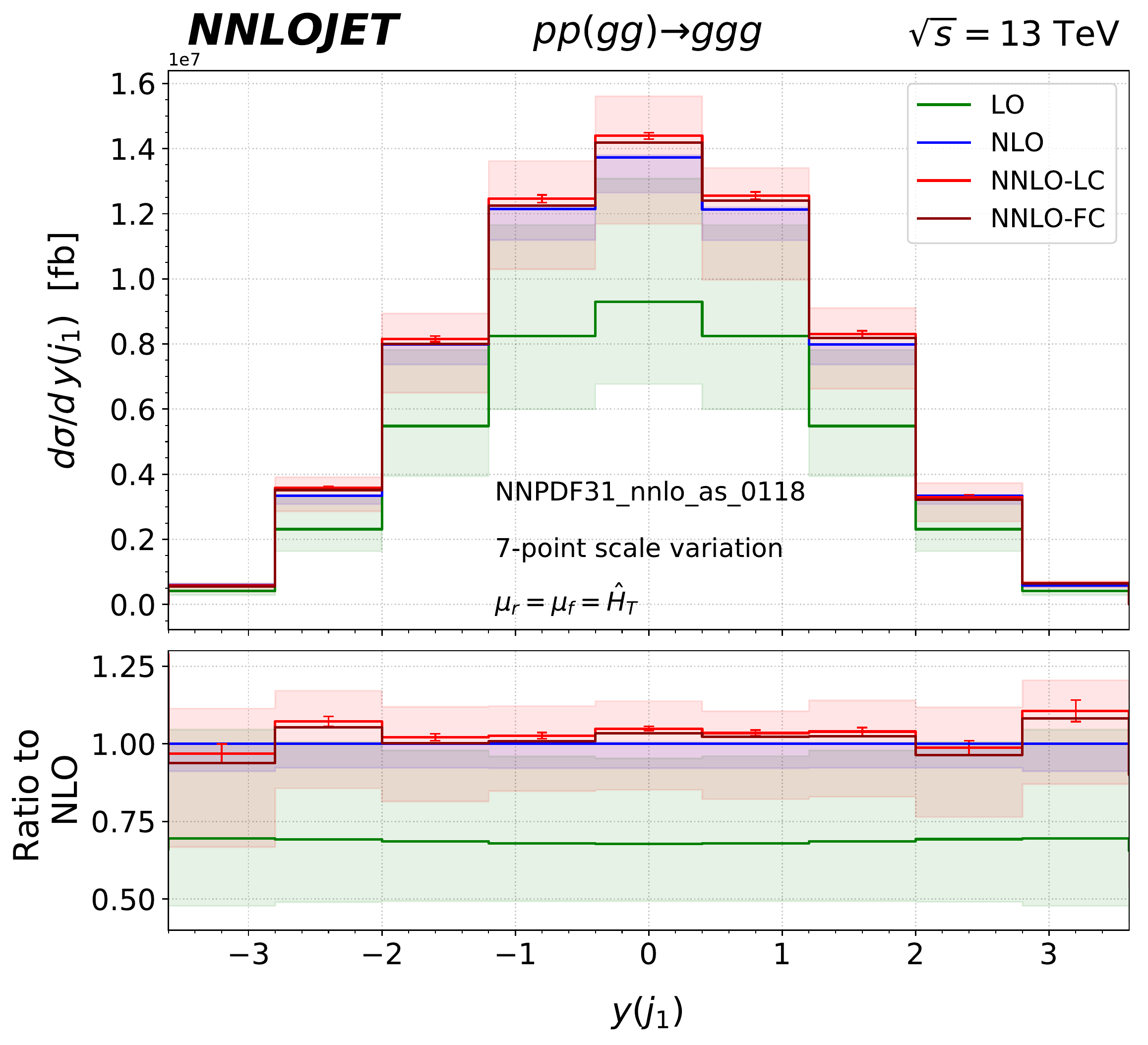}
		\includegraphics[width=0.32\textwidth]{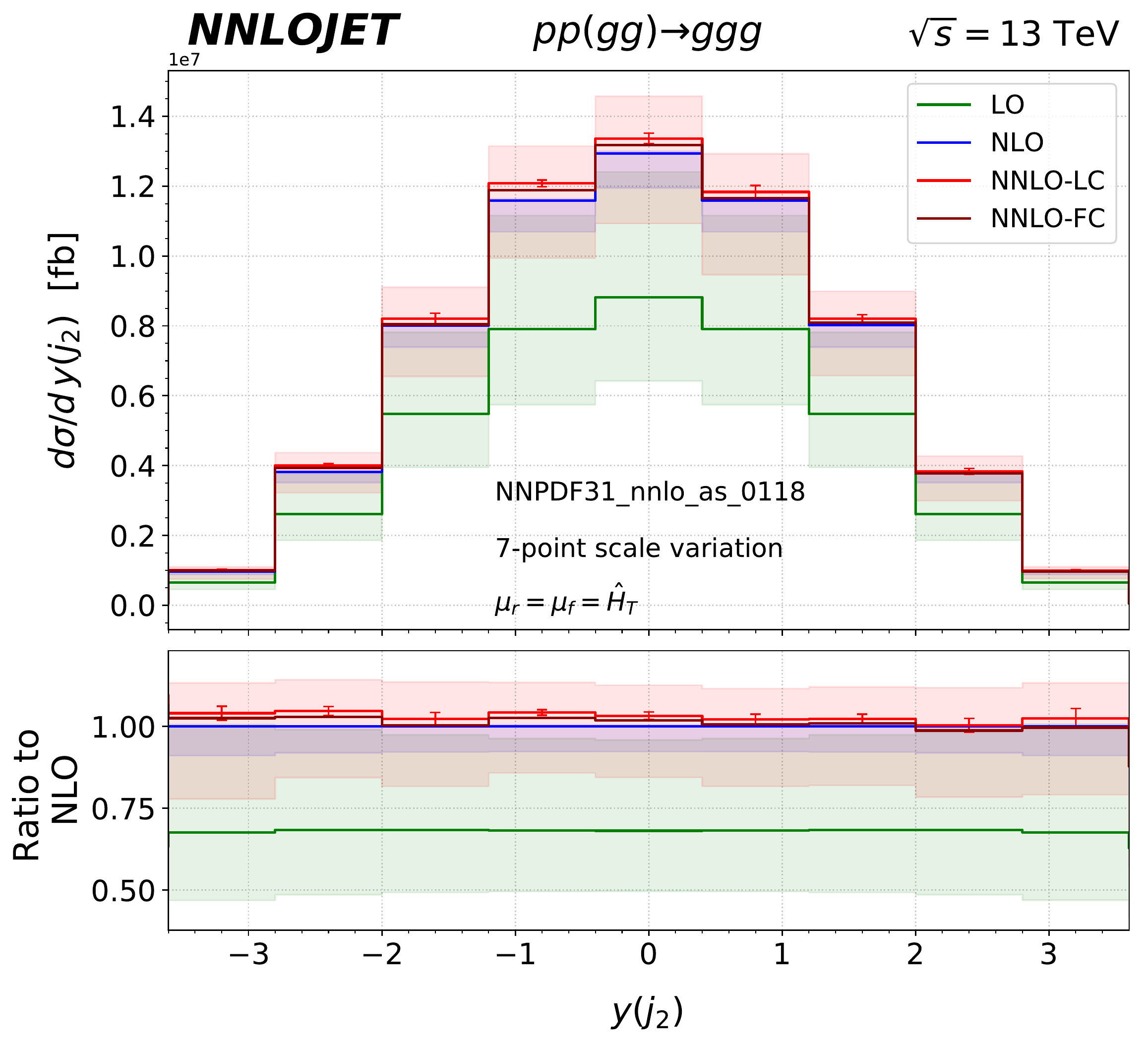}
		\includegraphics[width=0.32\textwidth]{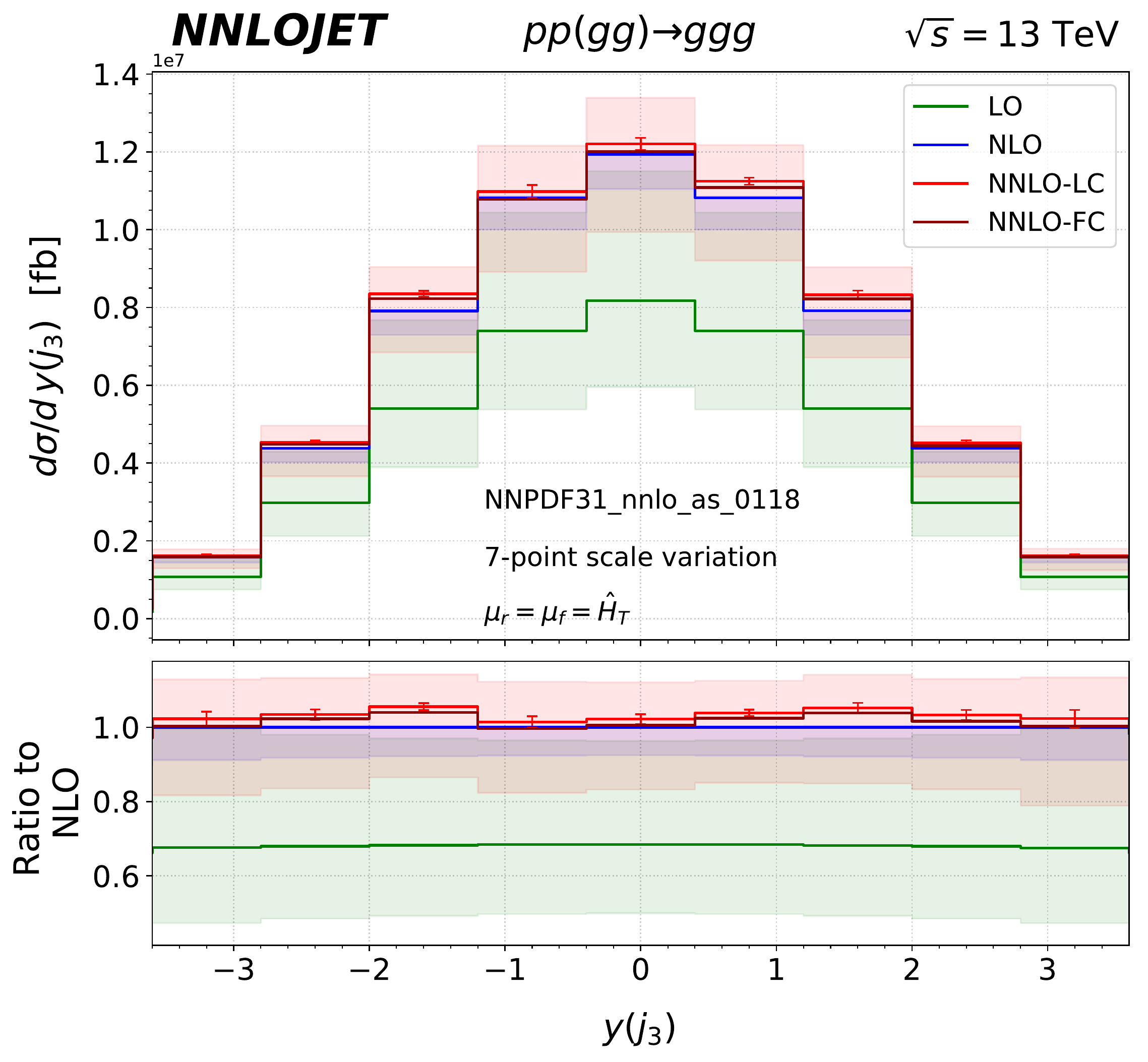}
	\caption{Differential distributions in individual jet transverse momenta (upper row) and 
	rapidities (lower row) for gluons-only three-jet production up to NNLO. Definition of NNLO-LC and  
 NNLO-FC as in Figure~\protect\ref{fig:ht_m123}.}	\label{fig:ptj_yj}
	\end{figure}
\begin{figure}
		\centering
		\includegraphics[width=0.45\textwidth]{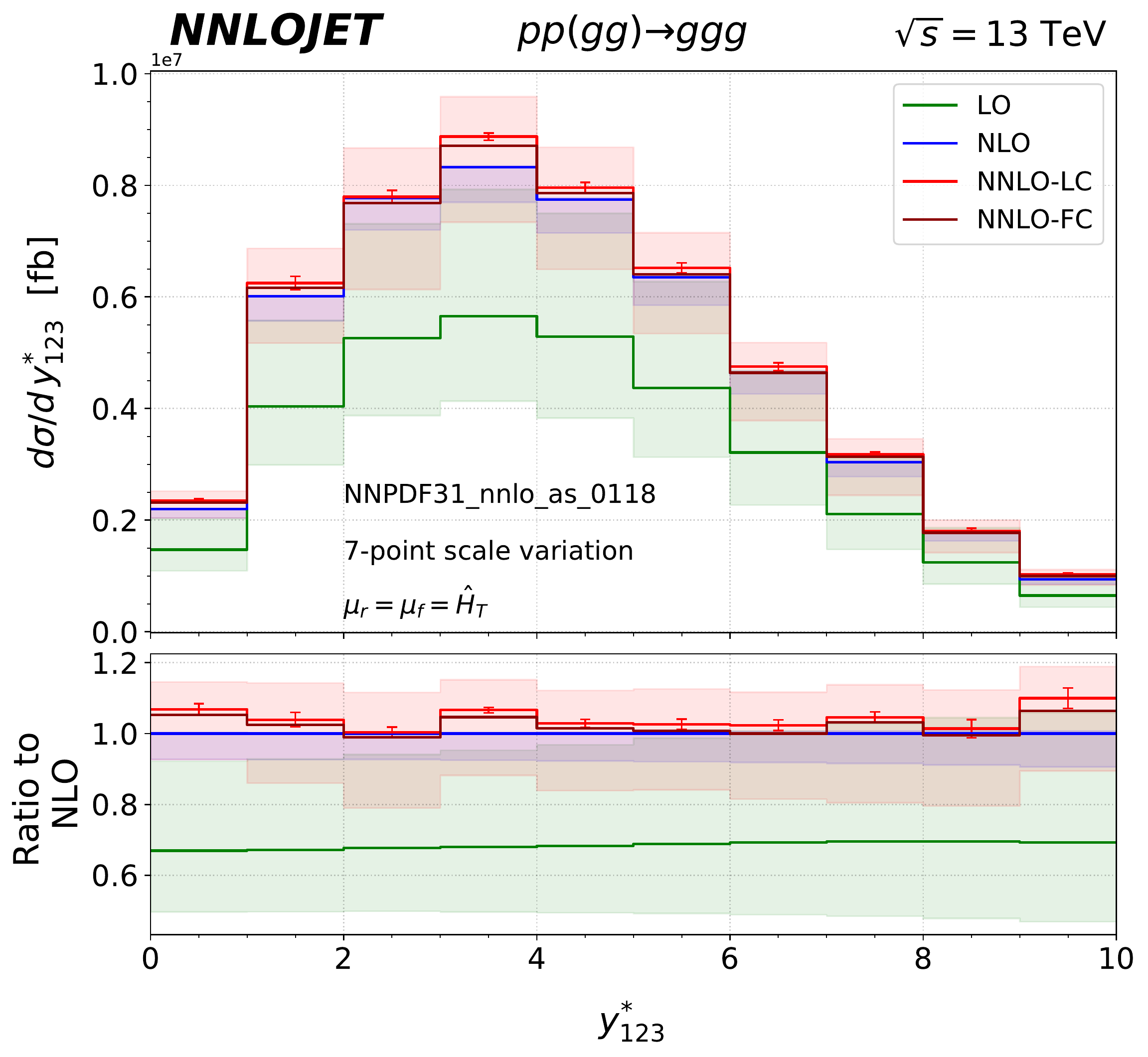}
		\includegraphics[width=0.45\textwidth]{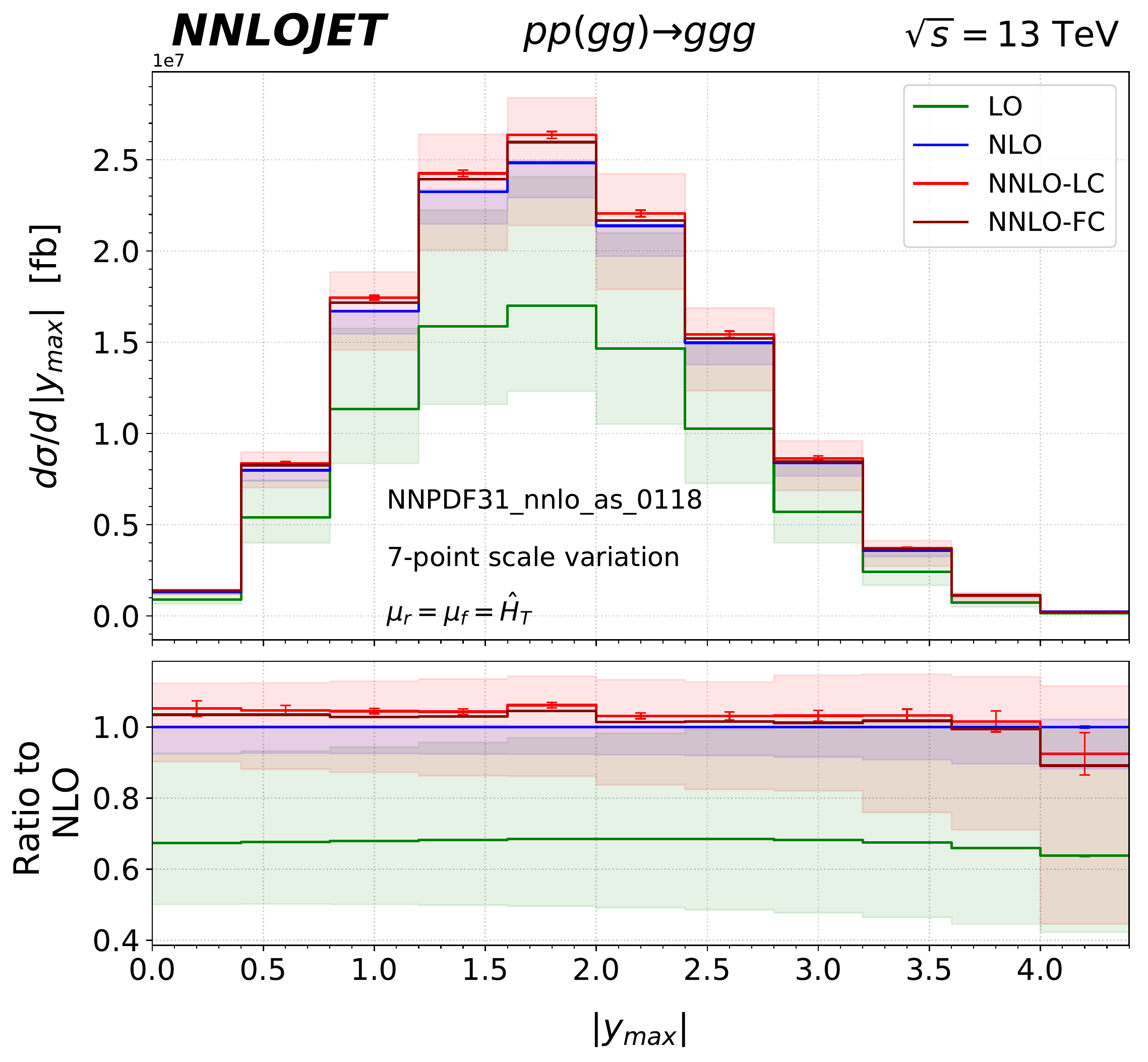}
\caption{Gluons-only three-jet production cross section differential in the combined rapidity variables
 $y^\star_{123}$ (left) and $|y_{{\rm max}}|$ (right) up to NNLO. Definition of 
NNLO-LC and  
 NNLO-FC as in Figure~\protect\ref{fig:ht_m123}.}\label{fig:yvar}
	\end{figure}
\begin{figure}
		\centering
		\includegraphics[width=0.32\textwidth]{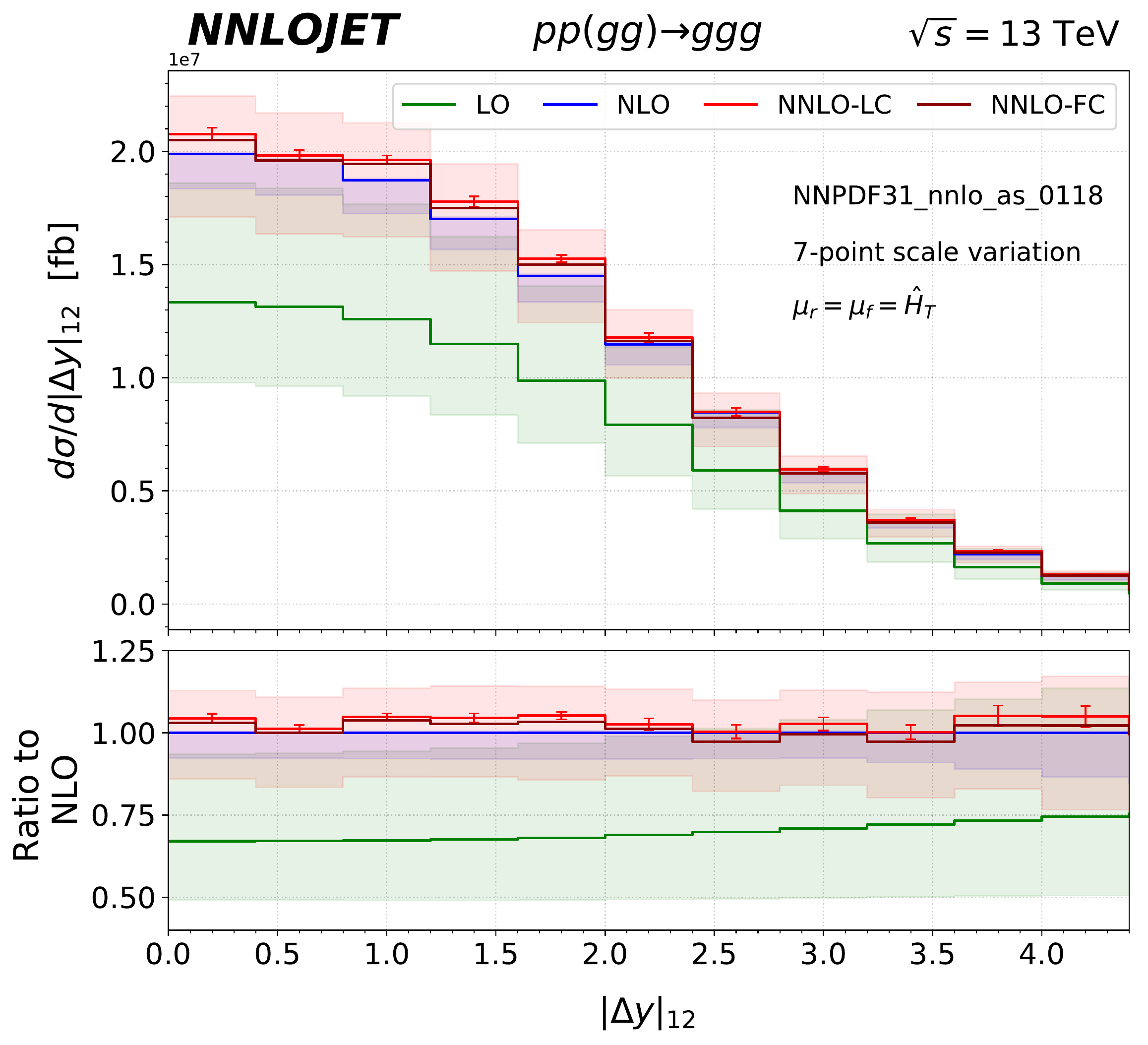}
		\includegraphics[width=0.32\textwidth]{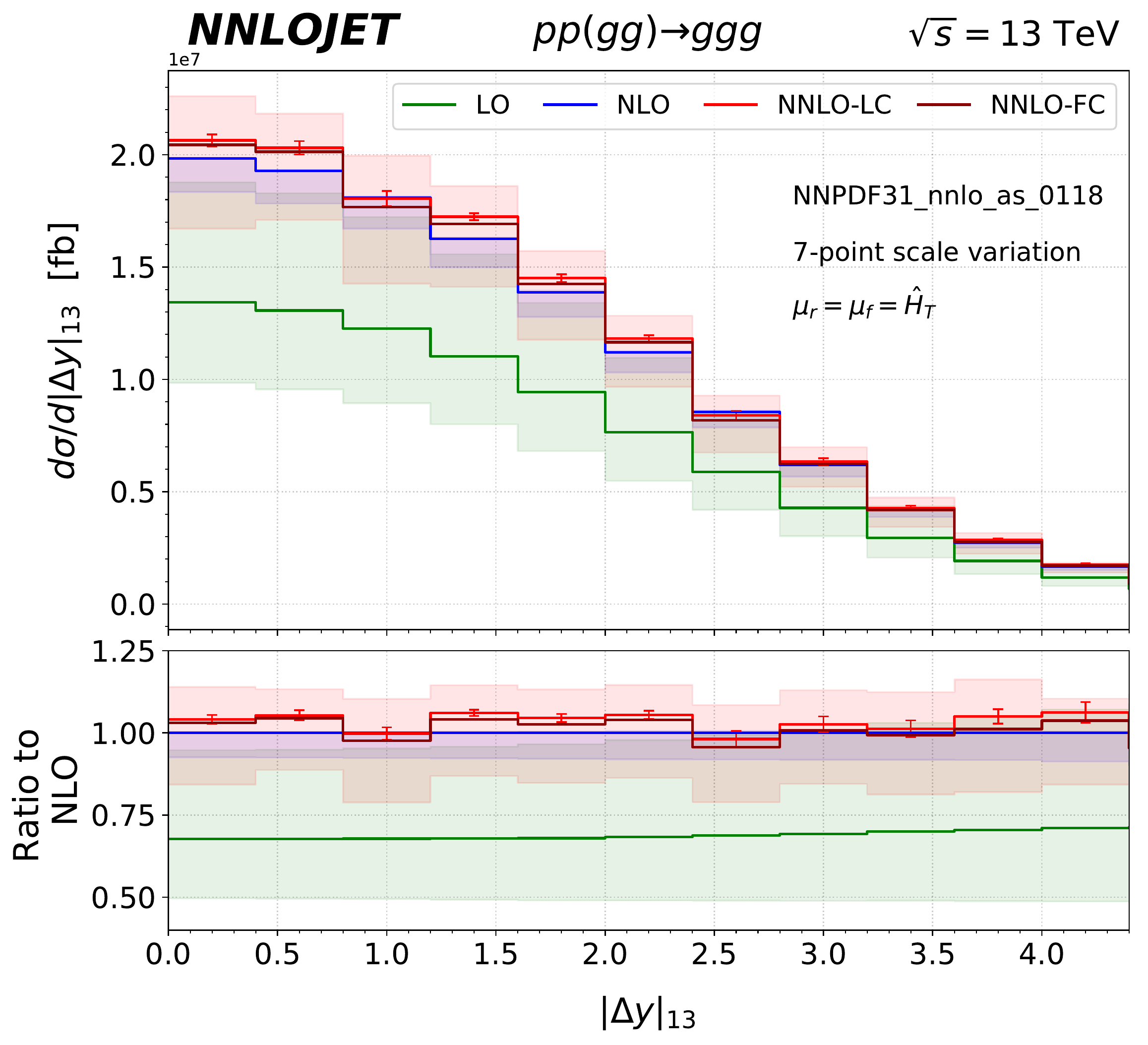}
		\includegraphics[width=0.32\textwidth]{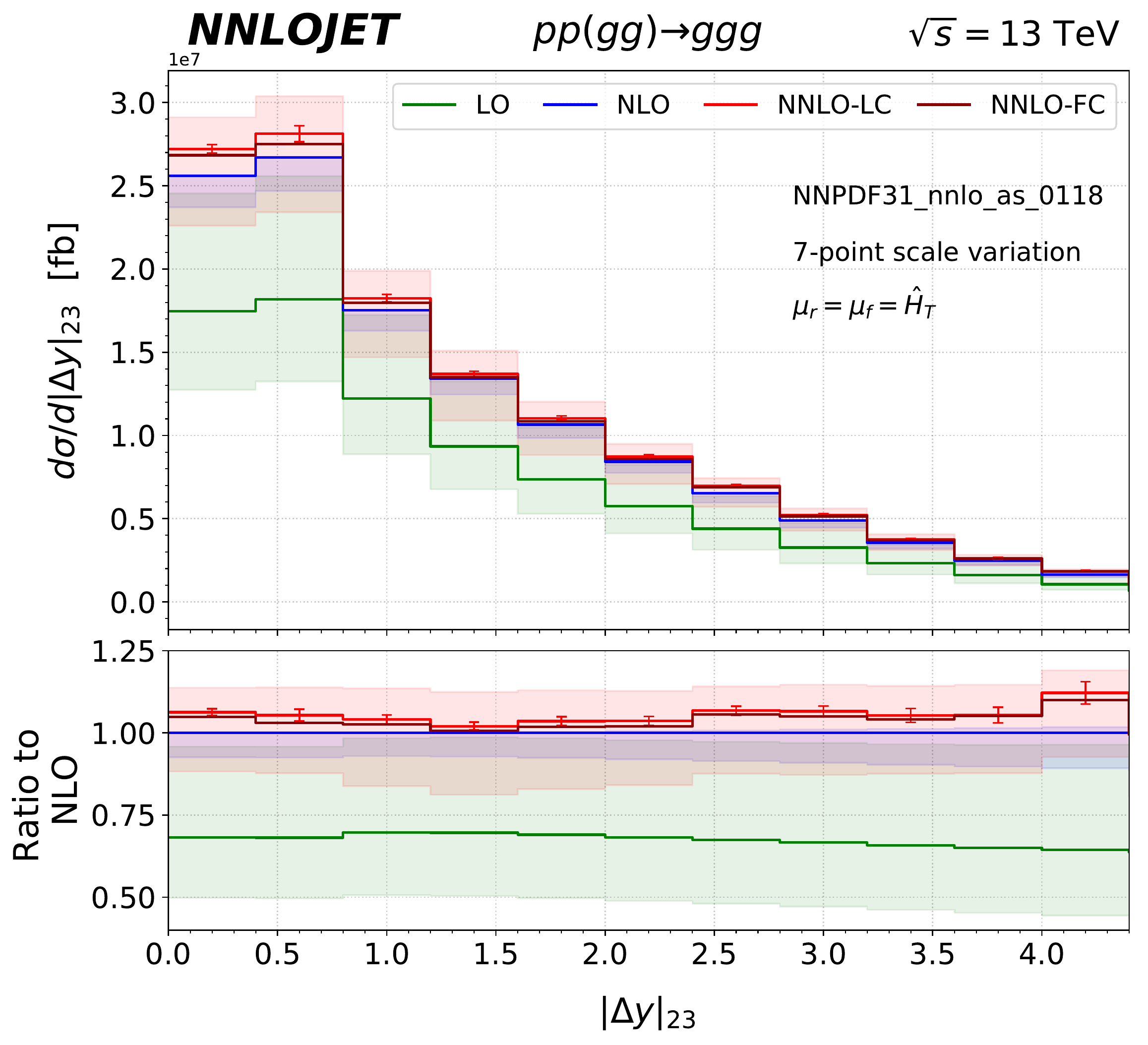}\\
		\includegraphics[width=0.32\textwidth]{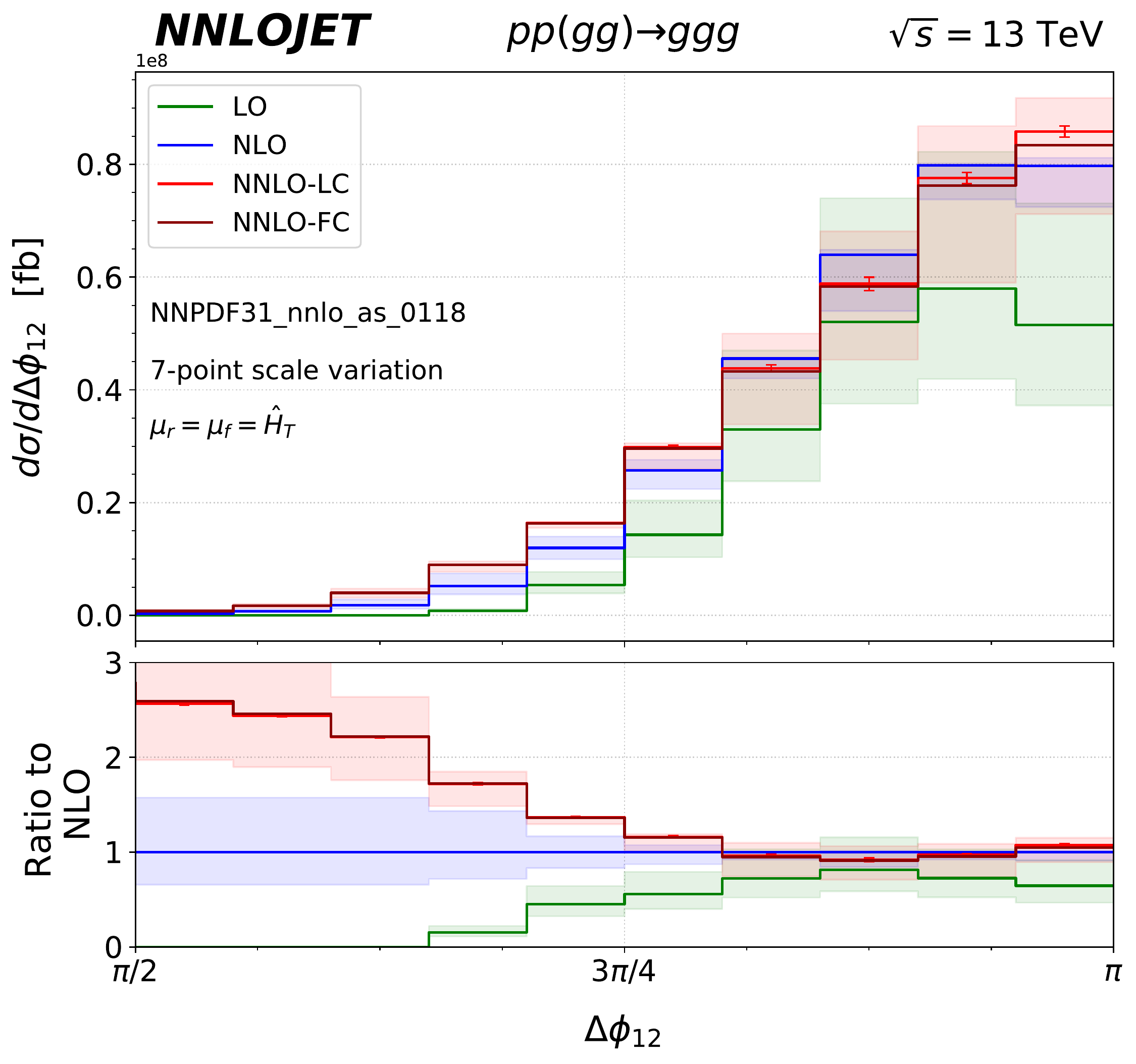}
		\includegraphics[width=0.32\textwidth]{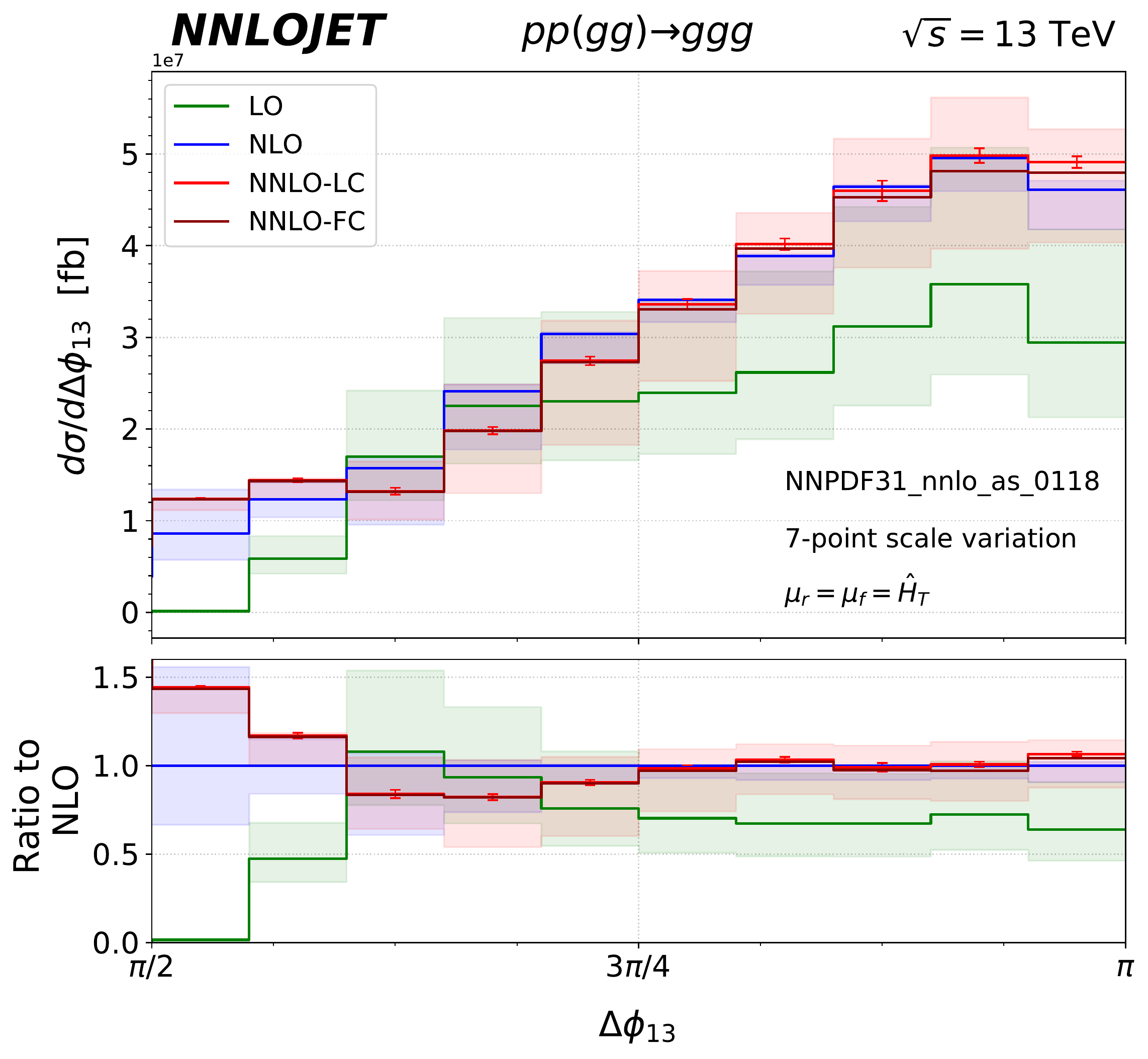}
		\includegraphics[width=0.32\textwidth]{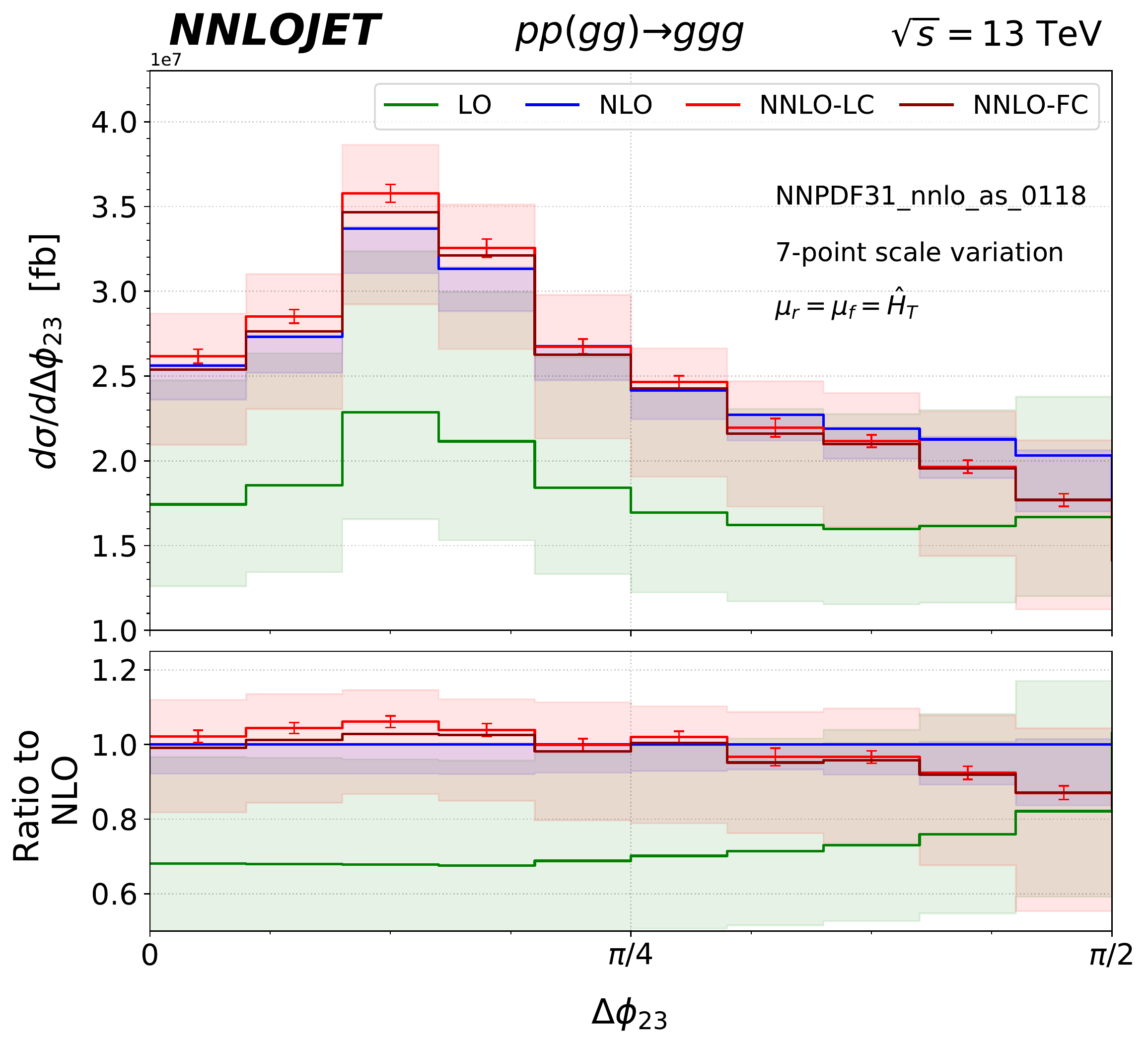}
	\caption{Differential distributions in geometrical jet-pair variables: rapidity differences 
	 (upper row) and 
	azimuthal angles (lower row) for gluons-only three-jet production up to NNLO. Definition of NNLO-LC and  
 NNLO-FC as in Figure~\protect\ref{fig:ht_m123}.}	\label{fig:geom}
	\end{figure}

Figures~\ref{fig:ht_m123}--\ref{fig:geom} show differential distributions related to gluons-only three-jet production in terms of various variables 
derived from the three-jet system (Figures \ref{fig:ht_m123} and \ref{fig:yvar}), from individual jets (Figure \ref{fig:ptj_yj}) and 
from jet pairs (Figure~\ref{fig:geom}) at LO, NLO and NNLO. The LO and NLO predictions are obtained at full colour and 
NNLO-LC corresponds to the NNLO predictions at leading colour, which is our default prediction that is also used to 
determine the scale uncertainty bands at NNLO. To estimate the potential impact of the yet incomplete 
subleading colour contributions, these were computed discarding the finite parts of the two-loop virtual corrections and added to the 
leading colour contributions (only for the central scale setting), yielding the NNLO-FC lines. In most distributions, the 
difference between NNLO-LC and NNLO-FC is at the per-cent level or below and can hardly be resolved, thereby indicating the small numerical 
impact of subleading colour effects at NNLO in gluonic jet production. 

	Except for the high-$H_T$ tail of the $H_T$ distribution, Figure~\ref{fig:ht_m123} (left), and 
the distributions in azimuthal opening angles, Figure~\ref{fig:geom} (lower panels), we observe the NNLO corrections to be quite moderate in all distributions, typically ranging between $\pm 10\%$. The error bars on the NNLO-LC central values indicate the numerical integration errors on 
the NNLO coefficients. 
Except for the very tails of the distributions, we managed to obtain quite small numerical integration errors on these NNLO coefficients, leading to 
total integration errors on the NNLO-LC predictions at the level of 2\% or below. These are already sufficient to establish the numerical convergence of
the predictions and  
could be lowered further with more Monte Carlo 
integration statistics.   

The differential three-jet distributions that we presented here demonstrate the applicability of the colourful antenna approach to the 
construction of NNLO subtraction terms for a highly non-trivial high-multiplicity process, and illustrate the quality of numerical convergence that 
can be obtained with these subtraction terms in the \textsc{NNLOjet} framework. 

As anticipated, a notable feature is the increase in the scale uncertainty bands from NLO to NNLO-LC that is observed across all distributions. This effect is an artefact of the gluons-only simplification and we further investigate it in the following section.

	\subsection{Comments on scale variation}\label{sec:scale_variation}
	
	In this section we investigate the anomalous scale variation behaviour observed in the NLO and NNLO correction to gluonic three-jet production. As pointed out above, the coupling constant $\alpha_s(\mu_r)$ sees the full content of QCD, since its determination relies on experimental measurements and scale evolution equations which incorporate the dependence on the number of light quark flavours. On the contrary, the matrix elements we implemented in our calculation only contain the gluonic degrees of freedom and are then evaluated for $N_f=0$. This mismatch breaks the interplay between the renormalization scale variation of the running coupling and the virtual corrections at higher orders. An analogous observation can be done for the factorization scale $\mu_f$. The scale dependence of the PDFs takes into account the presence of light quark flavours. To correctly compensate the PDF's factorization scale variation at higher orders in QCD, the inclusion of quark-induced channels is necessary. Therefore, the scale variation analysis should provide a consistent estimate of the uncertainty of theoretical predictions for a given process only when the entire set of subprocesses is considered and included in the final result. 
	
	We illustrate the effect of the inclusion of quarks on the scale variation analysis with a dedicated study on dijet production. We implement the same setup and cuts described in section~\ref{sec:results} and we compute the NNLO corrections with the traditional antenna subtraction 
	approach in the leading colour approximation~\cite{Currie:2013dwa,Currie:2017eqf}, for three different scenarios:
	
	\begin{itemize}
		\item $pp(gg) \to jj$ with $N_f=0$, gluons-only; 
		\item $pp(gg) \to jj$ with $N_f=5$, where we include the contribution coming from fermionic loops in virtual corrections as well as gluons splitting into quark-antiquark pairs in real emission corrections, but we keep the restriction to gluon-initiated subprocesses only;
		\item $pp \to jj$, where we consider the whole set of sub-processes that contribute to dijet production, both gluon- and quark-initiated.
	\end{itemize}
	
	The first scenario reflects the same implementation we used for gluonic three-jet production and we expect to observe a similar pattern, namely an anomalous size of the NNLO scale variation bands with respect to NLO. Allowing gluons to split into quark-antiquark pairs should restore the correct compensation at higher orders of the renormalization scale dependence of the strong coupling. Finally, with the inclusion of quark-induced sub-processes, we expect to correctly balance factorization scale variation effects due to the evolution of the PDFs.
	
	\begin{figure}
		\centering
		\includegraphics[width=0.5\textwidth]{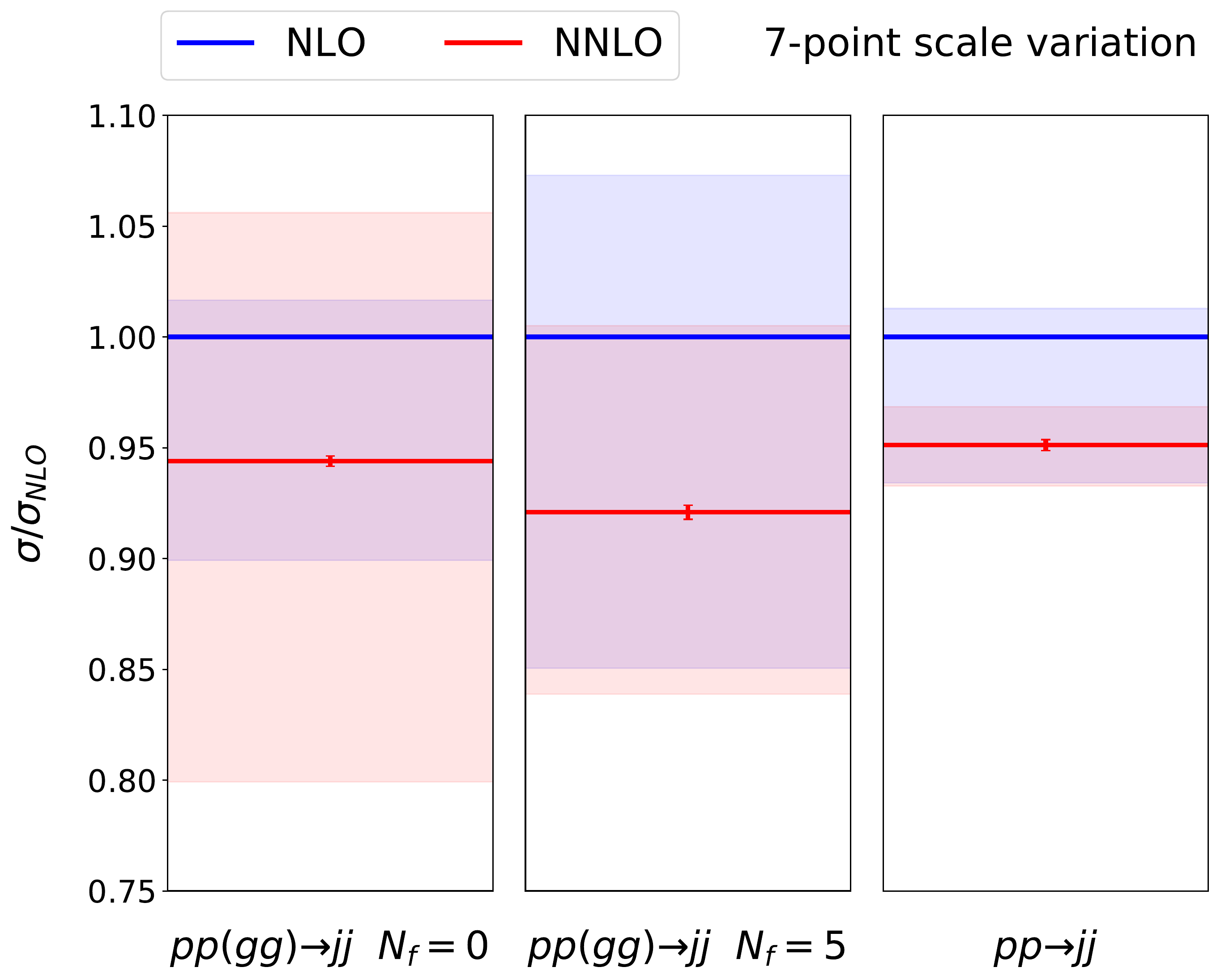}
		\caption{Total cross section for $pp(gg)\to jj$ with $N_f=0$ (left), $pp(gg)\to jj$ with $N_f=5$ (centre) and $pp\to jj$ (right). Each result is normalized with respect to the corresponding NLO total cross section.}\label{fig:tot_dijet}
	\end{figure} 
	
\begin{figure}[t]
		\centering
		\includegraphics[width=0.32\textwidth]{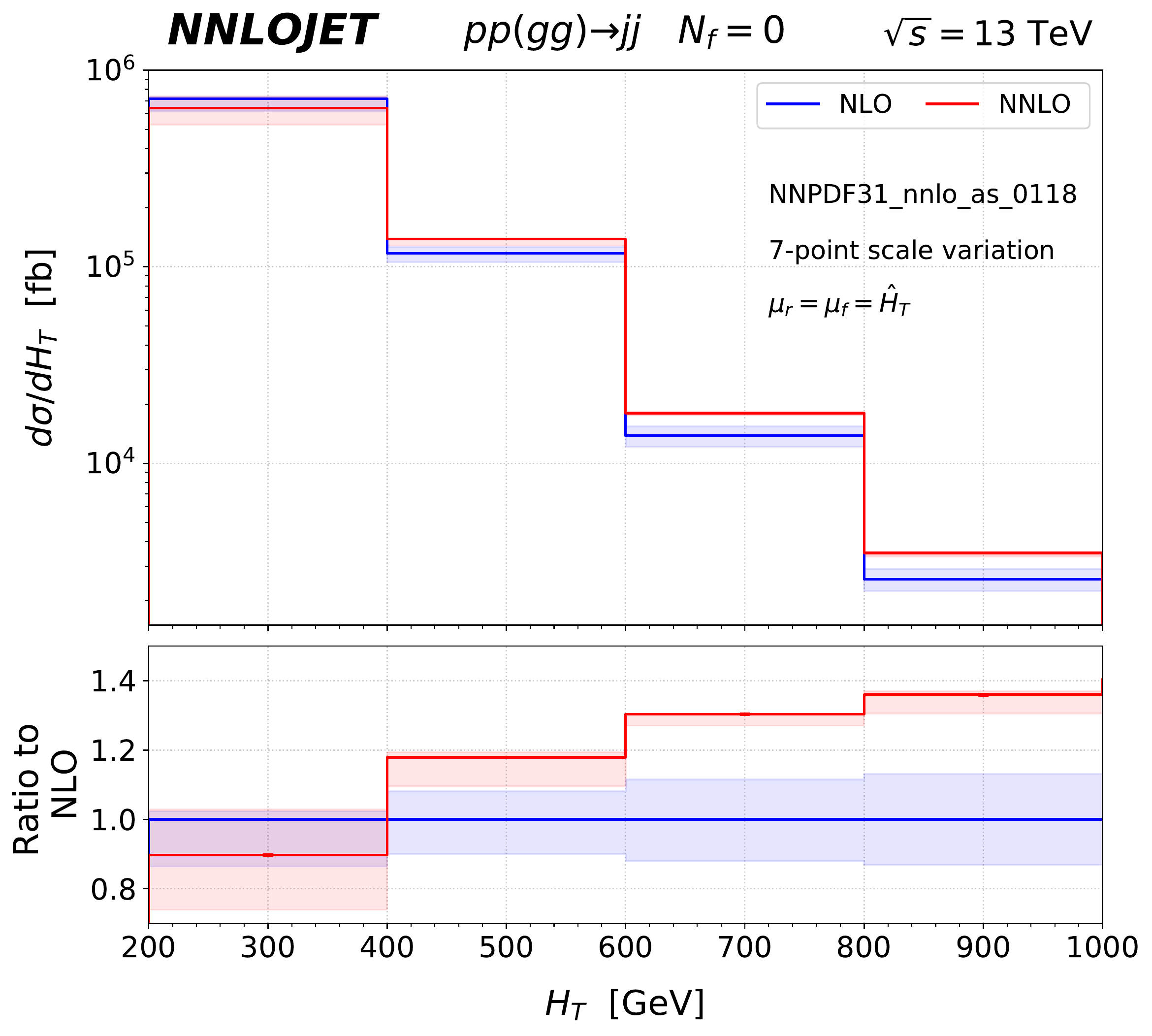}
		\includegraphics[width=0.32\textwidth]{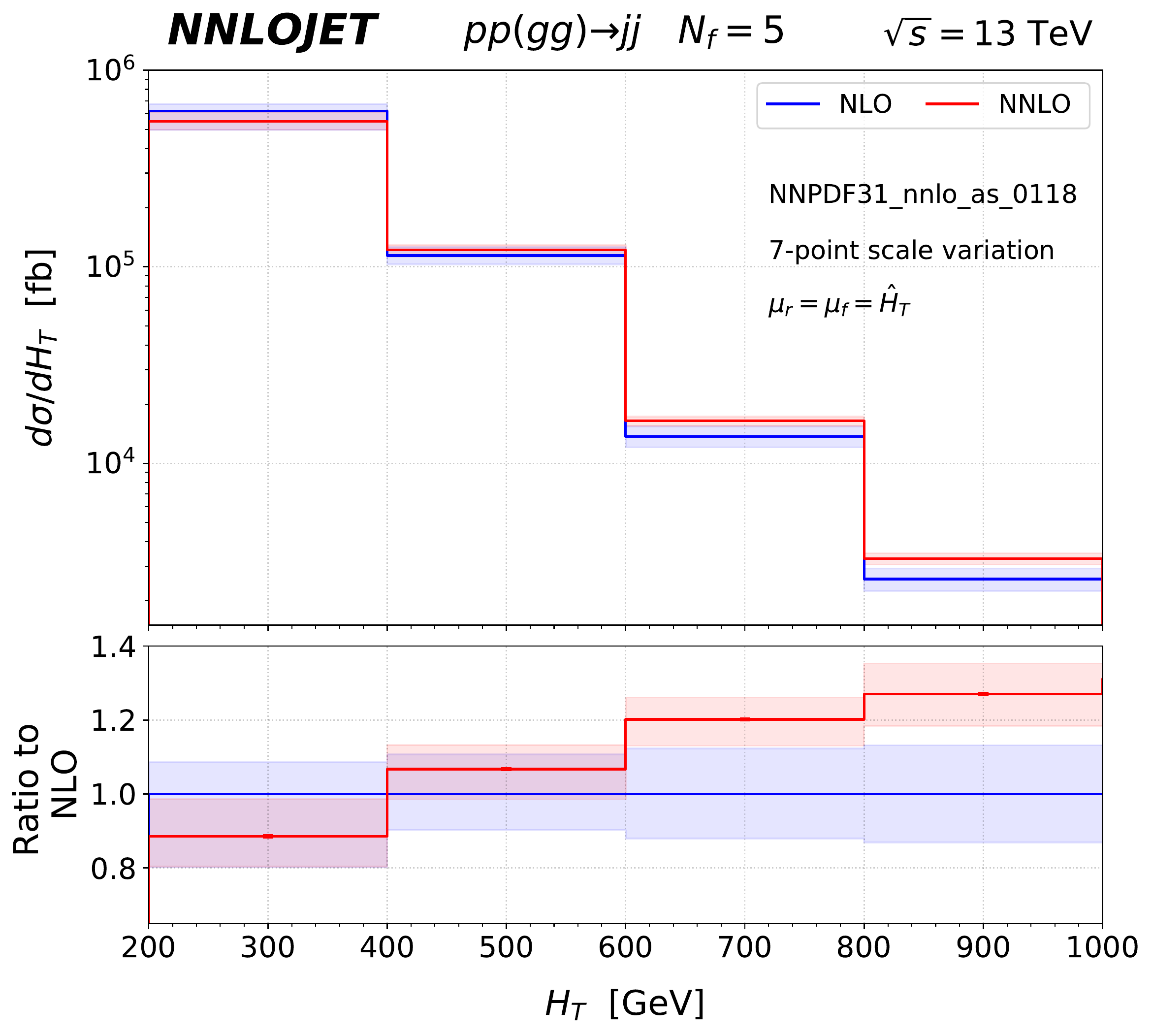}
		\includegraphics[width=0.32\textwidth]{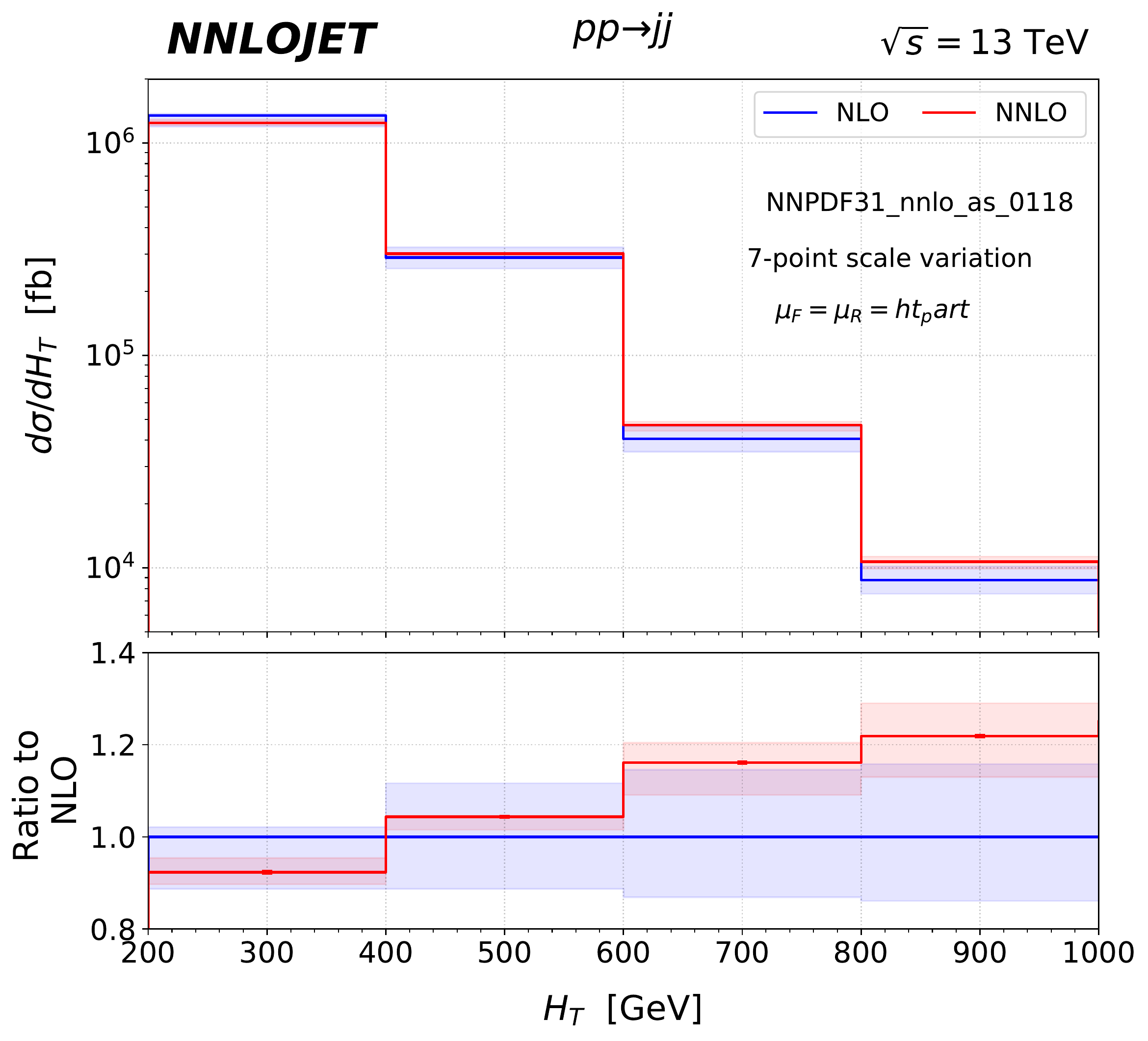}
		\caption{$H_T$ distribution for $pp(gg)\to jj$ with $N_f=0$ (left), $pp(gg)\to jj$ with $N_f=5$ (centre) and $pp\to jj$ (right).}\label{fig:ht_dijet}
	\end{figure}

	In Figure~\ref{fig:tot_dijet} we compare the total cross section in the three cases, normalized with respect to the corresponding NLO result. As expected, the size of the NNLO scale variation bands with respect to the NLO ones significantly reduces when the full QCD degrees of freedom are considered. This can also be noticed in the $H_T$ distributions in Figure~\ref{fig:ht_dijet}. The total cross section is dominated by the low-$H_T$ region, for which we observe the same scale variation pattern as in the total cross section. 
	
	\begin{figure}[t]
		\centering
		\includegraphics[width=0.32\textwidth]{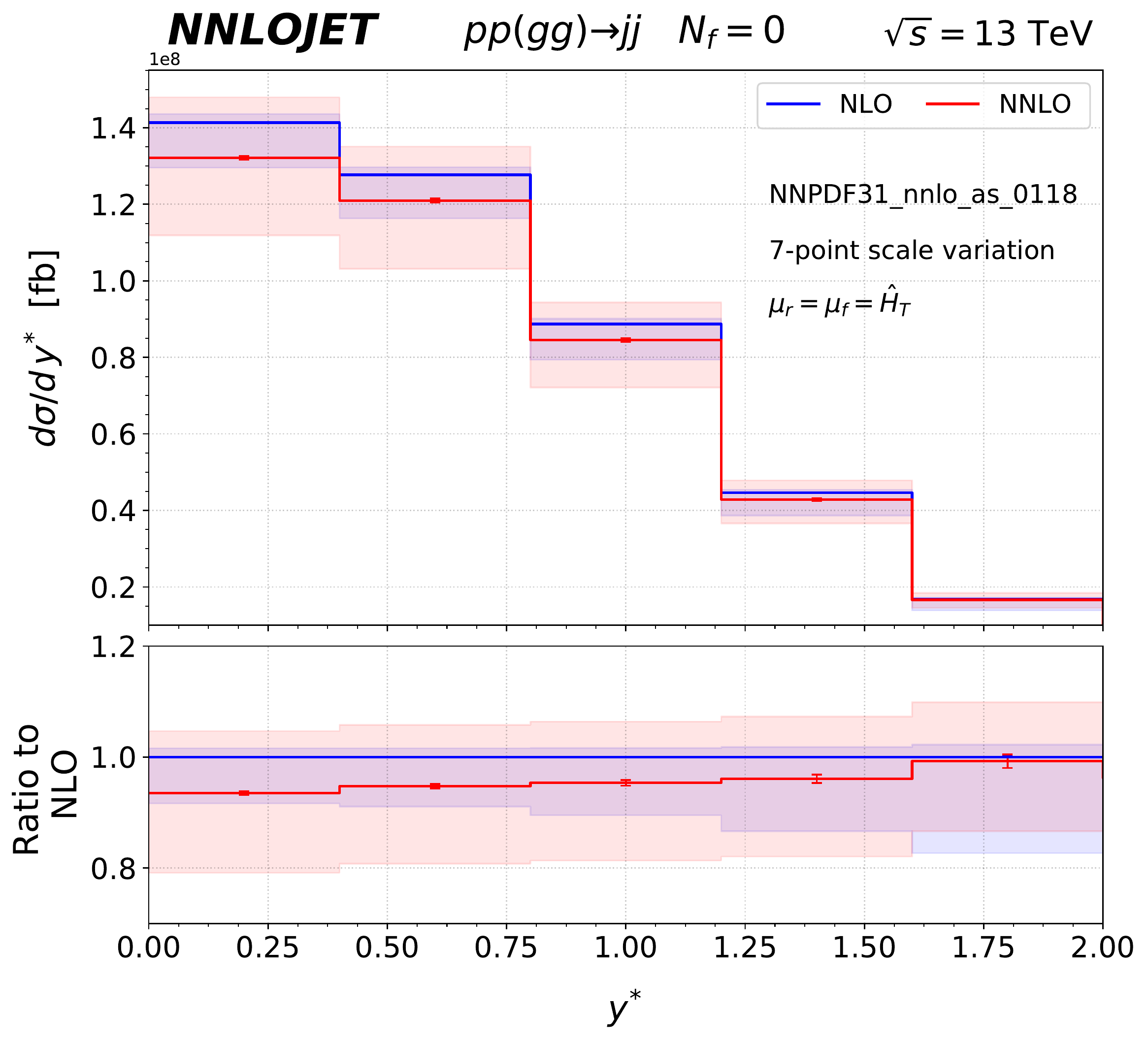}
		\includegraphics[width=0.32\textwidth]{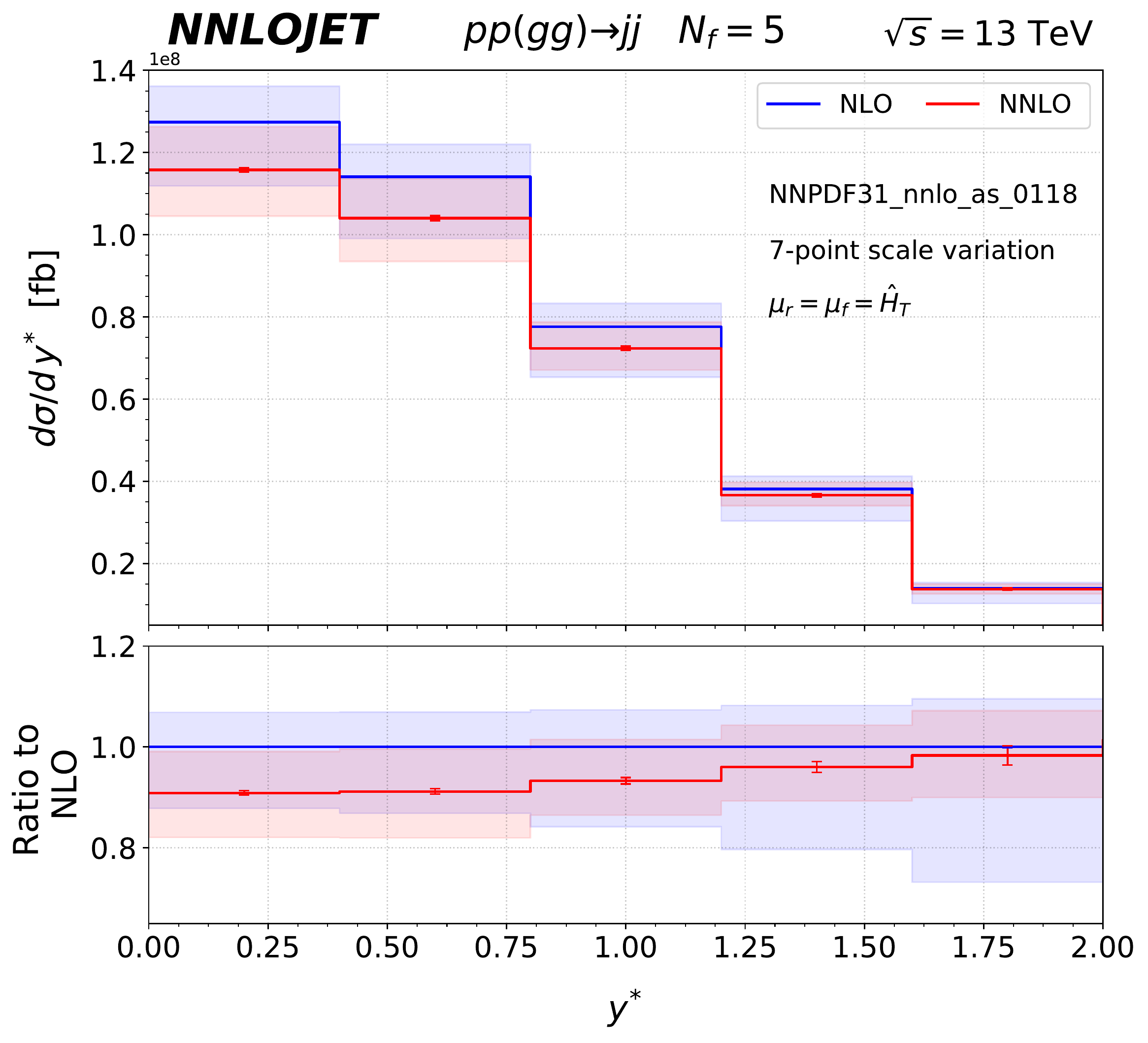}
		\includegraphics[width=0.32\textwidth]{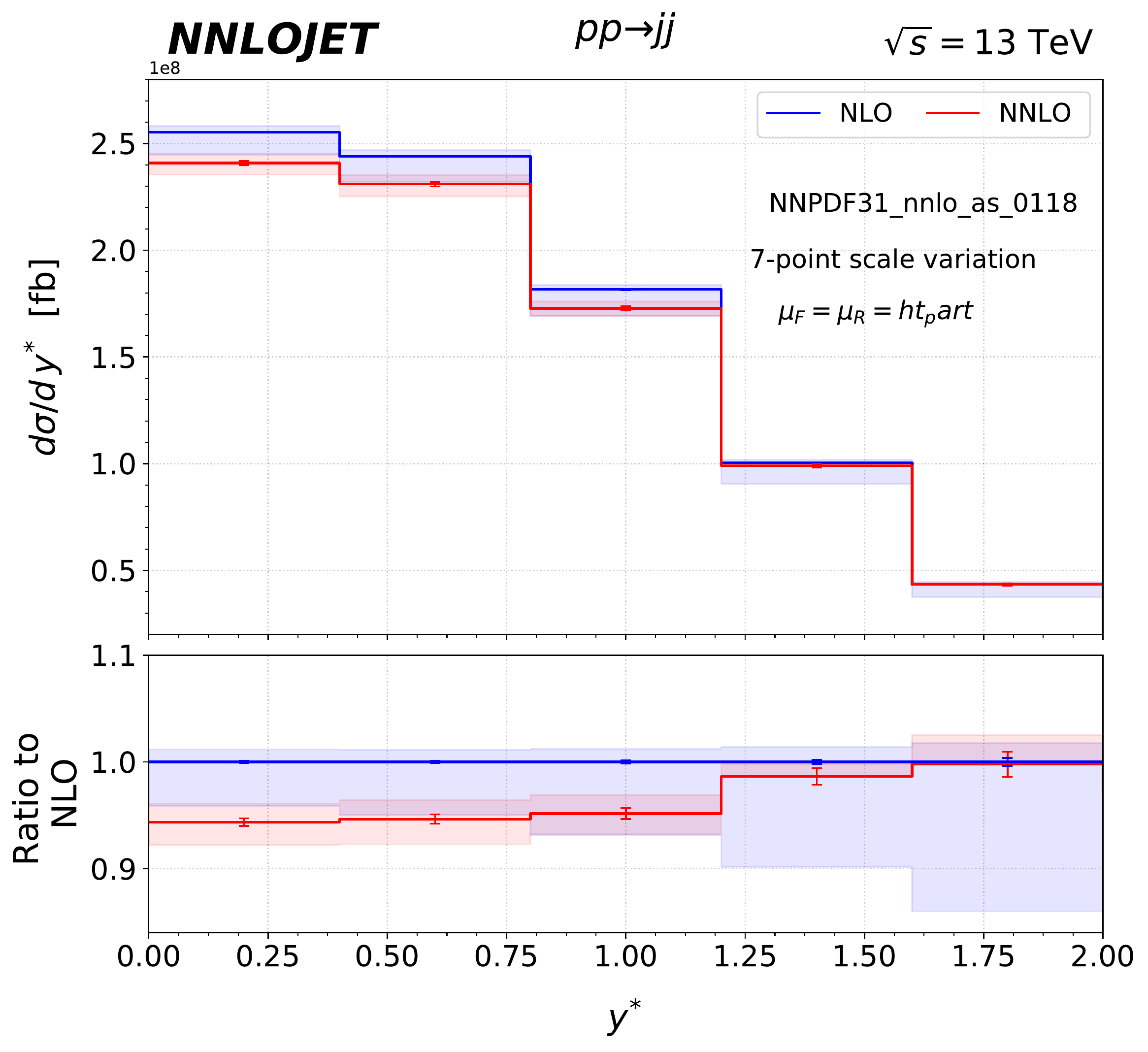}
		\caption{$y^*$ distribution for $pp(gg)\to jj$ with $N_f=0$ (left), $pp(gg)\to jj$ with $N_f=5$ (centre) and $pp\to jj$ (right).}\label{fig:ystar_dijet}
	\end{figure}
	\begin{figure}[t]
		\centering
		\includegraphics[width=0.32\textwidth]{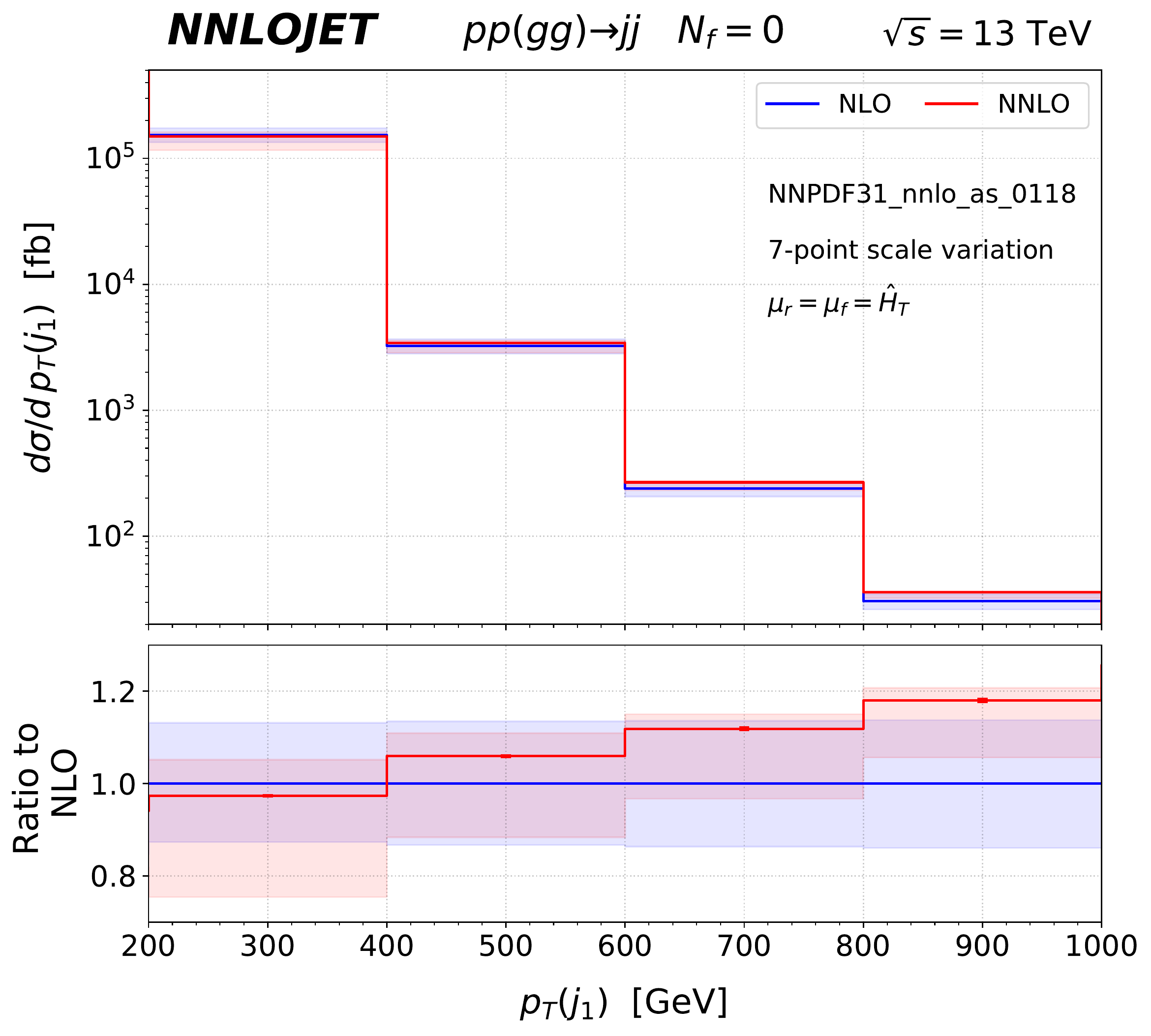}
		\includegraphics[width=0.32\textwidth]{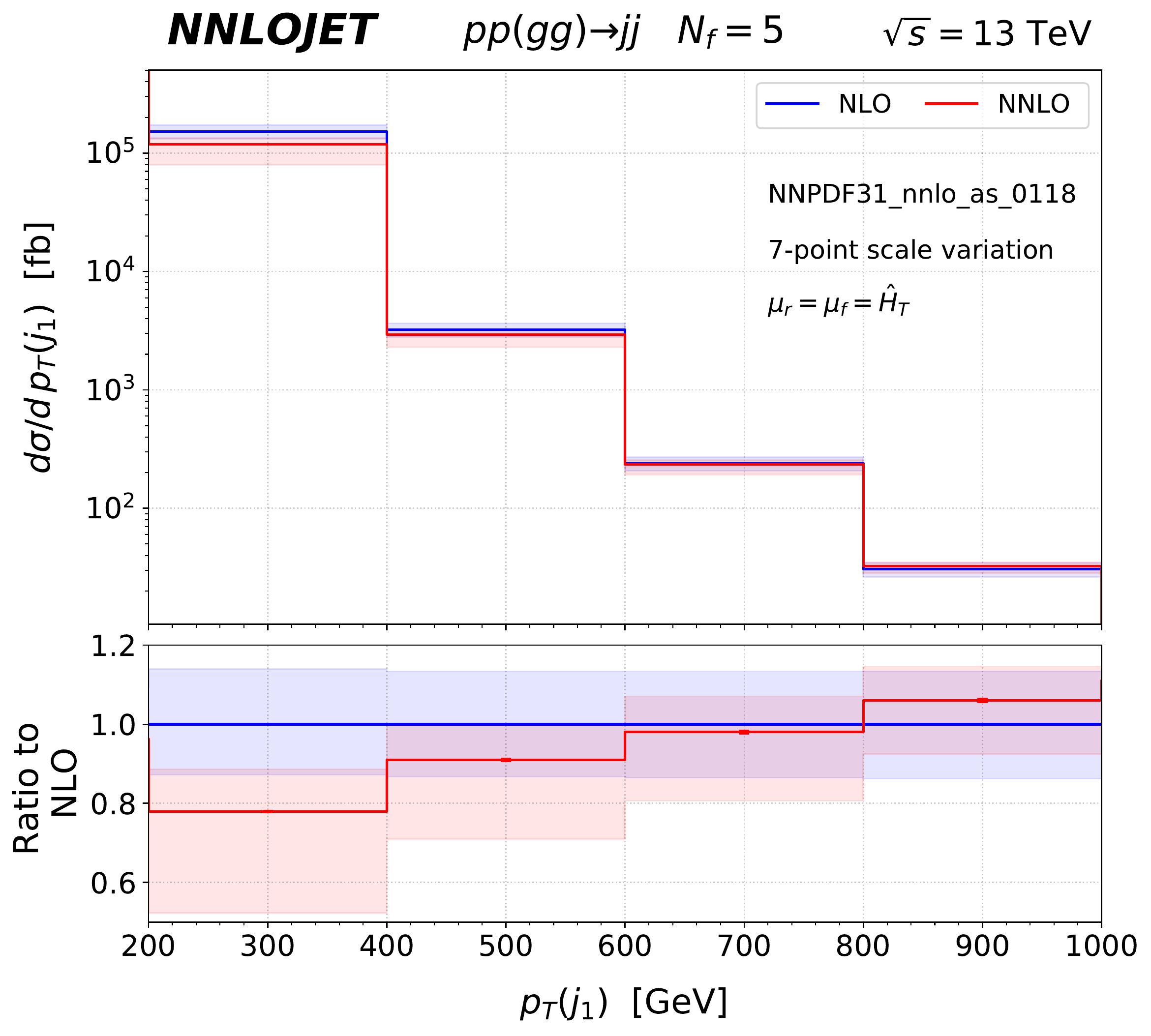}
		\includegraphics[width=0.32\textwidth]{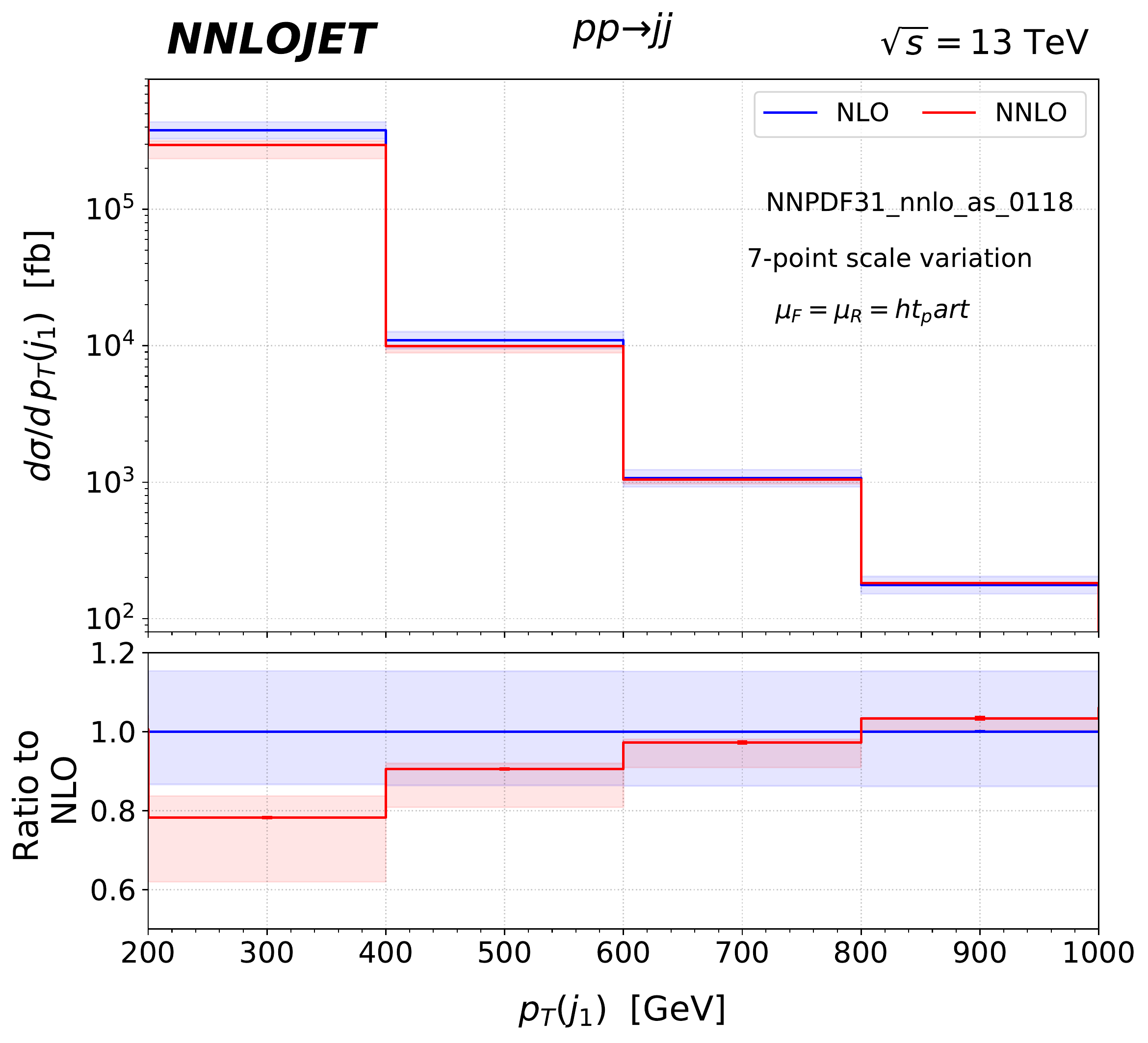}
		\caption{$p_T(j_1)$ distribution for $pp(gg)\to jj$ with $N_f=0$ (left), $pp(gg)\to jj$ with $N_f=5$ (centre) and $pp\to jj$ (right).}\label{fig:ptj1_dijet}
	\end{figure}

	Analogously, we have a similar adjustment of the scale variation bands in observables which are largely dominated by the low-$H_T$ region, such as $y^*=|y(j_1)-y(j_2)|/2$, as shown in Figure~\ref{fig:ystar_dijet}. For the $p_T(j_1)$ distribution in Figure~\ref{fig:ptj1_dijet}, the transition from $pp(gg)\to jj$ with $N_f=0$ to $pp(gg)\to jj$ with $N_f=5$ does not seem to reduce the relative size of the NLO and NNLO scale variation bands to the same extent observed in other distributions. However, this happens if the NNLO correction to the full process $pp \to jj$ is considered, which indicates that $\mu_f$ variation effects dominate in the studied range of $p_T(j_1)$.

	These results demonstrate the inconsistency of the scale variation analysis when specific sub-channels of a given process are considered. The effect on differential distributions is not uniform and may vary according to the considered observable, range, kinematical cuts and choice of central scale. We can confidently conclude that the considerably larger scale variation bands observed at NNLO in the differential distributions in section \ref{sec:results} are an artefact of missing fermionic contributions and quark-induced sub-processes.

	\section{Conclusions and outlook}
	\label{sec:conc}
	
	The antenna subtraction method for the construction of infrared subtraction terms at NLO and NNLO has originally been formulated 
	based on colour-ordered amplitudes and squared matrix elements. This formulation results in very compact subtraction terms for 
	low-multiplicity processes, but extends very poorly to higher multiplicities and requires complicated constructions beyond the leading-colour 
	contributions. This paper overcomes these constrictions by reformulating the antenna subtraction in a new colour space basis. The resulting 
	colourful antenna subtraction method allows for a systematic construction of antenna subtraction terms. Starting from an analysis 
	of the colour structure of purely virtual corrections at NLO and NNLO, we were able to devise an algorithm that translates 
	infrared poles of virtual corrections into real radiation dipole insertions that constitute the subtraction terms for single real radiation up to one-loop
	level and for double real radiation at tree level. We remark once again that the main advantage of this perspective consists in avoiding the direct treatment of the divergent behaviour of real emission corrections. In the case of an NNLO calculation, this feature represents a major simplification, since the double real subtraction term is obtained as the last step of a completely automatable procedure, with no need of dealing with the involved infrared structure of double real radiation matrix elements. 
	
	We fully formulated the colourful antenna subtraction method for gluons-only processes and automated its workflow. As proof-of-principle 
	applications, we rederived the NNLO antenna subtraction terms for gluons-only dijet production~\cite{Currie:2013dwa} and newly derived 
	these terms of gluons-only three-jet production. We verified the pointwise convergence of these subtraction terms towards the respective matrix elements and demonstrated the numerical stability of the predictions in computing various differential three-jet distributions. As a by-product, we
	also assessed the reliability of the gluons-only approximation to dijet production, demonstrating in particular the large impact of missing 
	fermionic contributions and quark-induced subprocesses onto the renormalization and factorization scale dependence of the predictions. The 
	gluons-only predictions for three jet production are thus not yet of phenomenological relevance, but should only be considered as 
	proof-of-principle application of the newly developed coloured antenna subtraction method and of its automated implementation. 
	We also were able to quantify the numerical impact of subleading colour contributions at NNLO, which were found to be 
	small throughout most distributions, thus lending support to a recent computation of NNLO corrections to three-jet production at leading 
	colour~\cite{Czakon:2021mjy} obtained with a residue subtraction method.

	The next step towards a complete, process-independent subtraction scheme at NNLO consists in the inclusion of quarks in the presented formalism. To treat these contributions in the context of the colourful antenna subtraction approach, a crucial distinction must be made between the identity-preserving (IP) and identity-changing (IC) sectors of the subtraction infrastructure. The IP sector contains infrared limits for which the real emission corrections factorize onto a lower multiplicity process with the same initial-state parton species. On the contrary, the IC sector refers to the configurations where a final-state quark becomes collinear to an initial-state gluon or same-flavour quark, effectively changing the initial-state parton species of the associated reduced matrix element in the collinear limit. 
	
	After integration over the unresolved radiation, the IP contributions generate $\e$-poles which cancel against the explicit singularities of the corresponding virtual contributions and IP mass factorization counterterms. Therefore, the whole IP sector would be systematically constructed at NNLO following the procedure summarized in Figure~\ref{fig:colant_scheme}, in complete analogy to what was done in the simplified case of pure gluonic scattering. Of course, appropriate colour stripped one- and two-loop integrated dipoles need to be defined to include quark-antiquark and quark-gluon configurations, as well as the correct translation between integrated and unintegrated antenna functions. Nevertheless, these steps do not represent a major issue and most of the ingredients for the extension to the IP sector of sub-processes involving quarks have already been constructed and assessed. 
	
	The IC infrared divergences cancel against $\e$-poles in IC mass factorization counterterms and do not communicate with the singularity structure of the virtual corrections. For this reason, these contributions can not be straightforwardly generated with the method we described in this paper, 
	thus requiring an extension of the approach.  The starting point for the construction of the IC sector will be the IC mass factorization structure at one  and two loops, which can be written down in a general way from~\eqref{MF},~\eqref{MFVV} and~\eqref{MFRV}. The main issue is then to dress the splitting kernels with the appropriate IC integrated antenna functions, in such a way suitable universal IC integrated dipoles can be constructed and the entire generation of the subtraction layers can proceed in the same way as it does for the IP sector. At NLO, IC integrated dipoles have already been defined, for example in~\cite{Currie:2013vh}, while at NNLO this task still requires additional work.

	\section*{Acknowledgements}
	
	We thank James Currie, Aude Gehrmann-De Ridder, Jonathan Mo and Joao Pires for valuable discussions and contributions to the jet production processes in \textsc{NNLOjet}. We are grateful to Jean-Nicolas Lang and Jonas Lindert for the support with \textsc{OpenLoops} and to Vasily Sotnikov for the assistance in the usage of the C\texttt{++} library for the five-point two-loop finite remainder.
	This work has received funding from the Swiss National Science Foundation (SNF) under contract 200020-204200 and from the European Research Council (ERC) under the European Union's Horizon 2020 research and innovation programme grant agreement 101019620 (ERC Advanced Grant TOPUP). This work is also supported in part by the Deutsche Forschungsgemeinschaft (DFG, German Research Foundation) under grant 396021762-TRR 257.



	\bibliographystyle{JHEP}
	\bibliography{bib_col_ant}

\end{document}